\documentclass{article}
\usepackage[utf8]{inputenc}

\title{Frequentist-Bayes Hybrid Covariance Estimation for Unfolding Problems}
\author{Pim Jordi Verschuuren}
\date{October 2021}

\usepackage[square,sort,comma,numbers]{natbib}
\usepackage{graphicx}
\usepackage{amsmath}
\usepackage{hyperref}
\usepackage{subcaption}
\usepackage{bbold}
\usepackage{tabularx}
\usepackage{hhline}
\usepackage{siunitx}
\begin{document}

\maketitle

\begin{abstract}
	In this paper we present a frequentist-Bayesian hybrid method for estimating covariances of unfolded distributions using pseudo-experiments. The method is compared with other covariance estimation methods using the unbiased Rao-Cramér bound (RCB) and frequentist pseudo-experiments. We show that the unbiased RCB method diverges from the other two methods when regularization is introduced. The new hybrid method agrees well with the frequentist pseudo-experiment method for various amounts of regularization. However, the hybrid method has the added advantage of not requiring a clear likelihood definition and can be used in combination with any unfolding algorithm that uses a response matrix to model the detector response.
\end{abstract}

\section{Introduction}
In High Energy Physics (HEP) and in many other fields one often measures distributions of quantities such as particle energies or other characteristics of observed events. Because the experimental apparatus (the “detector”) inevitably has a limited resolution, the measured (or “reconstructed”) value of the quantity in question will differ in general from its true value. This results in a distortion or smearing of the measured distribution relative to what would be obtained if the detector had perfect resolution. The statistical procedure of estimating the true distribution from the directly measured one is usually called unfolding in HEP, or deconvolution in many other fields. Unfolding algorithms and their implementation in software have been widely discussed in HEP (see, e.g., Refs. \cite{cowan1998statistical,Blobel:2011fih,Behnke:2013pga,Cowan:2002in,Kuusela:2015xqa,spanoreview:2013,Zech:2016gca}).\\
\\
\noindent
Estimating a true distribution will result in one or more estimators which have an uncertainty and non-trivial correlations between them. Estimating these uncertainties and correlations is an important part of any unfolding framework. Many unfolding algorithms supply an estimator for statistical covariances \cite{2010arXiv1010.0632D, Schmitt:2012kp, Hocker:1995kb,Malaescu:2011yg,Bozson:2018asz} i.e. covariances as a result of the stochastic nature of the data and the bin-to-bin migrations of events. However, in HEP it is very common to also have systematic sources of error which induce additional uncertainties and correlations. These systematic sources of error originate from estimates or assumptions that enter the unfolding framework, eg. energy resolution or reconstruction efficiencies, but have limited accuracy. The goal is to include all statistical and systematic sources of error in unfolding covariance estimation.\\
\\
\noindent
In Sec. \ref{sec:math} we provide a mathematical description of the unfolding problem and the inclusion of nuisance parameters to model systematic sources of error. In Sec. \ref{sec:cov} we present two conventional and one novel approach to covariance matrix estimation. In Sec. \ref{sec:comp} a comparison study between the three methods is presented and conclusions are given in Sec. \ref{sec:conc}.

\section{Mathematics of unfolding}\label{sec:math}
A description of the mathematics behind unfolding is given here. The definition of the unfolding problem is given in Sec. 2.1 and the inclusion of nuisance parameters is described in Sec. 2.2. For a more detailed description of the unfolding problem see  book and for more information on nuisance parameters and their use in HEP see Ref. \cite{Cranmer:2015nia,Cousins:2018tiz,Cowan:2013pha}. 

\subsection{Definition of the unfolding problem}
Lets assume we have a counting experiment where each event has some variable $x$ that differs from its true value because of detector effects. One can construct a histogram $\vec{n} = (n_{1},...,n_{N})$ from the measured values. These bin values follow a probability distribution, often taken to be Poisson with expectation values $E[\vec{n}] = \vec{\nu} = (\nu_{1},...,\nu_{N})$. The relationship between these and their corresponding true values is given by 

\begin{equation}\label{eq:nu}
    \nu_{i} = \sum_{j=1}^{M} R_{ij}\mu_{j} + \beta_{i}
\end{equation}

\noindent
with $\vec{\mu}=(\mu_{1},...,\mu_{M})$ is the histogram filled with values of variable $x$ if it would be measured perfectly i.e. without any detector effects. The response matrix $R_{ij}$ gives the conditional probability to measure an event in bin $i$ of the measured histogram if the true value was in bin $j$ of the true histogram. Additionally, the data consists of events from the \textit{signal process} and from one or more irreducible \textit{background processes}. The expected bin values for the total background are denoted by $\vec{\beta}=(\beta_{1},...,\beta_{N})$. The goal of unfolding is to estimate the parameters $\vec{\mu}$. The likelihood is a product of p.d.f.s $f(n_{i}|\vec{\mu})$ that define how the data $\vec{n}$ is distributed under the assumption of parameters $\vec{\mu}$. A common p.d.f. choice is the Poisson distribution which results in 

\begin{equation}\label{eq:likmu}
    L(\vec{\mu}) =  \prod_{i=1}^{N}f(n_{i}|\vec{\mu}) = \prod_{i=1}^{N}\frac{\nu_{i}^{n_{i}}}{n_{i}!}e^{-\nu_{i}}
\end{equation}

\noindent
with $\nu_{i}$ depending on $\vec{\mu}$ according to equation \ref{eq:nu}. However, other p.d.f. choices such as a multivariate Gaussian are also possible. Maximizing Eq. \ref{eq:likmu} w.r.t. $\vec{\mu}$ will give the Maximum-Likelihood Estimators(MLE) of the true histogram. However, these solutions can have very large variances. One can suppress this by imposing some form of regularization by making a linear combination of the likelihood function and some regularization function $S(\vec{\mu})$.

\begin{equation}\label{eq:phimu}
    \Phi(\vec{\mu}) = L(\vec{\mu}) + \tau S(\vec{\mu})
\end{equation}

\noindent
Maximizing Eq. \ref{eq:phimu} w.r.t. $\vec{\mu}$ will result in the Regularized Maximum-Likelihood Estimators(RMLE). For any two estimators, the covariance is defined as 

\begin{equation}\label{eq:covar}
    U_{ij} \equiv \mbox{cov}[\hat{\mu}_{i}, \hat{\mu}_{j}] = E[(\hat{\mu}_{i} - E[\hat{\mu}_{i}])(\hat{\mu}_{j} - E[\hat{\mu}_{j}])]
\end{equation}

\noindent
and the bias as

\begin{equation}\label{eq:bias}
    b_{i} = E[\hat{\mu}_{i}] - \mu_{i}
\end{equation}

\noindent
with $E[x]$ denoting the expected value of a random variable $x$ and the diagonal elements $U_{ii}$ being the variances of $\hat{\mu}_{i}$. The RMLEs will have smaller variance $U_{ii}$ but larger bias $b_{i}$ w.r.t. the MLEs caused by the regularization function.

\subsection{Nuisance parameters}
In general, the response matrix $R$ and backgrounds $\vec{\beta}$ depend on additional parameters $\vec{\theta} = (\theta_{1},...,\theta_{K})$ introduced by properties of the detector response also known as \textit{nuisance parameters}. One influence this will have is that the increased likelihood model flexibility will reduce the bias but increase the variance. One should therefore take care when removing, also known as pruning, or introducing nuisance parameters into the model. Up until now, we assumed that these nuisance parameters, and therefore the response matrix and background distribution, were known with negligible uncertainty. However, in practice this is not valid and one needs to propagate these into the uncertainty on the unfolded distribution. By expressing the response matrix and background distribution as a function of $\vec{\theta}$ one includes the nuisance parameters in the likelihood function and thus incorporating systematic uncertainties in the model. Their correlations with the parameters of interests $\vec{\mu}$ will inflate the variance of the estimators $\hat{\vec{\mu}}$.  Additionally, the best estimates $\tilde{\theta}_{k}$ of $\theta_{k}$ are treated as an \textit{auxiliary measurement} which follow some probability distribution function $g(\tilde{\theta}_{k}|\theta_{k})$ that assumes some value $\theta_{k}$. A common choice for these p.d.f.s is a Gaussian which will result in

\begin{equation}\label{eq:likmutheta}
    L(\vec{\mu},\vec{\theta}) = \prod_{i=1}^{N} f(n_{i}|\vec{\mu},\vec{\theta}) \prod_{k=1}^{K} g(\tilde{\theta}_{k}|\theta_{k}) = \prod_{i=1}^{N}\frac{\nu_{i}^{n_{i}}}{n_{i}!}e^{-\nu_{i}}\prod_{k=1}^{K}\frac{1}{\sigma_{\tilde{\theta}_{k}}\sqrt{2\pi}}e^{-\frac{1}{2}\frac{(\theta_{k} - \tilde{\theta}_{k})^{2}}{\sigma^{2}_{\tilde{\theta}_{k}}}}
\end{equation}

\noindent
with $\nu_{i}$ now depending on both $\vec{\mu}$ and $\vec{\theta}$. However, other choices such as a log-normal or Student's t distribution are also possible as p.d.f. for the auxiliary measurements. One can construct a new linear combination of Eq. \ref{eq:likmutheta} and some regularization function $S(\vec{\mu})$. 

\begin{equation}\label{eq:phimutheta}
    \Phi(\vec{\mu},\vec{\theta}) = L(\vec{\mu},\vec{\theta}) + \tau S(\vec{\mu})
\end{equation}

\noindent
A common choice of regularization function $S(\vec{\mu})$ is a discretized measure of smoothness of the true distribution also known as Tikhonov regularization \cite{PhiReg,TikReg}. 

\begin{equation}\label{eq:tik}
    S(\vec{\mu}) = - \sum_{i=1}^{M-2}(-\mu_{i} + 2\mu_{i+1} -\mu_{i+2})^{2}
\end{equation}

\noindent
In the rest of the paper we will assume that we have constructed estimators $\hat{\vec{\mu}}$ and $\hat{\vec{\theta}}$ by maximizing Eq. \ref{eq:phimutheta} with regularization function Eq. \ref{eq:tik} w.r.t. $\vec{\mu}$ and $\vec{\theta}$.

\section{Covariance Estimation}\label{sec:cov}
This section describes three different methods to estimate the covariance matrix $U$ of the estimators $\hat{\vec{\mu}}$.

\subsection{Inverse Hessian}
Let us put both the parameters of interest $\vec{\mu}$ and the nuisance parameters $\vec{\theta}$ into one single vector $\vec{\lambda}=(\vec{\mu},\vec{\theta})$. The Cramer-Rao Bound(RCB) \cite{rao,cramer}, also known as the Minimum Variance Bound(MVB), states that the covariance between two estimators $\hat{\lambda}_{i}$ and $\hat{\lambda}_{j}$ has a lower bound set by the following inequality

\begin{equation}\label{eq:rcb}
    V[\hat{\lambda}_{i}, \hat{\lambda}_{j}] \geq \Big((\mathbb{1} - B)I^{-1}(\mathbb{1} - B)^{T}\Big)_{ij}
\end{equation}

\noindent
with $\mathbb{1}$ being the identity matrix and $B$ being the bias gradient matrix with $B_{ij}=\frac{\partial b_{\hat{\mu}_{i}}}{\partial\mu_{j}}$. The Fisher information matrix $I$ is defined by

\begin{equation}\label{eq:fisher}
    I(\lambda_{i},\lambda_{j}) = E \Bigg[ \frac{\partial^{2} \log L}{\partial \lambda_{i} \partial \lambda_{j}} \Bigg]
\end{equation}

\noindent
with $L$ being the likelihood function. In the case of negligible bias we see that Eq. \ref{eq:rcb} reduces to the unbiased RCB.

\begin{equation}\label{eq:unbiasrcb}
    V[\hat{\lambda}_{i}, \hat{\lambda}_{j}] \geq \Big(I^{-1}\Big)_{ij}
\end{equation}

\noindent
Under the large sample approximation the covariance is assumed to equal the zero-bias RCB. If one can estimate the matrix of second order derivatives of the log-likelihood, also known as the \textit{Hessian matrix}, one can take the inverse of this matrix as an estimate for the covariance. Under the large sample approximation one can also assume the log-likelihood is parabolic shaped around its maximum. In this case one can numerically approximate the second derivatives with finite differences.

\begin{equation}\label{eq:inverse}
    V[\hat{\lambda}_{i}, \hat{\lambda_{j}}] = \Bigg( \frac{\partial^{2} \log L}{\partial \lambda_{i} \partial \lambda_{j}} \Bigg|_{\hat{\vec{\lambda}}}\Bigg)^{-1}
\end{equation}

\noindent
However, this approach has two important caveats. The first is that this method is only suitable for the special case of no regularization i.e. $\tau=0$. In general, regularization is needed which introduces non-zero bias and thus calls for the non-trivial task of estimating the bias gradient matrix $B$.\\
\\
\noindent
Secondly, a common misconception is assuming Eq. \ref{eq:phimutheta} can be treated as a likelihood and its Hessian matrix can be used as a covariance matrix estimate, i.e.,

\begin{equation}\label{eq:phiinverse}
    V[\hat{\lambda}_{i}, \hat{\lambda_{j}}] \neq \Bigg( \frac{\partial^{2} \Phi}{\partial \lambda_{i} \partial \lambda_{j}} \Bigg|_{\hat{\vec{\lambda}}}\Bigg)^{-1}.
\end{equation}

\noindent
Again, only in the special case of no regularization can Eq. \ref{eq:phiinverse} be used as a covariance matrix estimate as it will reduce back to Eq. \ref{eq:inverse}. In the upcoming sections we will denote Eq. \ref{eq:phiinverse} as the \textit{inverse Hessian} method and show when this method can hold or break.

\subsection{Frequentist Pseudo-Experiments}\label{sec:freq}
An alternative approach would be to use pseudo-experiments to estimate the covariance. We assumed that the measured bin values follow a Poisson distribution $n_{i} \sim f(n_{i}|\vec{\mu},\vec{\theta})$ and the auxiliary measurements a Gaussian distribution $\tilde{\theta}_{k} \sim g(\tilde{\theta}_{k}|\theta_{k})$. One can set the p.d.f. parameters to the estimates of $\hat{\vec{\mu}}$ and $\hat{\vec{\theta}}$ constructed with the observed data $\vec{n}$ i.e. $f(n_{i}|\hat{\vec{\mu}},\hat{\vec{\theta}})$ and $g(\tilde{\theta}_{k}|\hat{\theta}_{k})$. From these it is possible to sample new values $\vec{n}^{t}$ and $\vec{\tilde{\theta}}^{t}$ for the data and auxiliary measurements \cite{parmbootstrap,parmbootstrap2,parmbootstrap3}. Each new sample is what is known as a \textit{pseudo-experiment} or \textit{toy MC}. For each $t$-th pseudo-experiment one can then evaluate Eq. \ref{eq:likmutheta} with the sampled values and construct new estimators $\hat{\vec{\lambda}}^{t}$ by maximizing w.r.t $\vec{\mu}$ and $\vec{\theta}$. For $T$ pseudo-experiments one can use the set of many estimators to estimate the covariance matrix

\begin{equation}\label{eq:sample}
    V[\hat{\lambda}_{i}, \hat{\lambda_{j}}] = \frac{1}{T-1}\sum_{t=1}^{T}(\hat{\lambda}^{t}_{i} - \bar{\hat{\lambda}}_{i} )(\hat{\lambda}^{t}_{j} -  \bar{\hat{\lambda}}_{j}) 
\end{equation}

\noindent
with

\begin{equation}\label{eq:sampleav}
    \bar{\hat{\lambda}}_{i} = \frac{1}{T}\sum_{t=1}^{T}(\hat{\lambda}^{t}_{i})
\end{equation}

\noindent
This covariance estimation approach will work for both biased and unbiased estimators even if the log-likelihood function is not parabolic.

\subsection{Frequentist-Bayes Hybrid Pseudo-Experiments}\label{sec:hybrid}
For some unfolding algorithms an explicit definition of a likelihood is not obvious, e.g. like for some iterative unfolding algorithms \cite{2010arXiv1010.0632D,Malaescu:2011yg}). This makes the definition of nuisance parameters and thus the previous two covariance estimation methods nonviable. In this section a covariance estimation method is presented that includes both statistical and systematic effects but does not need an explicit likelihood definition or an alteration of the chosen unfolding algorithm to include nuisance parameters. The algorithm consists out of the following steps: 

\begin{enumerate}
    \item Sample new nuisance parameter values from a prior e.g. $\theta_{k} \sim \mbox{Gaus}(\tilde{\theta}_{k},\sigma_{\tilde{\theta}_{k}})$.
    \item Compute a new response matrix $R(\vec{\theta})$ and consequently new expected values $\vec{\nu}(\vec{\theta})$.
    \item Sample new data $n_{i} \sim \mbox{Pois}(\nu_{i})$ with the newly calculated means $\vec{\nu}(\vec{\theta})$.
    \item Repeat many times and use the set of evaluated estimators to calculate the sample covariance.
\end{enumerate}

\noindent
Note that instead of sampling auxiliary measurements $\vec{\tilde{\theta}}$ one samples parameter values $\vec{\theta}$ from a prior which introduces the Bayesian aspect of this treatment. However, we expect the methods proposed in Sec. \ref{sec:freq} and \ref{sec:hybrid} to be equivalent in certain scenarios. 

\section{Comparison Study}\label{sec:comp}
This section presents the results of a comparison study between the three before mentioned covariance estimation methods. The first section introduces the example unfolding scenarios used to test the methods. The second section shows how (dis-)similar the three methods are with the use of several metrics.

\subsection{Unfolding Test Scenarios}
In this experimental setup two different underlying physics models are chosen as truth distribution corresponding to realistic unfolding scenarios in HEP. For some variable measurable $x$ we defined the following underlying physics models.

\subsubsection{Double Gaussian Model}

\begin{multline*}
    f_{sig}(x_{true}|\mu_{1},\mu_{2},\sigma_{1},\sigma_{2}) = \\ 
    \frac{0.5}{\sigma_{1}\sqrt{2\pi}}\exp\Bigg[-\frac{(x_{true}-\mu_{1})^{2}}{2\sigma_{1}^{2}}\Bigg] + \frac{0.5}{\sigma_{2}\sqrt{2\pi}}\exp\Bigg[-\frac{(x_{true}-\mu_{2})^{2}}{2\sigma_{2}^{2}}\Bigg]
\end{multline*}

\noindent
with $\mu_{1}=1.5$, $\mu_{2}=-1.5$, $\sigma_{1}=\sigma_{2}=0.12$ and $N_{sig}=50000$ number of sampled events.

\begin{equation}
  f_{bkg}(x_{true}|a,b)=\begin{cases}
    \frac{1}{b-a}, & \text{if $a<x<b$}.\\
    0, & \text{otherwise}.
  \end{cases}
\end{equation}

\noindent
with $a = -4$, $b=4$ and $N_{bkg}=5000$ number of sampled events. The filled truth and reconstructed histograms range between $[-4,4]$ and both have constant bin size $\Delta x=1.6$.

\subsubsection{Exponential Model}

\begin{equation}
    f_{sig}(x_{true}|\lambda) = \lambda \exp \Big[-\lambda x_{true} \Big]
\end{equation}

\noindent
with $\lambda = 0.14$ and $N_{sig}=10000$.

\begin{equation}
    f_{bkg}(x_{true}|\gamma) =  \gamma \exp \Big[-\gamma x_{true} \Big]
\end{equation}

\noindent
with $\gamma = 0.15$ and $N_{sig}=40000$. The filled truth and reconstructed histograms range between $[0,60]$ and variable bin widths with bin edges $\Delta x_{truth}=\{0,2,4,6,8,10,12,14,18,25,35,60\}$ and $\Delta x_{reco}=\{0,1,2,3,4,5,6,7,8,9,10,11,$ $12,13,14,15,16,17,18,20,25,30,35,45,60\}$.

\subsubsection{Detector Function \& Nuisance Parameters} 
To simulate a detector response a piece-wise function is applied to simulate the reconstruction efficiency and smearing loosely inspired on a calorimeter response. This function also introduces the nuisance parameters that will simulate the systematic sources of error included in the covariance estimation.

\begin{equation}
    x_{reco}=\begin{cases}
    x_{true} + \theta_{1} \cdot \mbox{Gaus}(0,\theta_{2} + a \cdot \sqrt{x_{true}/300}, & \text{if $0 < \epsilon <\theta_{3} - b \cdot \frac{|x_{true}|}{600}$}\\
    \text{None}, & \text{otherwise}
  \end{cases}
\end{equation}

\noindent
The efficiency is simulated by evaluating the condition of the if-statement for each value $x_{true}$ and $\epsilon$ sampled from a uniform distribution between $[0,1]$. In case of passing efficiency, an additive Gaussian smearing function is applied. The whole detector function depends on the nuisance parameters $\theta_{1}$, $\theta_{2}$, $\theta_{3}$ and two constants $a$ and $b$. For the double Gaussian model the constants $a$ and $b$ are set to 0. For the exponential model they are set to 1 to include variable dependency of the efficiency and smearing functions. The nuisance parameters are set to their auxiliary measurements which are taken to be $\tilde{\theta}_{1}=1.0$, $\tilde{\theta}_{2}=0.3$, $\tilde{\theta}_{3}=0.95$ with corresponding uncertainties $\sigma_{\tilde{\theta}_{1}}=0.01$, $\sigma_{\tilde{\theta}_{2}}=0.05$ and $\sigma_{\tilde{\theta}_{1}}=0.02$.

\vspace{10mm}

\begin{figure}[h!]
\centering
\begin{subfigure}{.5\textwidth}
  \centering
  \includegraphics[width=1\linewidth]{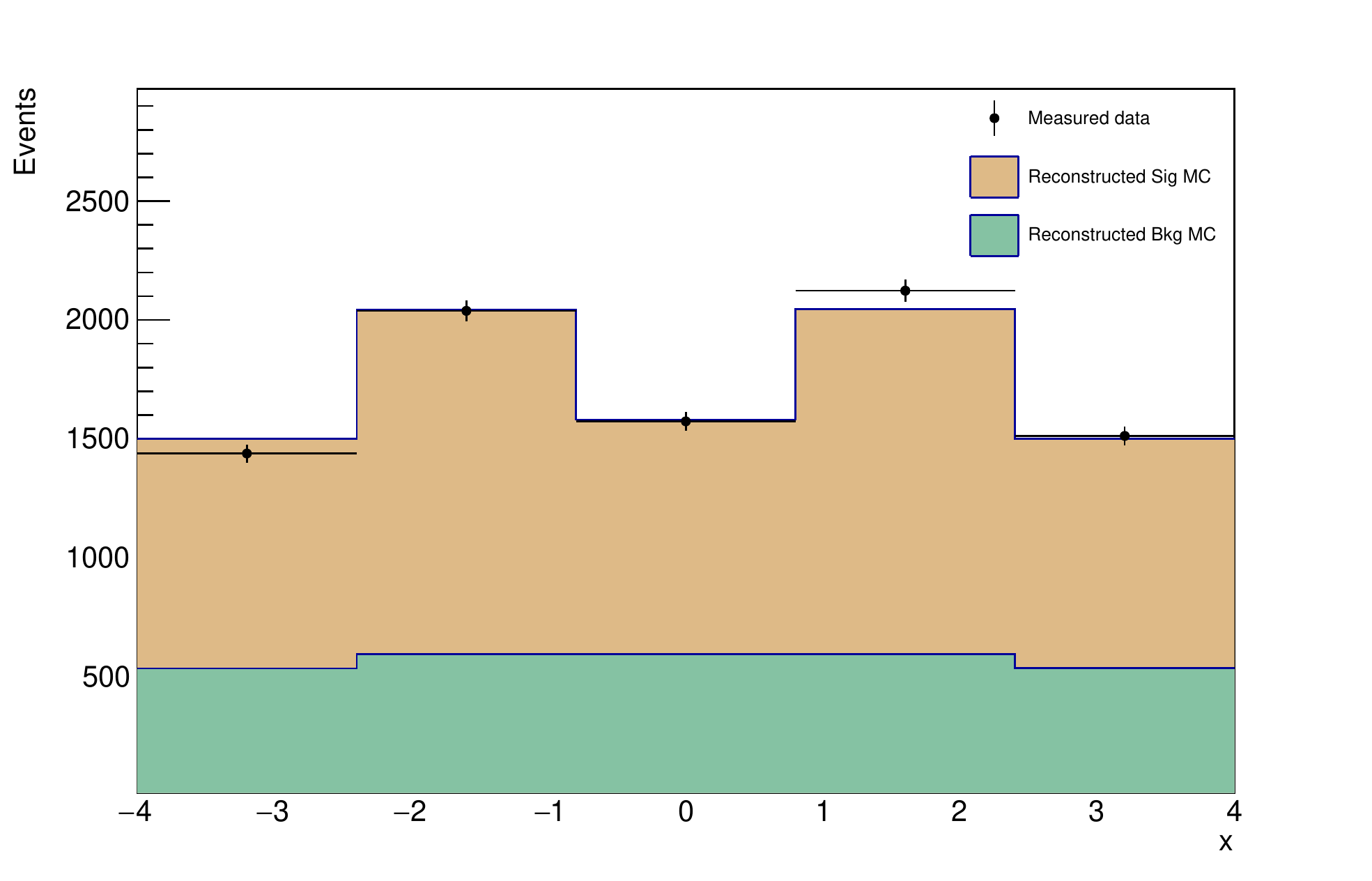}
  \label{fig:bimodalhists}
  \caption{ }
\end{subfigure}%
\begin{subfigure}{.5\textwidth}
  \centering
  \includegraphics[width=1\linewidth]{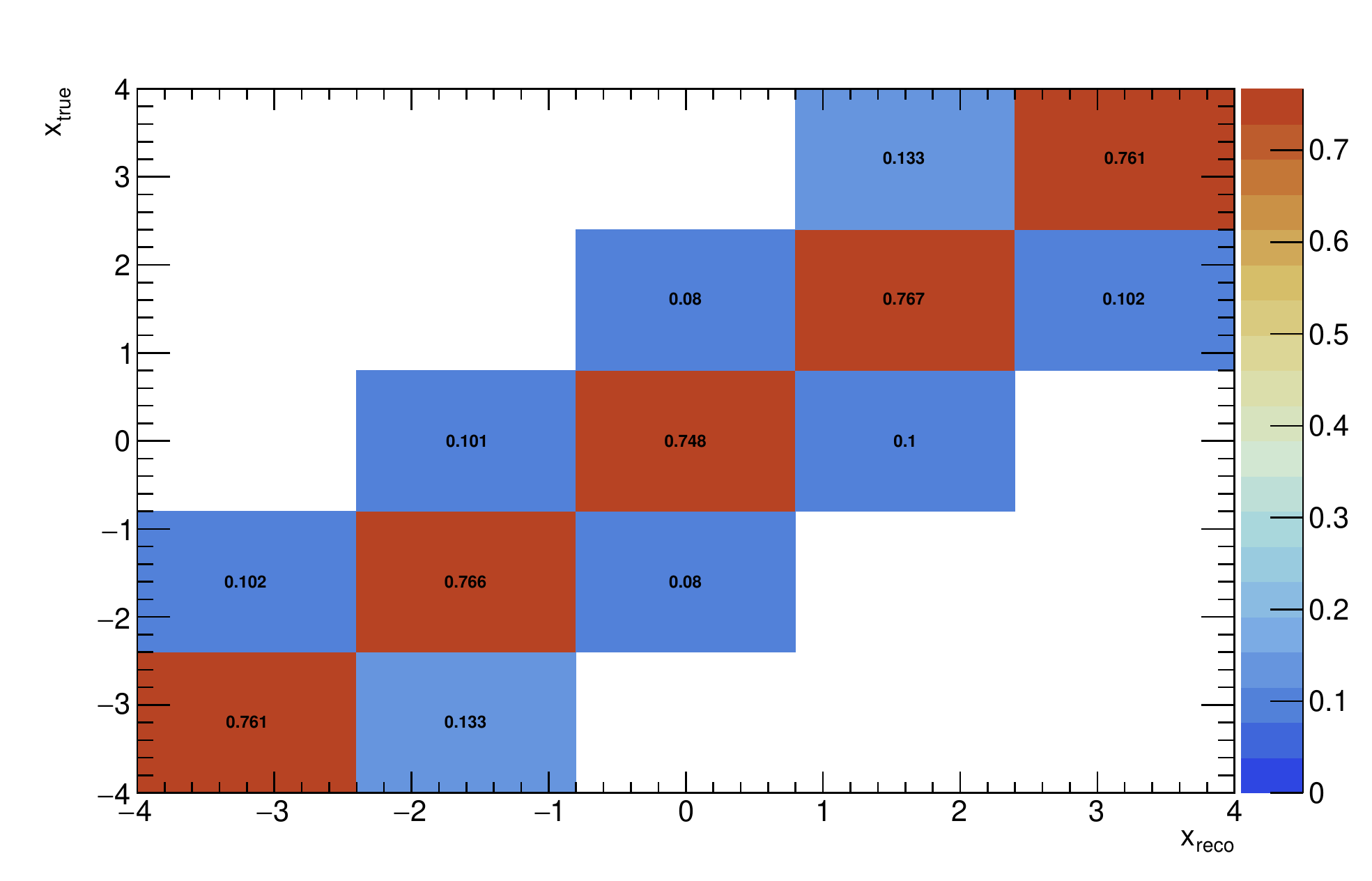}
  \label{fig:bimodalresp}
  \caption{ }
\end{subfigure}
\caption{A plot of a) the input distributions on reconstructed level and b) corresponding response matrix of the double Gaussian model}
\label{fig:bimodalinput}
\end{figure}

\begin{figure}[h!]
\centering
\begin{subfigure}{.5\textwidth}
  \centering
  \includegraphics[width=1\linewidth]{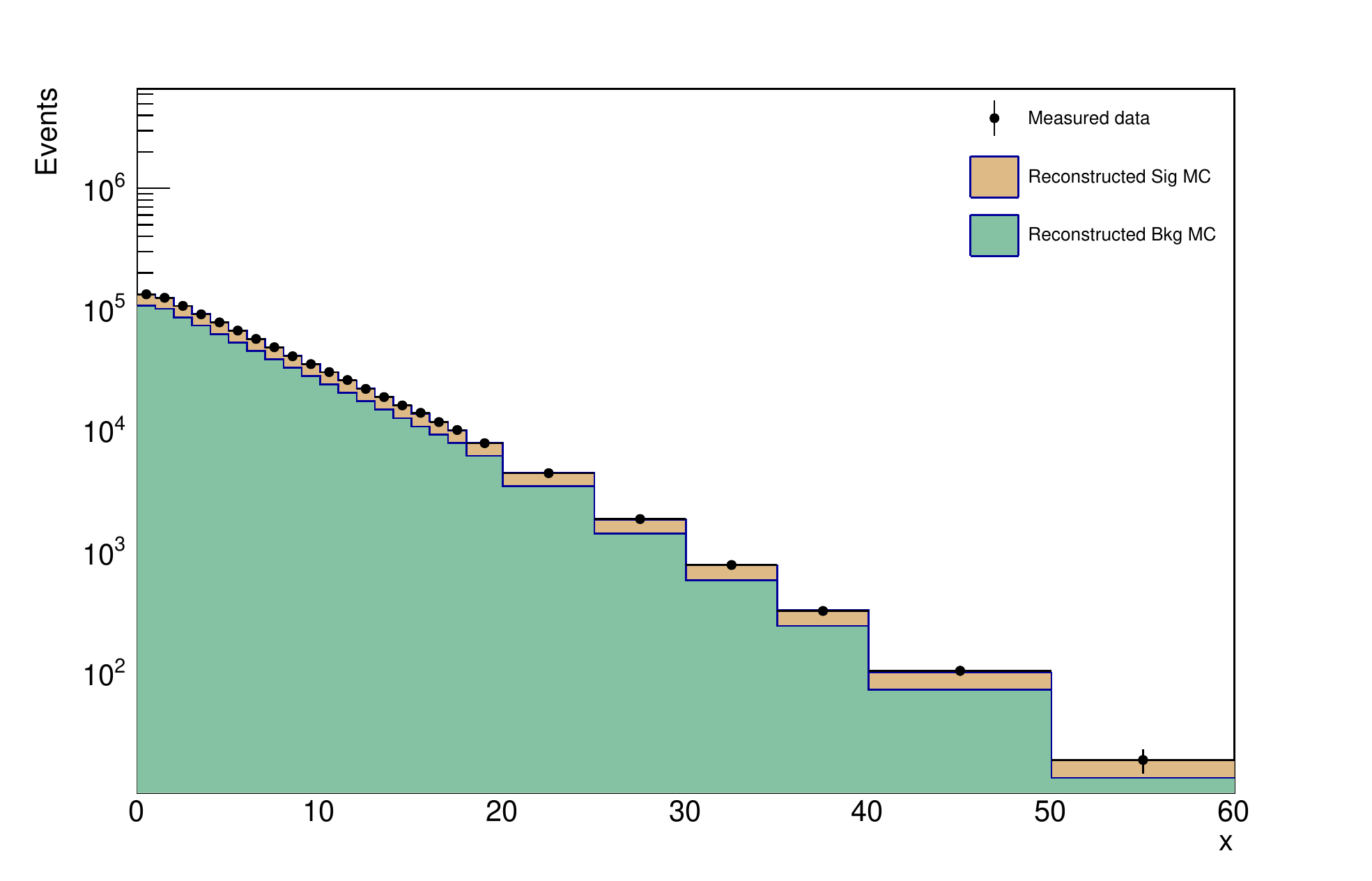}
  \label{fig:exphists}
  \caption{ }
\end{subfigure}%
\begin{subfigure}{.5\textwidth}
  \centering
  \includegraphics[width=1\linewidth]{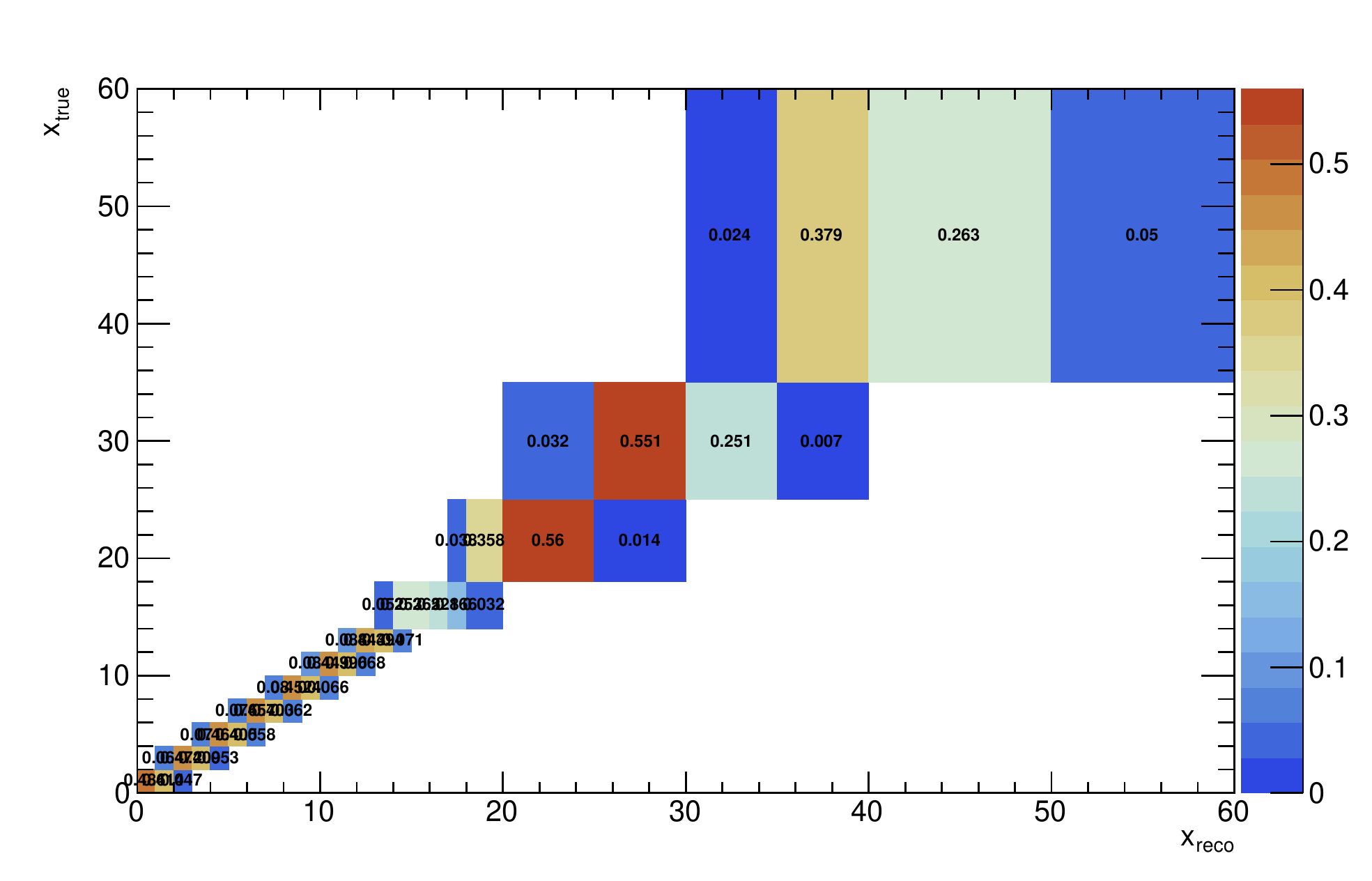}
  \label{fig:expresp}
  \caption{ }
\end{subfigure}
\caption{A plot of a) the input distributions on reconstructed level and b) corresponding response matrix of the exponential model}
\label{fig:expinput}
\end{figure}

\newpage
\subsection{Covariance Estimates}
Both the double Gaussian and exponential distributions were unfolded by maximizing the regularized full likelihood w.r.t. $\vec{\mu}$ and $\vec{\theta}$ for different amounts of regularization $\tau\in\lbrace 0, 10^{-6}, 10^{-5}, 5\times 10^{-5} \rbrace$. For each unfolded distribution the covariance matrix was estimated once with each of the three methods. We would like to stress here that the inverse Hessian method is only valid to use for $\tau=0$. We show its estimations for all values of $\tau$ just to illustrate this. 

\begin{figure}[h!]
\centering
\begin{subfigure}{.33\textwidth}
  \includegraphics[width=1\linewidth]{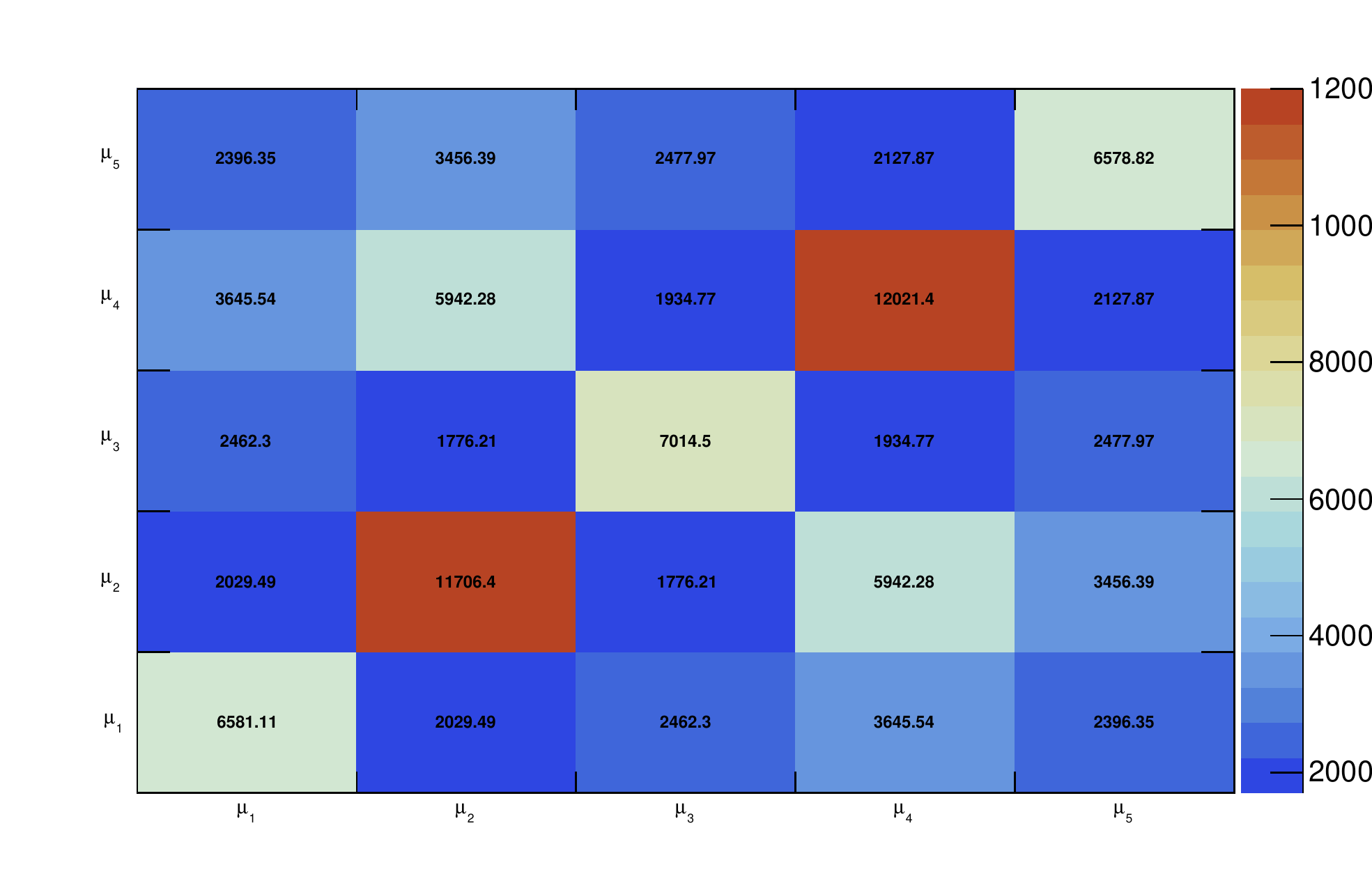}
  \caption{Frequentist $\tau=0.$}
\end{subfigure}%
\begin{subfigure}{.32\textwidth}
  \includegraphics[width=1\linewidth]{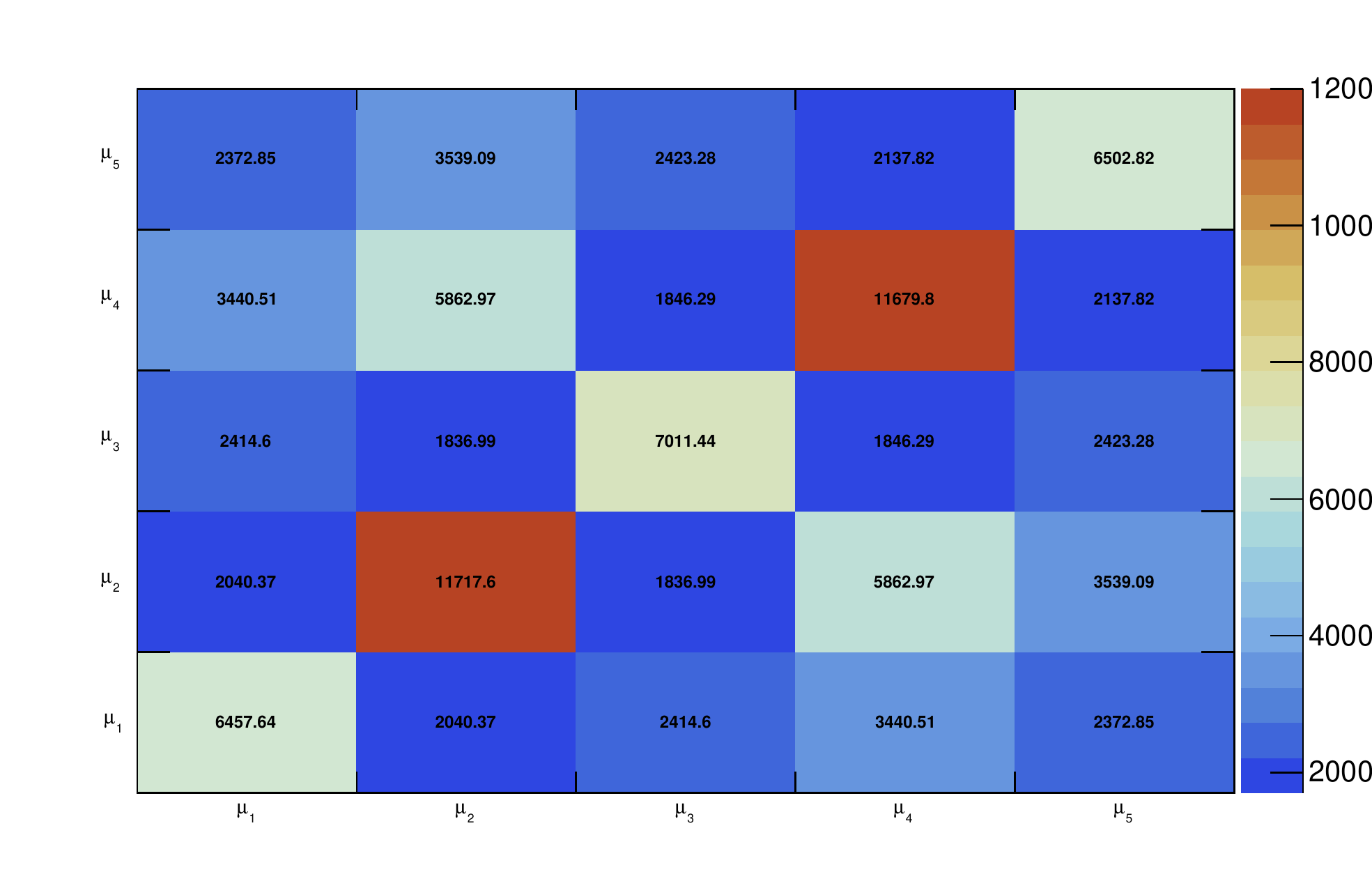}
  \caption{Hybrid $\tau=0.$}
\end{subfigure}
\begin{subfigure}{.32\textwidth}
  \includegraphics[width=1\linewidth]{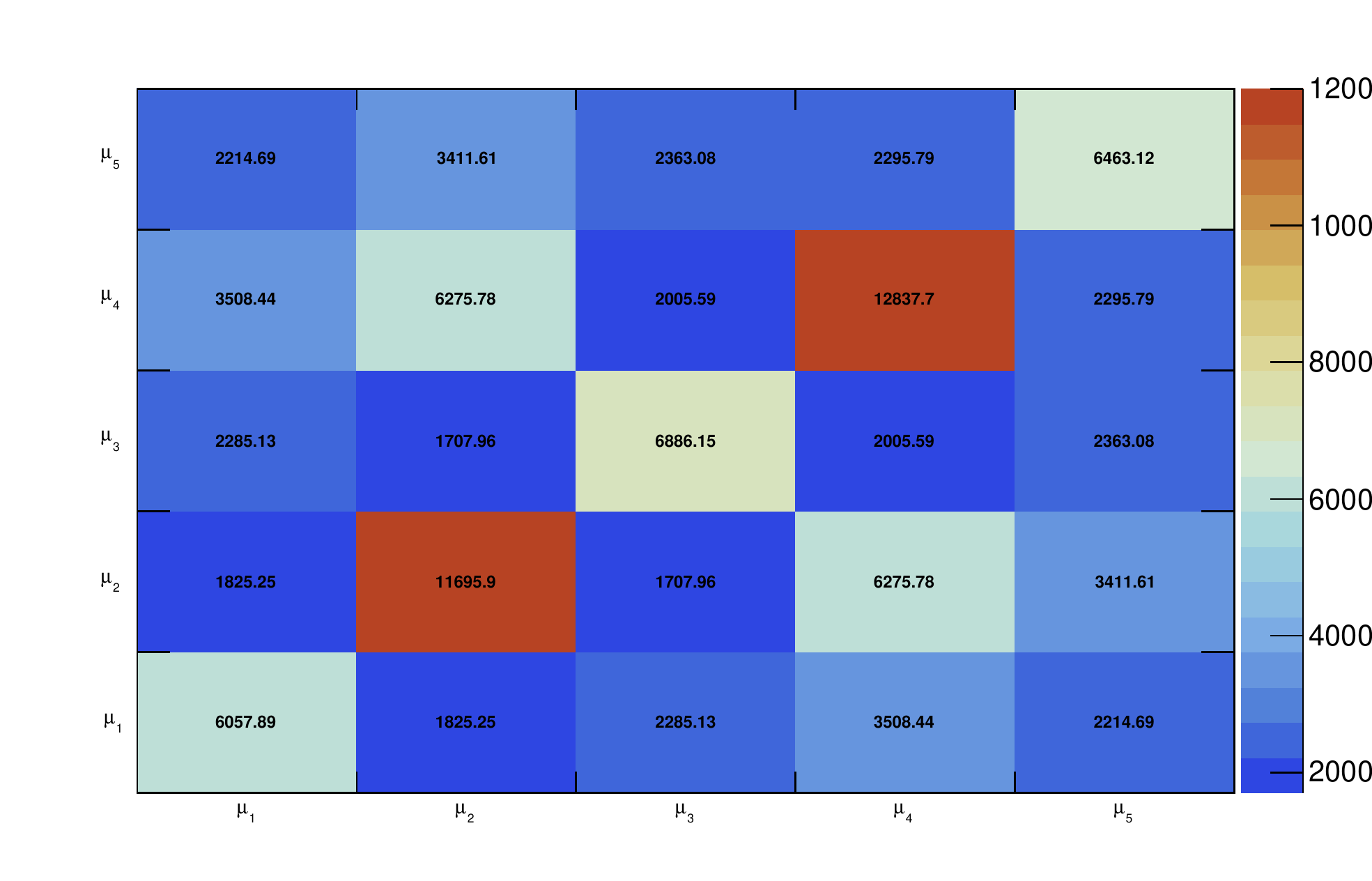}
  \caption{Hessian $\tau=0.$}
\end{subfigure}

\vskip\baselineskip

\begin{subfigure}{.33\textwidth}
  \includegraphics[width=1\linewidth]{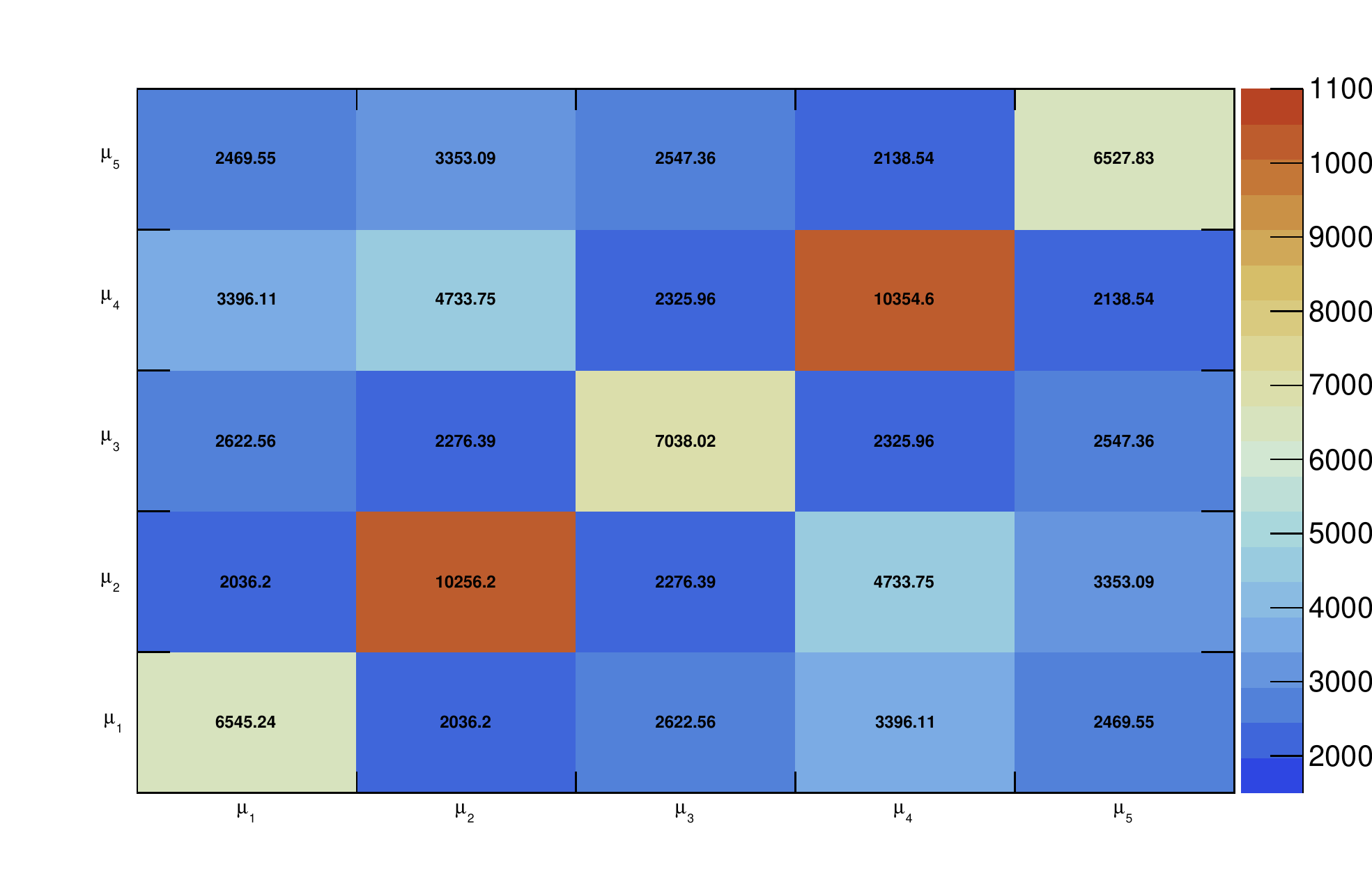}
  \caption{Frequentist $\tau=10^{-6}$}
\end{subfigure}%
\begin{subfigure}{.33\textwidth}
  \includegraphics[width=1\linewidth]{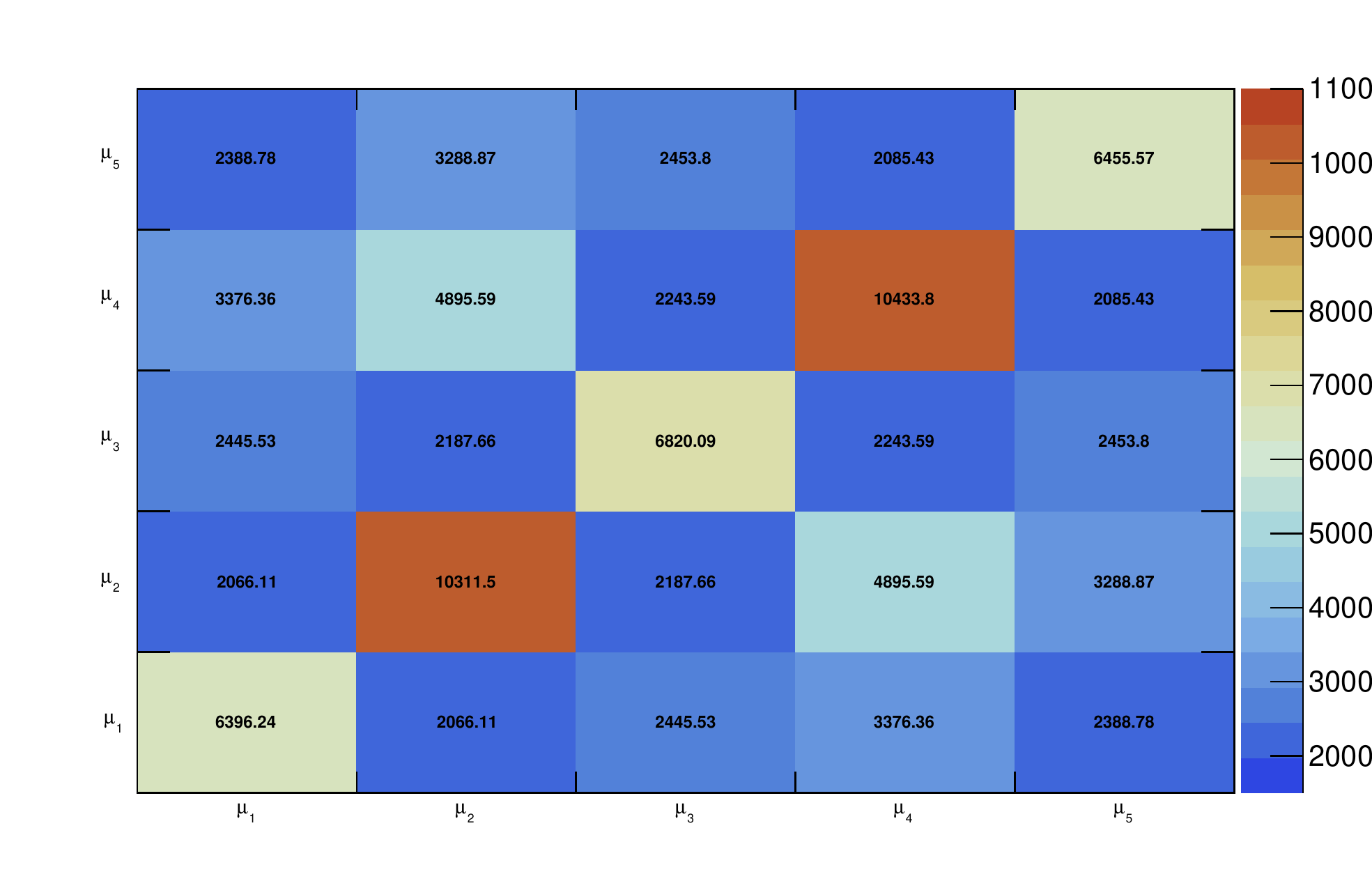}
  \caption{Hybrid $\tau=10^{-6}$}
\end{subfigure}
\begin{subfigure}{.33\textwidth}
  \includegraphics[width=1\linewidth]{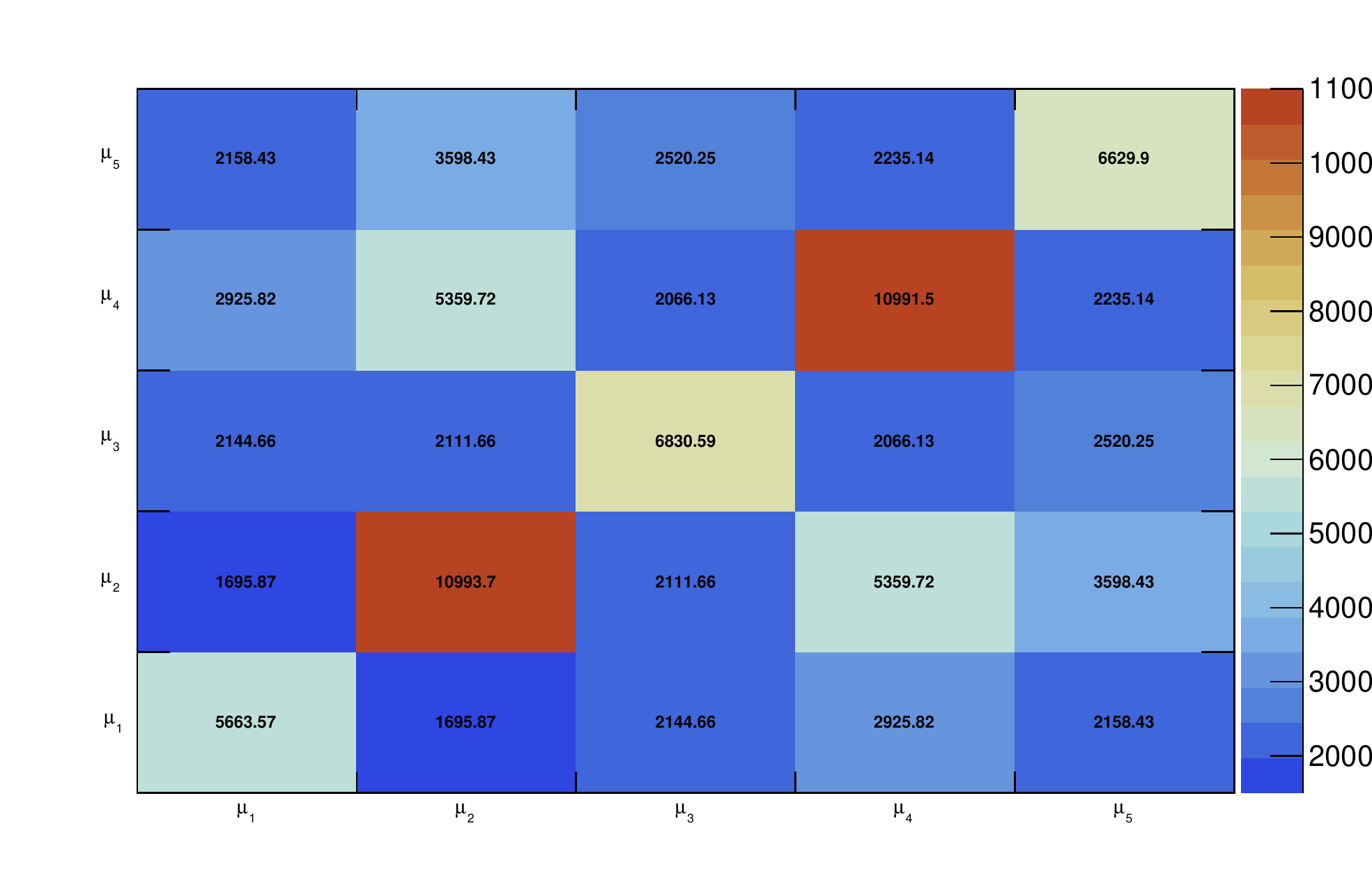}
  \caption{Hessian $\tau=10^{-6}$}
\end{subfigure}

\vskip\baselineskip

\begin{subfigure}{.33\textwidth}
  \includegraphics[width=1\linewidth]{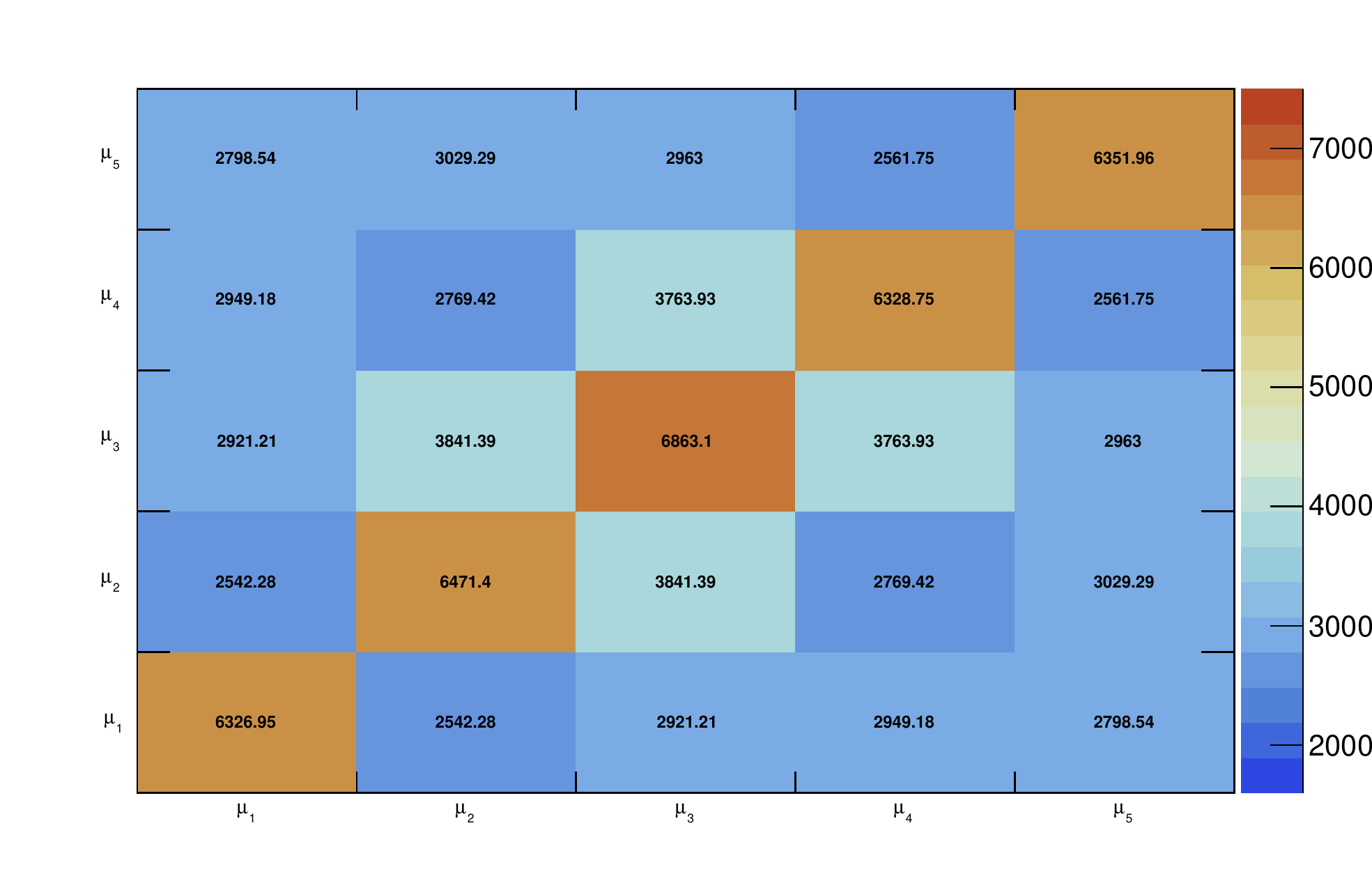}
  \caption{Frequentist $\tau=10^{-5}$}
\end{subfigure}%
\begin{subfigure}{.33\textwidth}
  \includegraphics[width=1\linewidth]{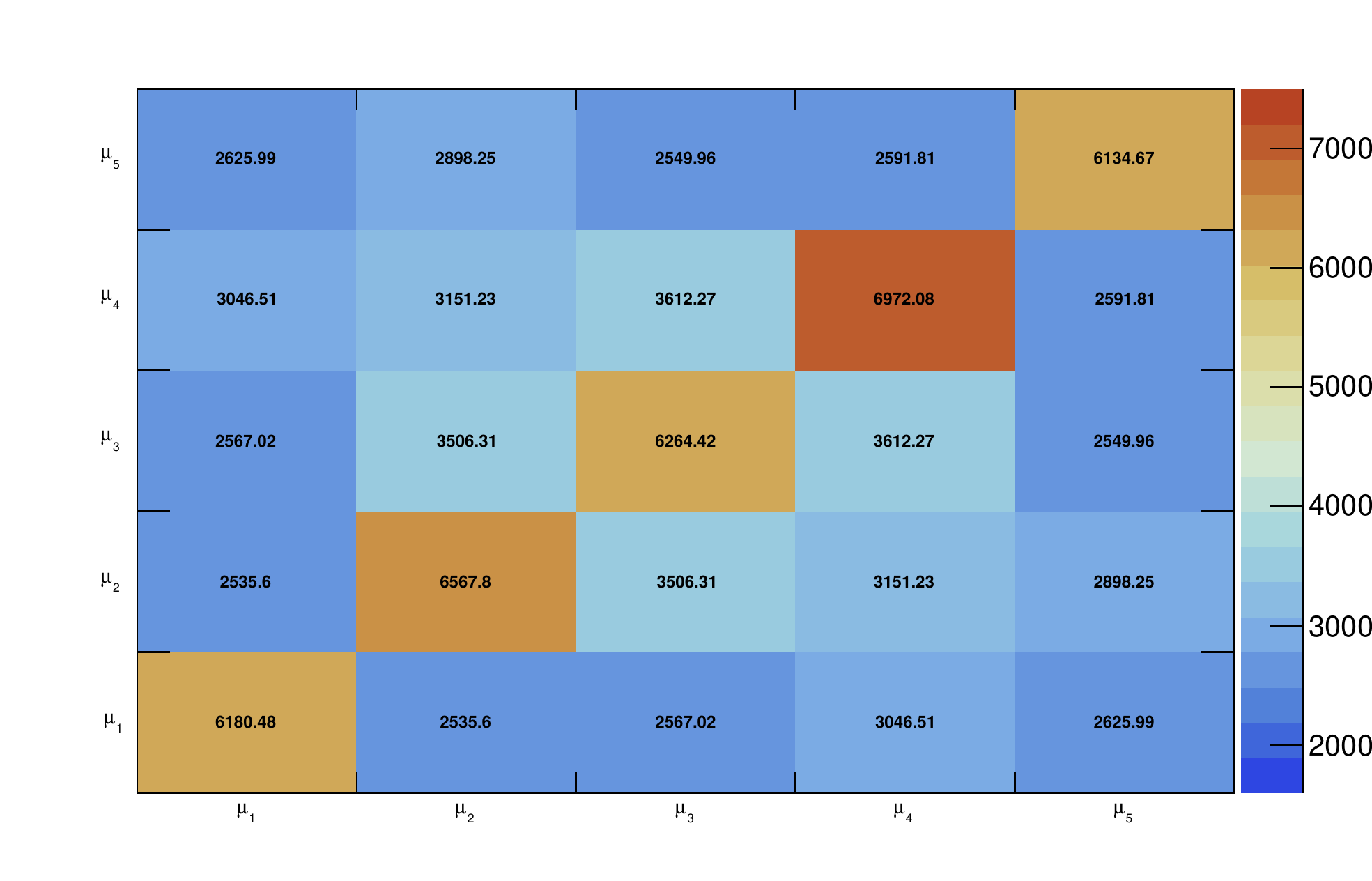}
  \caption{Hybrid $\tau=10^{-5}$}
\end{subfigure}
\begin{subfigure}{.33\textwidth}
  \includegraphics[width=1\linewidth]{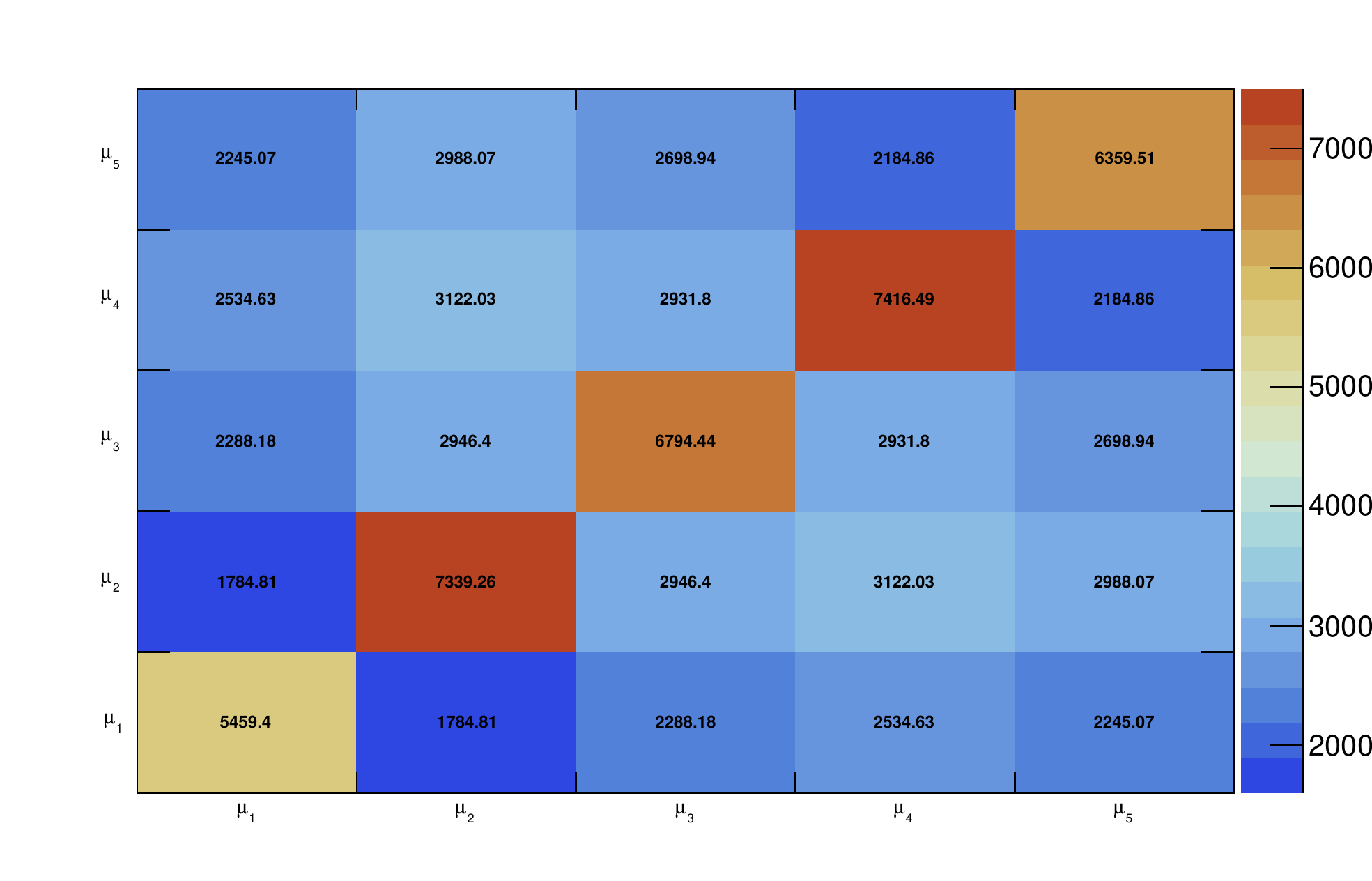}
  \caption{Hessian $\tau=10^{-5}$}
\end{subfigure}

\vskip\baselineskip

\begin{subfigure}{.33\textwidth}
  \includegraphics[width=1\linewidth]{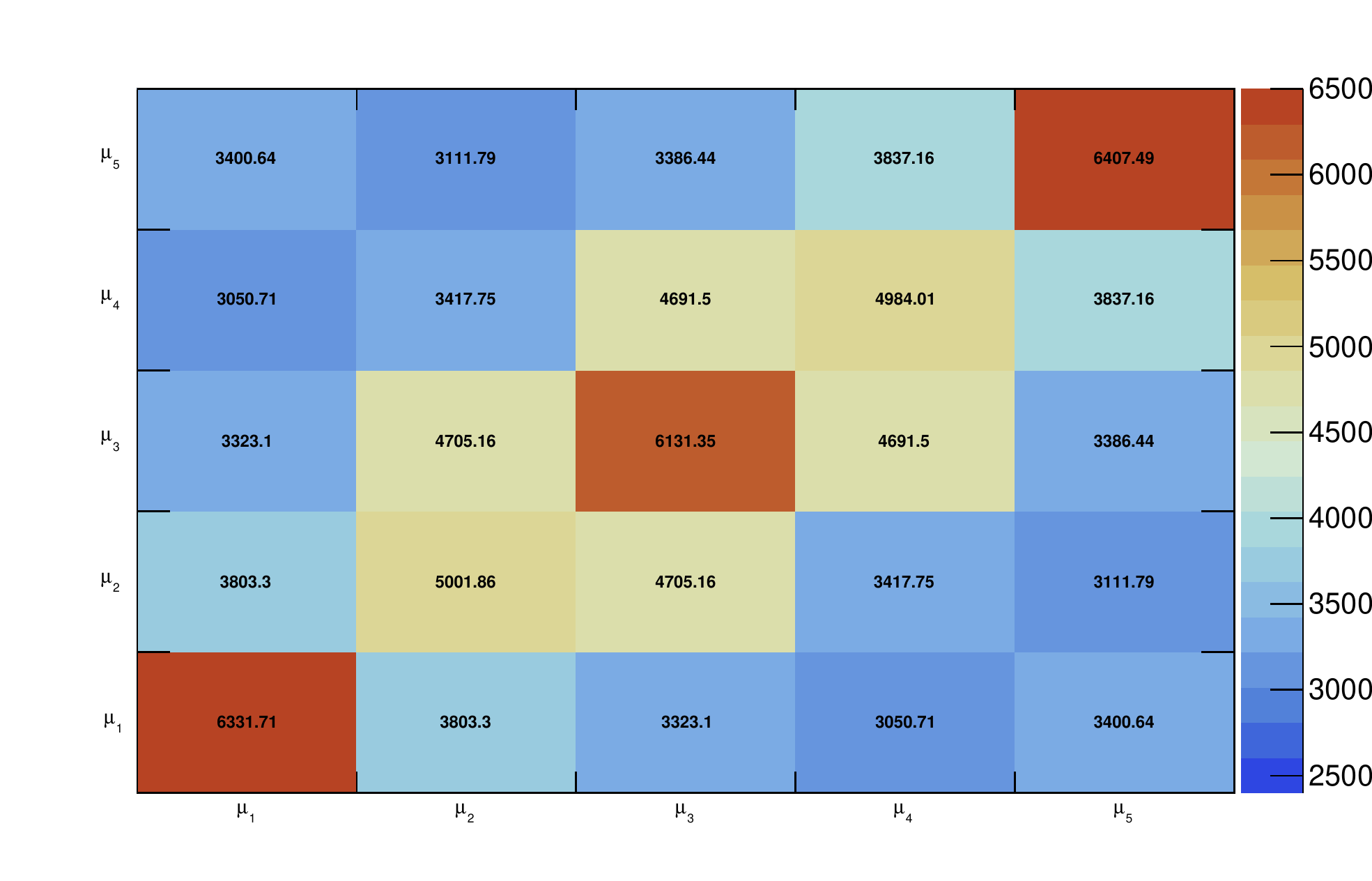}
  \caption{Frequentist $\tau=5\times 10^{-5}$}
\end{subfigure}%
\begin{subfigure}{.33\textwidth}
  \includegraphics[width=1\linewidth]{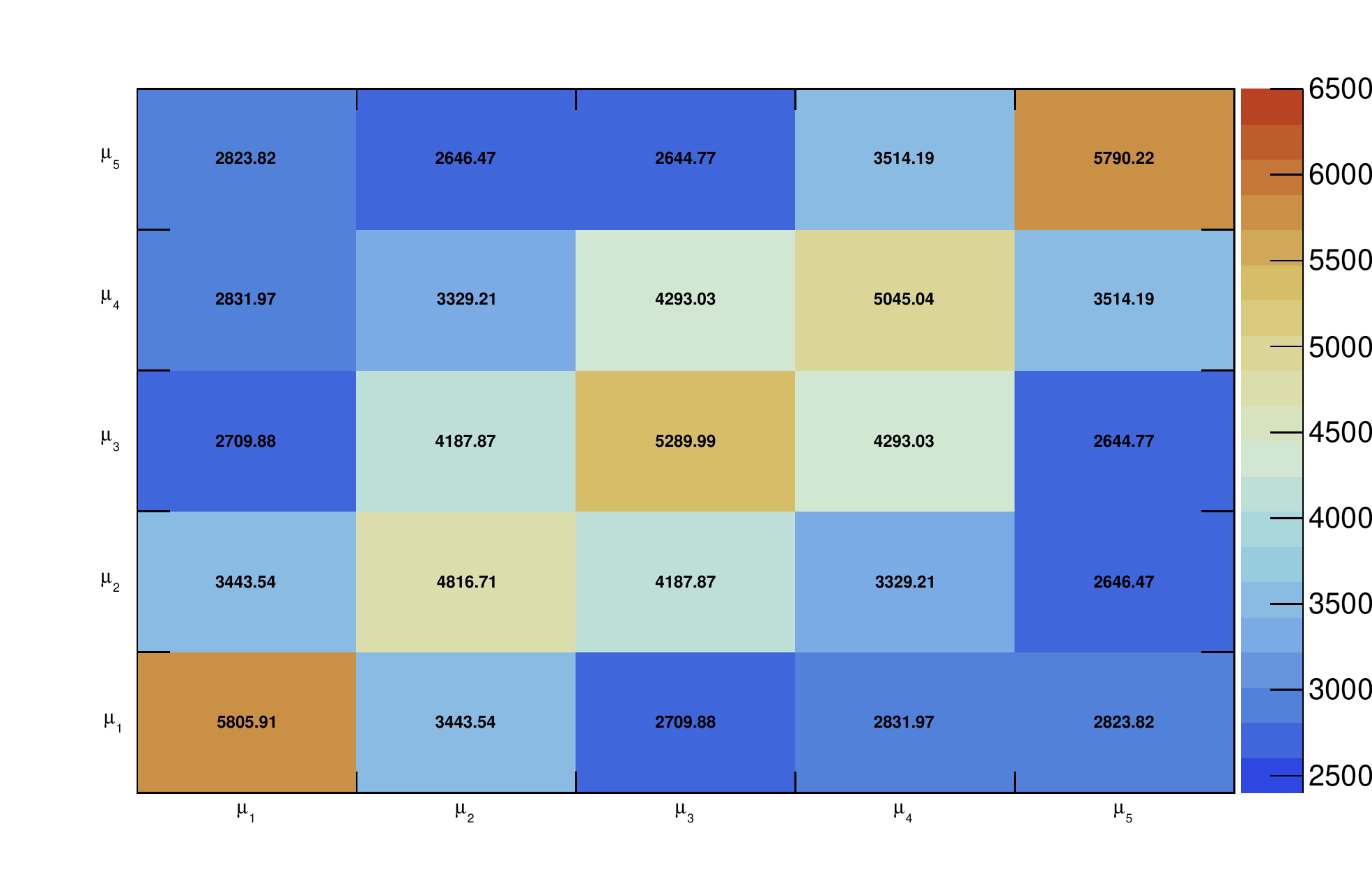}
  \caption{Hybrid $\tau=5\times 10^{-5}$}
\end{subfigure}
\begin{subfigure}{.33\textwidth}
  \includegraphics[width=1\linewidth]{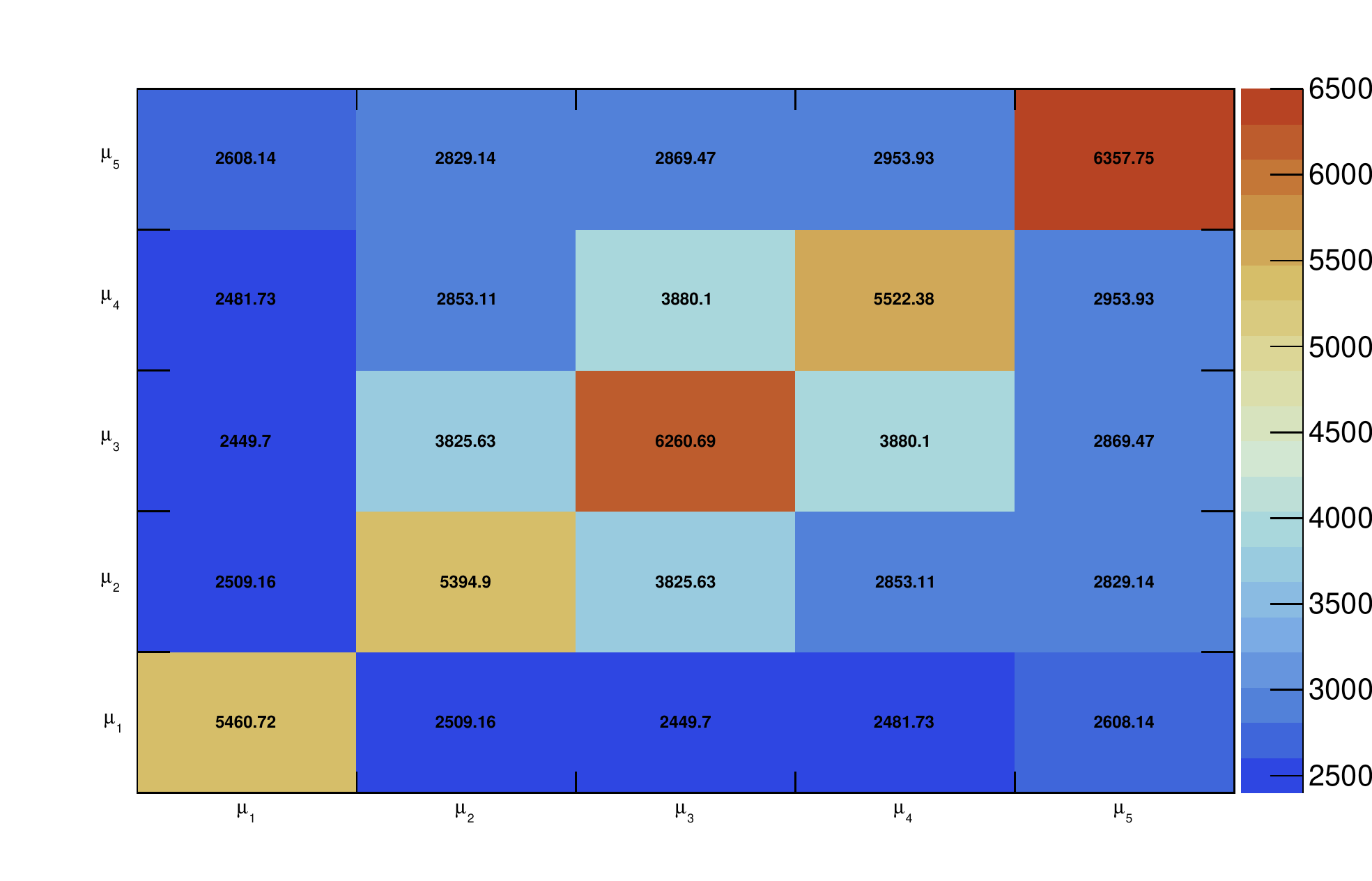}
  \caption{Hessian $\tau=5\times 10^{-5}$}
\end{subfigure}

\caption{Covariance matrix for the double Gaussian distribution estimated with the frequentist pseudo-experiments, inverse hessian and frequentist-bayes hybrid pseudo-experiments method for various regularization strengths $\tau$}
\label{fig:bimodalcovariances}
\end{figure}

\newpage

\begin{figure}[h!]
\centering
\begin{subfigure}{.33\textwidth}
  \includegraphics[width=1\linewidth]{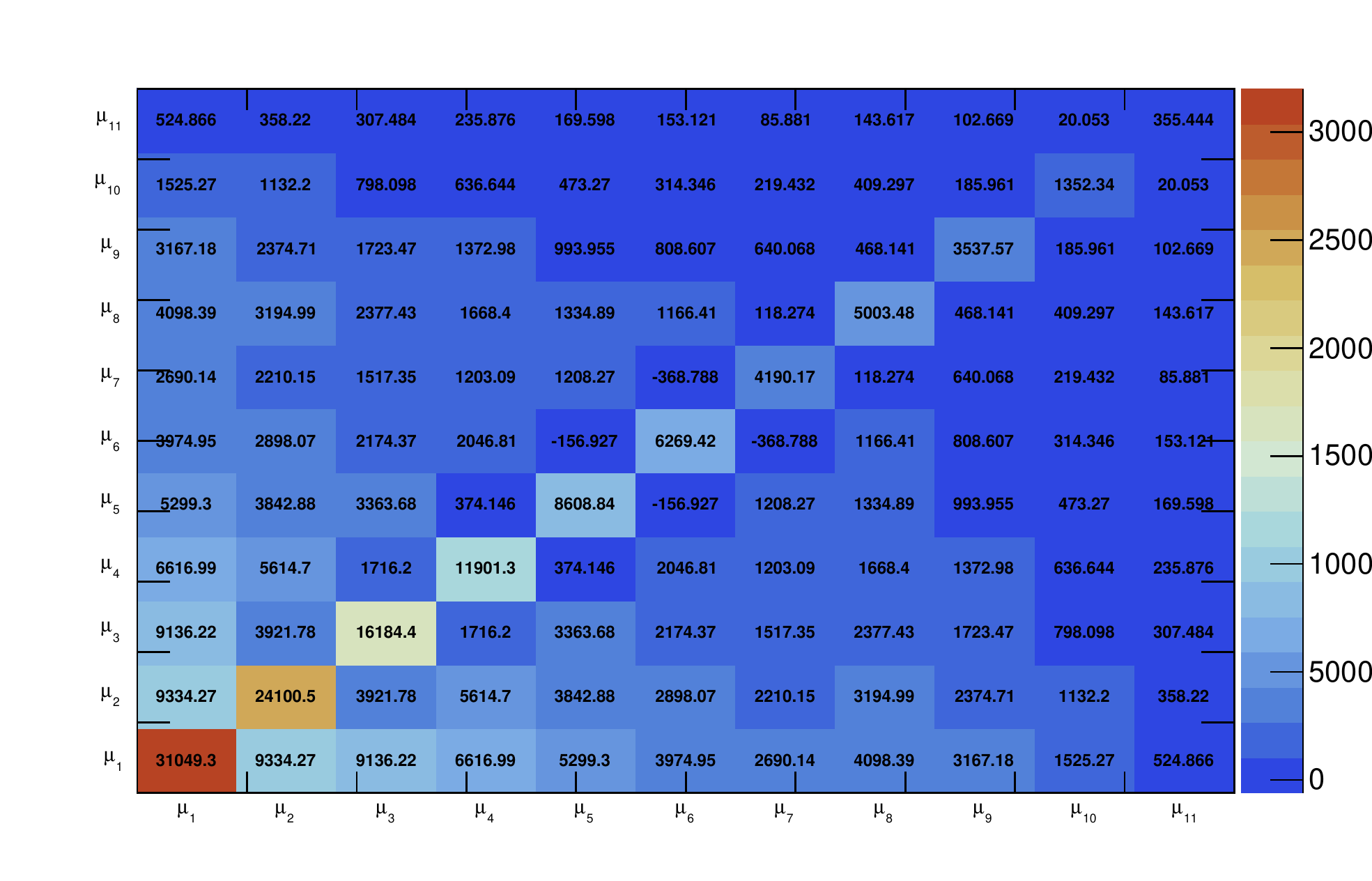}
  \caption{Frequentist $\tau=0.$}
\end{subfigure}%
\begin{subfigure}{.33\textwidth}
  \includegraphics[width=1\linewidth]{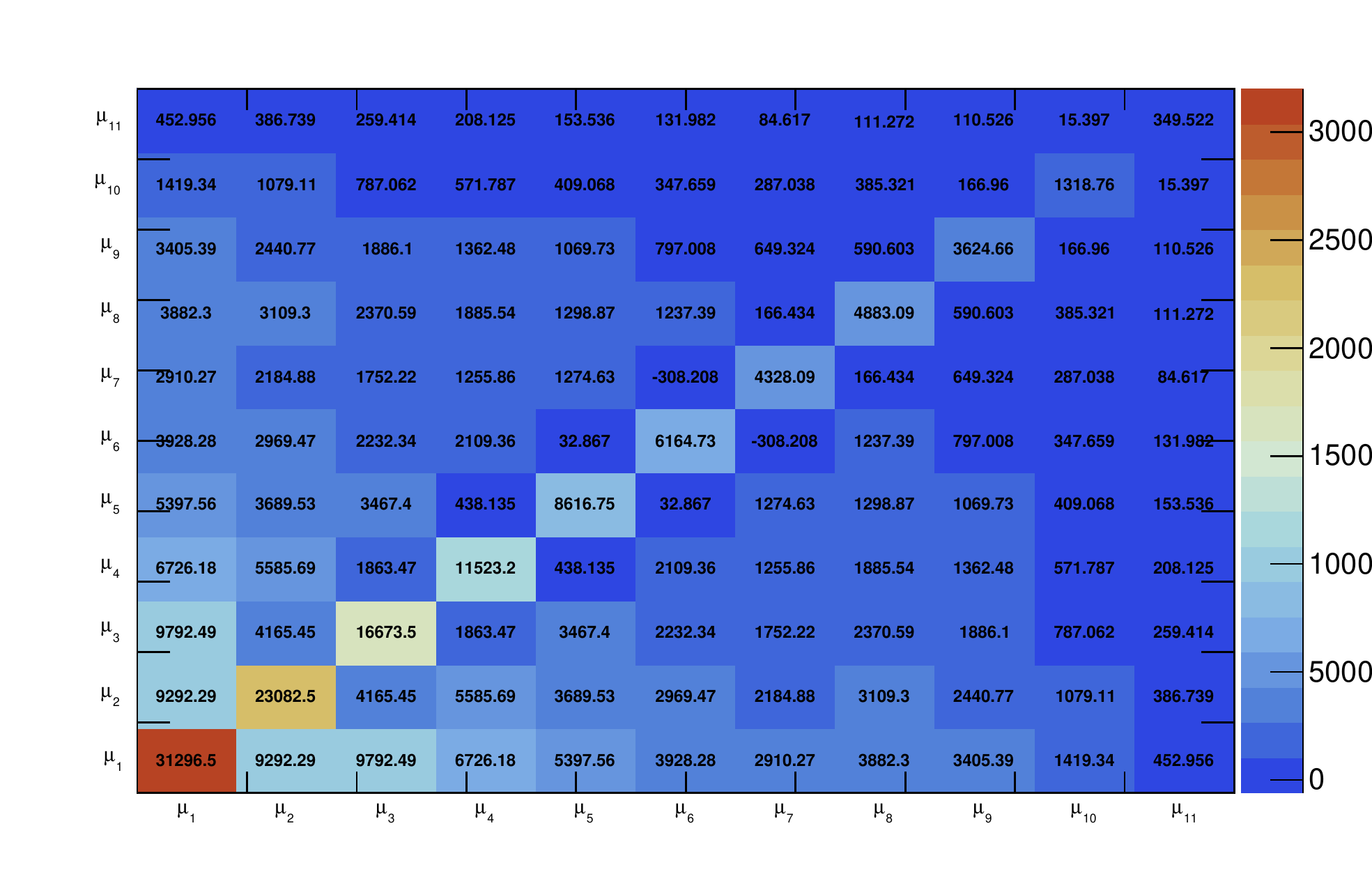}
  \caption{Hybrid $\tau=0.$}
\end{subfigure}
\begin{subfigure}{.33\textwidth}
  \includegraphics[width=1\linewidth]{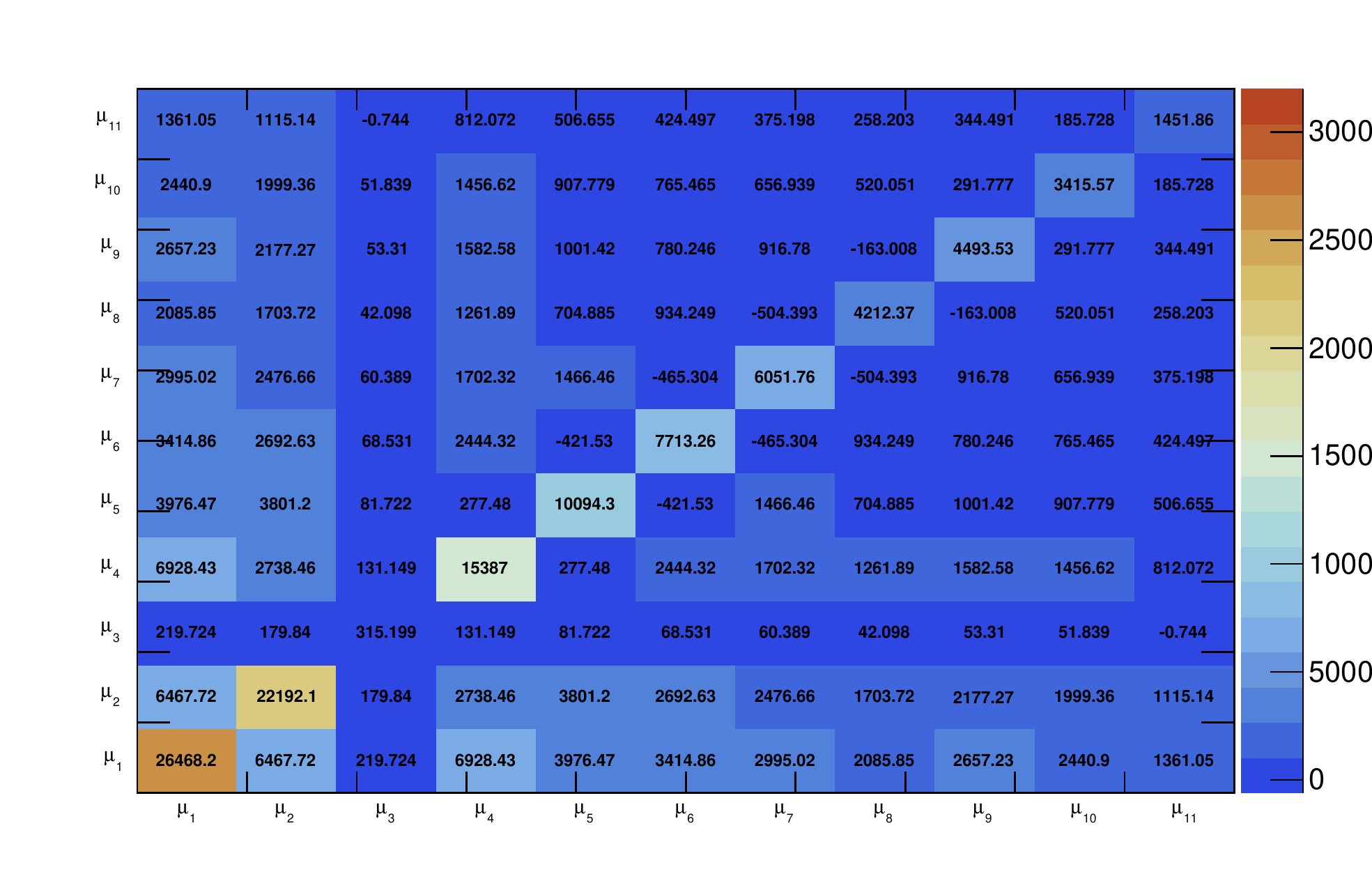}
  \caption{Hessian $\tau=0.$}
\end{subfigure}

\vskip\baselineskip

\begin{subfigure}{.33\textwidth}
  \includegraphics[width=1\linewidth]{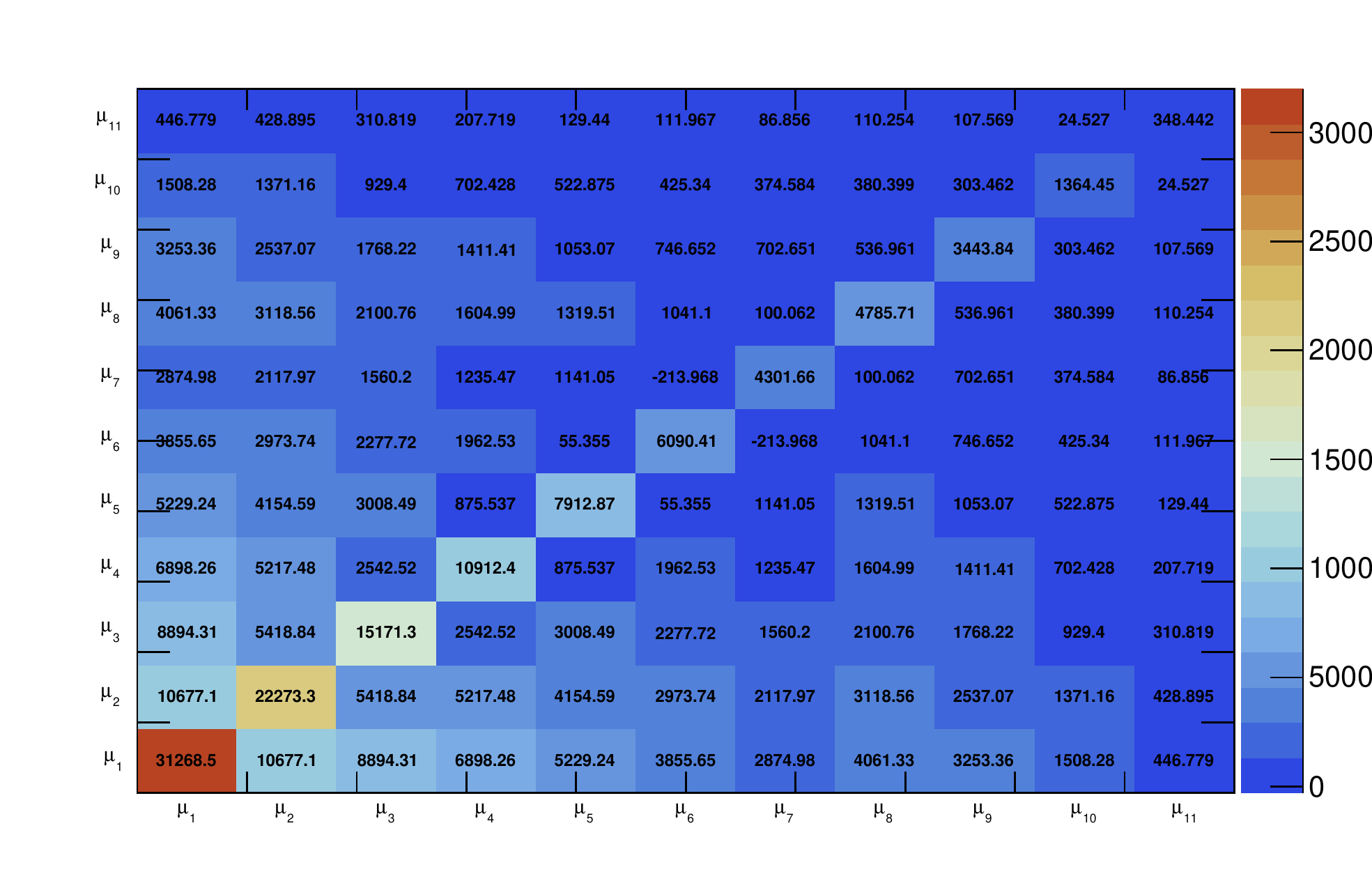}
  \caption{Frequentist $\tau=10^{-6}$}
\end{subfigure}%
\begin{subfigure}{.33\textwidth}
  \includegraphics[width=1\linewidth]{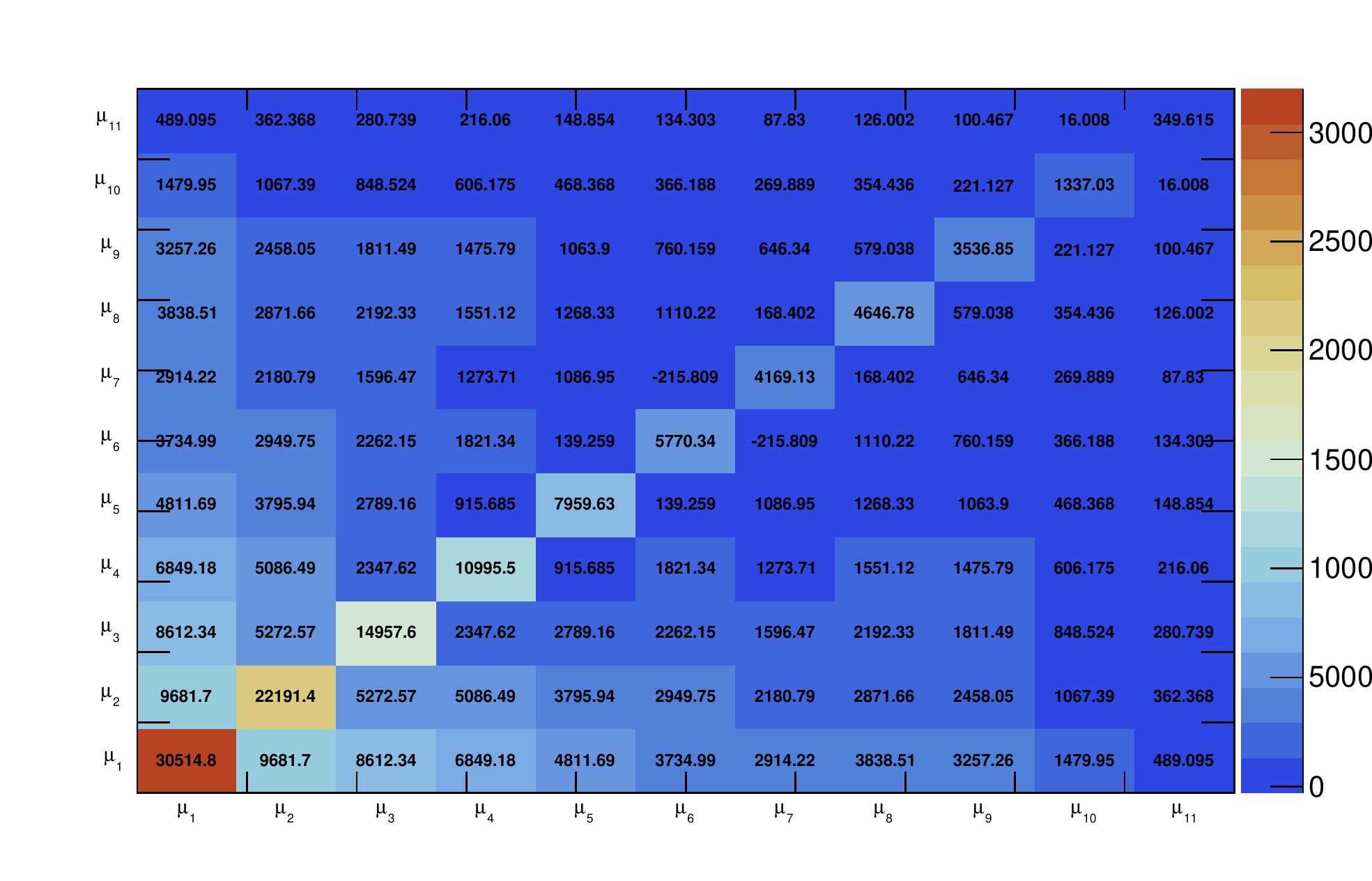}
  \caption{Hybrid $\tau=10^{-6}$}
\end{subfigure}
\begin{subfigure}{.33\textwidth}
  \includegraphics[width=1\linewidth]{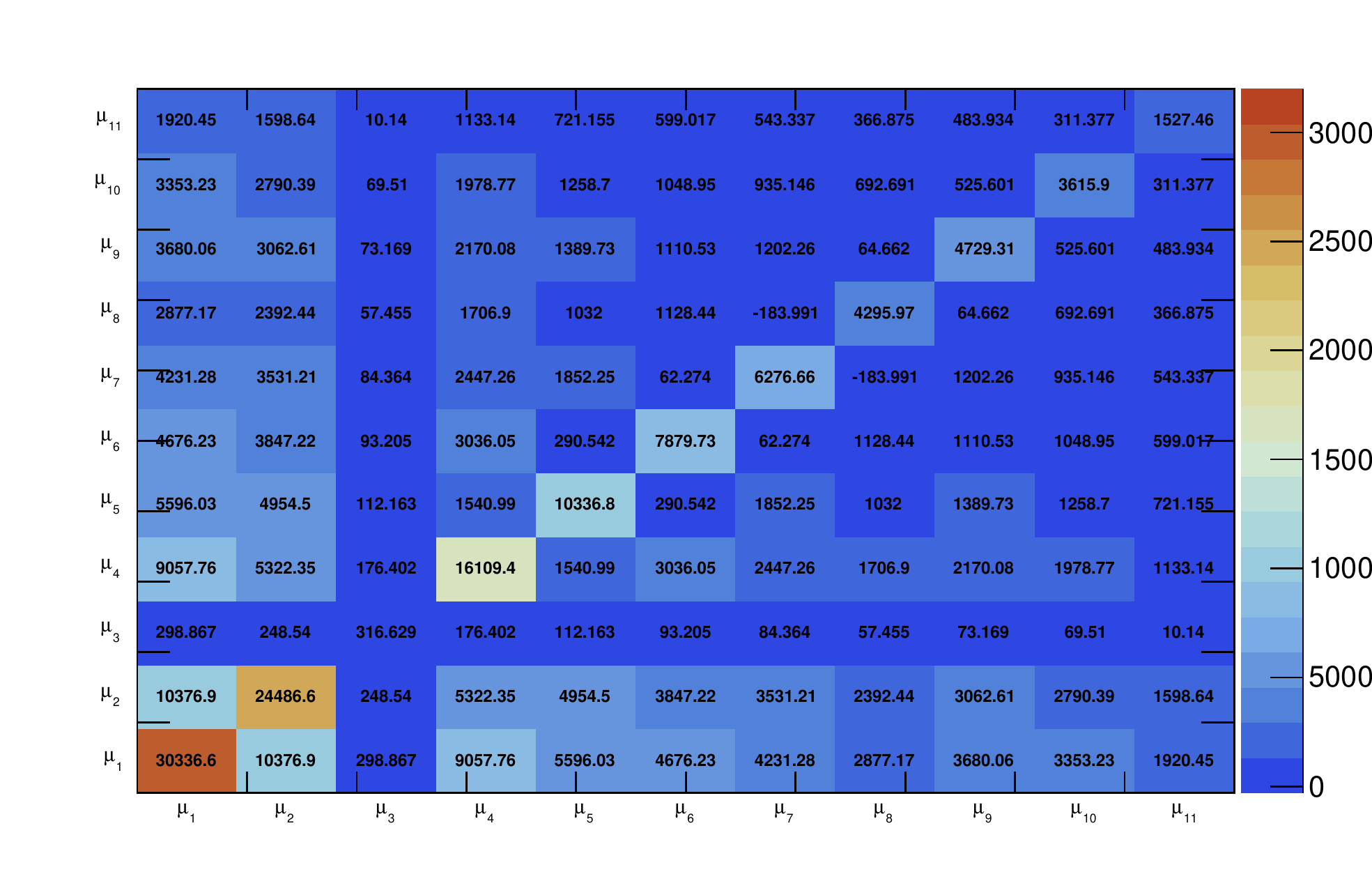}
  \caption{Hessian $\tau=10^{-6}$}
\end{subfigure}

\vskip\baselineskip

\begin{subfigure}{.33\textwidth}
  \includegraphics[width=1\linewidth]{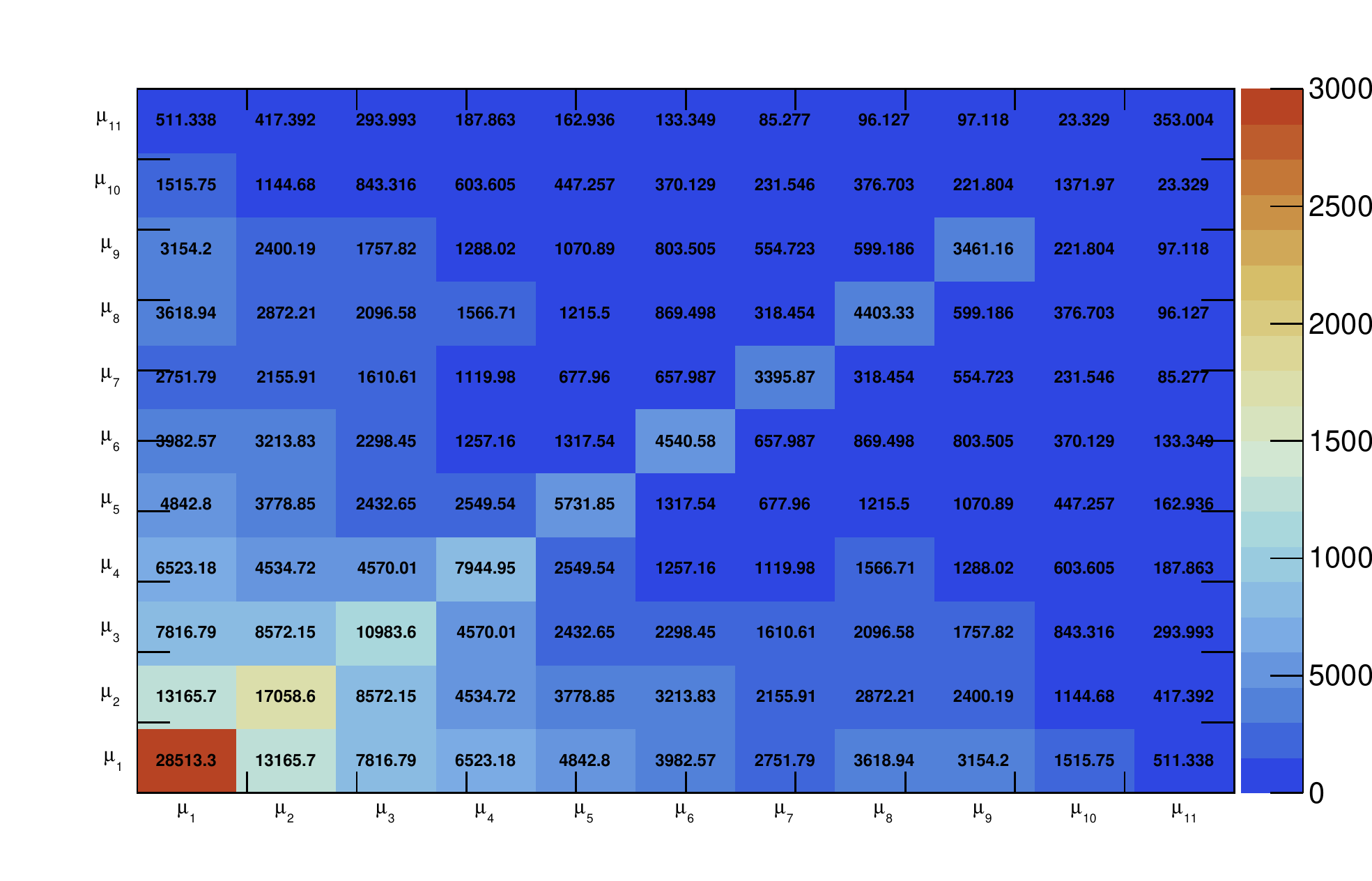}
  \caption{Frequentist $\tau=10^{-5}$}
\end{subfigure}%
\begin{subfigure}{.33\textwidth}
  \includegraphics[width=1\linewidth]{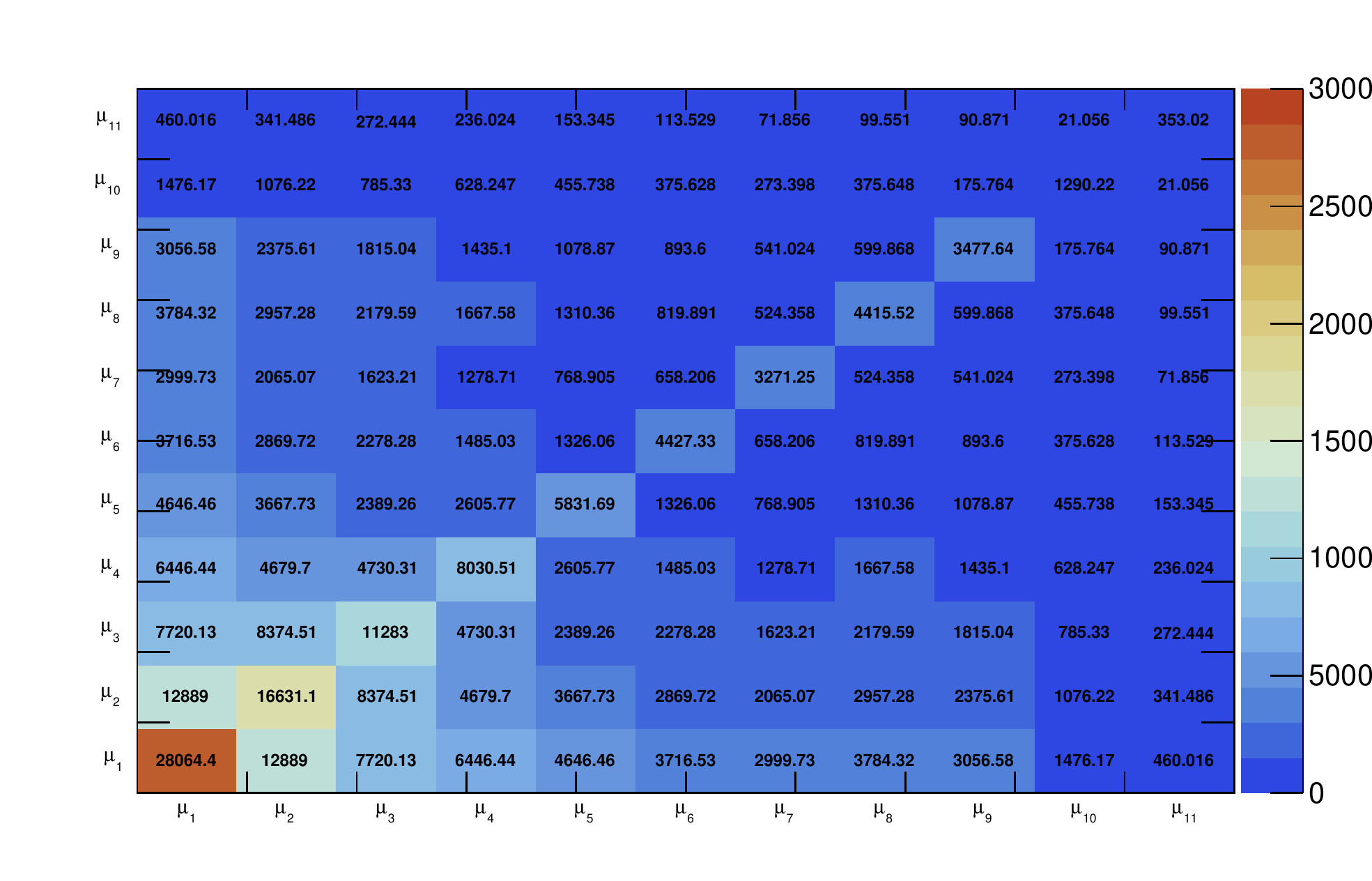}
  \caption{Hybrid $\tau=10^{-5}$}
\end{subfigure}
\begin{subfigure}{.33\textwidth}
  \includegraphics[width=1\linewidth]{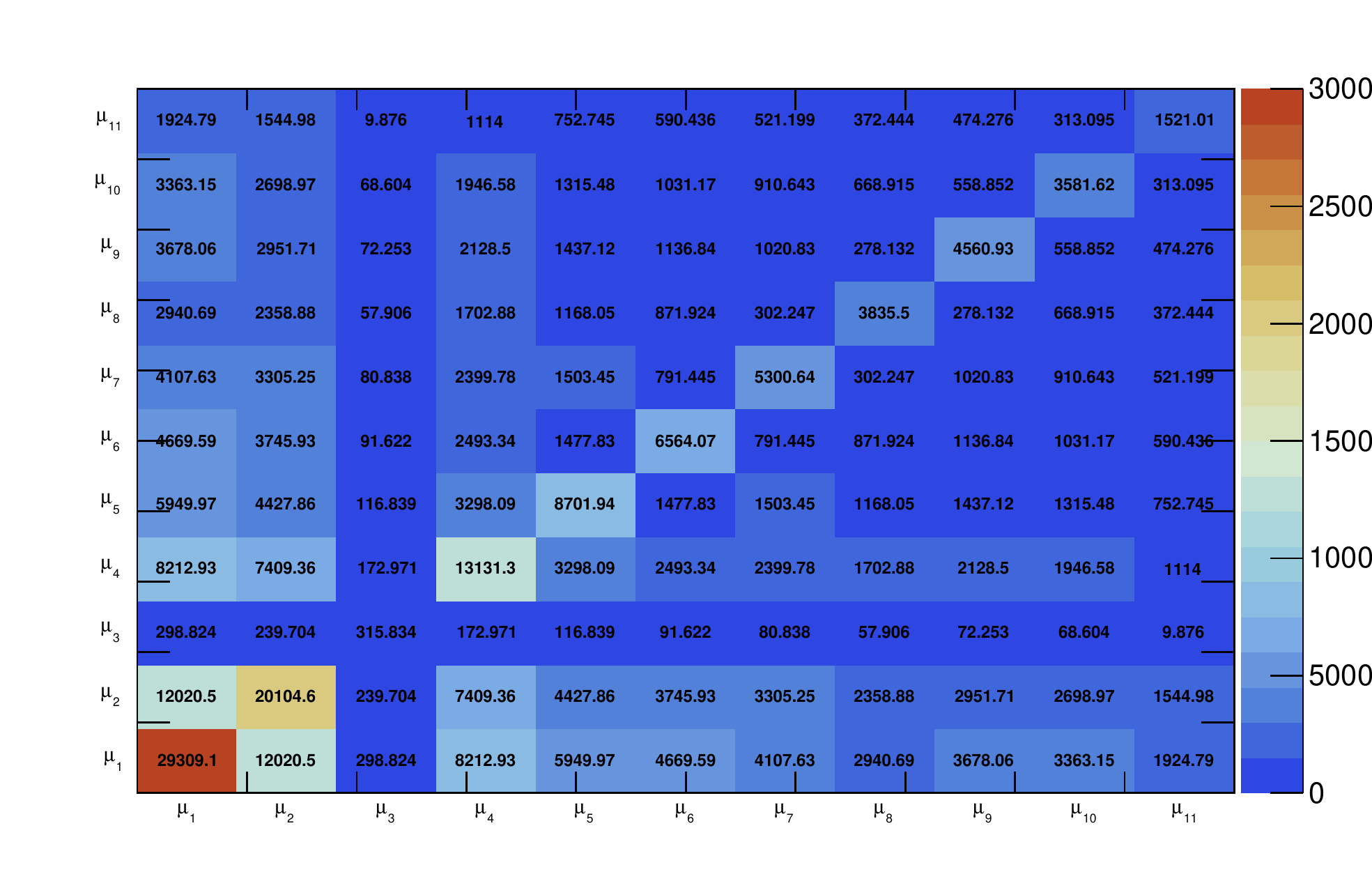}
  \caption{Hessian $\tau=10^{-5}$}
\end{subfigure}

\vskip\baselineskip

\begin{subfigure}{.33\textwidth}
  \includegraphics[width=1\linewidth]{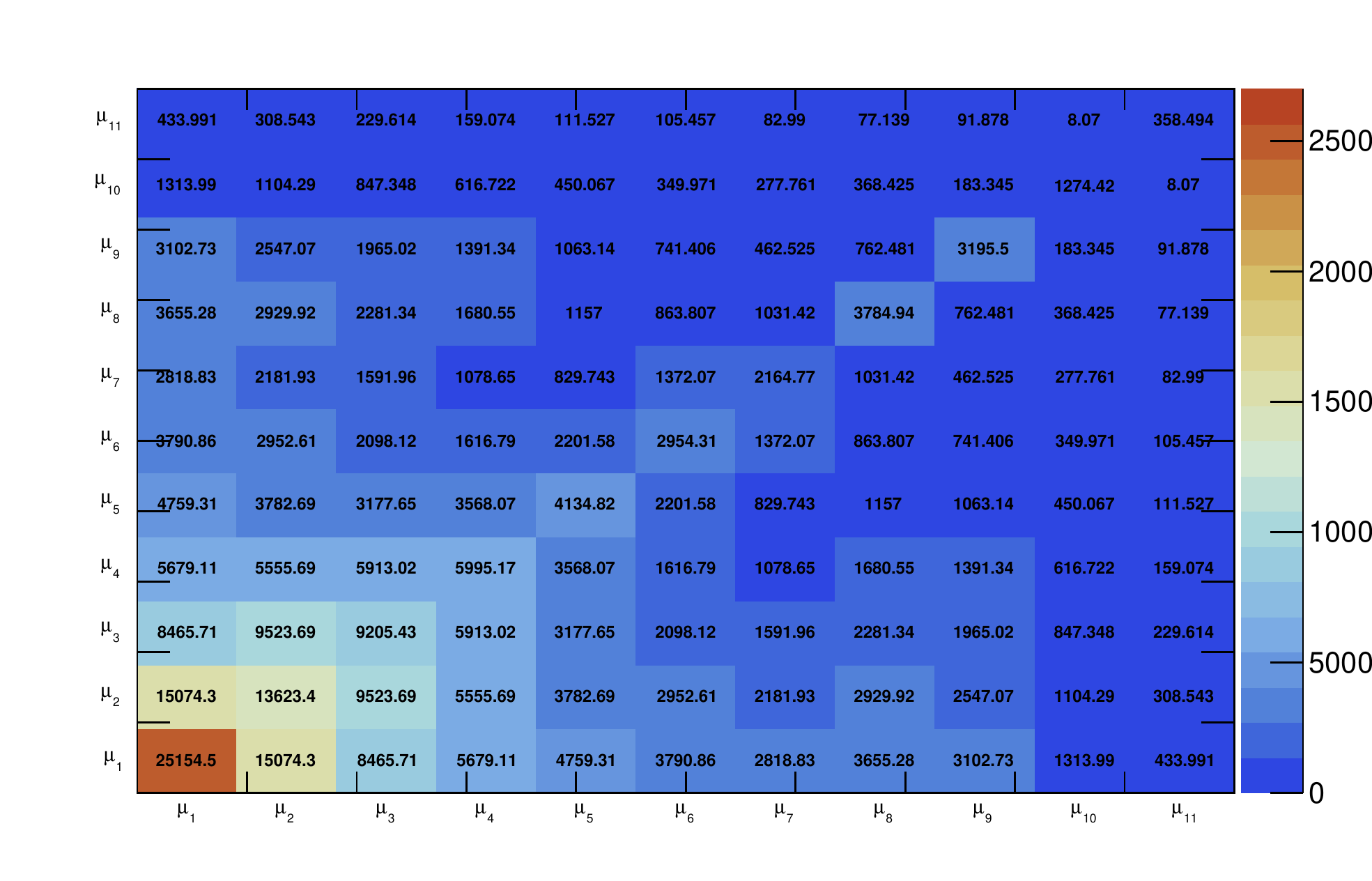}
  \caption{Frequentist $\tau=5\times 10^{-5}$}
\end{subfigure}%
\begin{subfigure}{.33\textwidth}
  \includegraphics[width=1\linewidth]{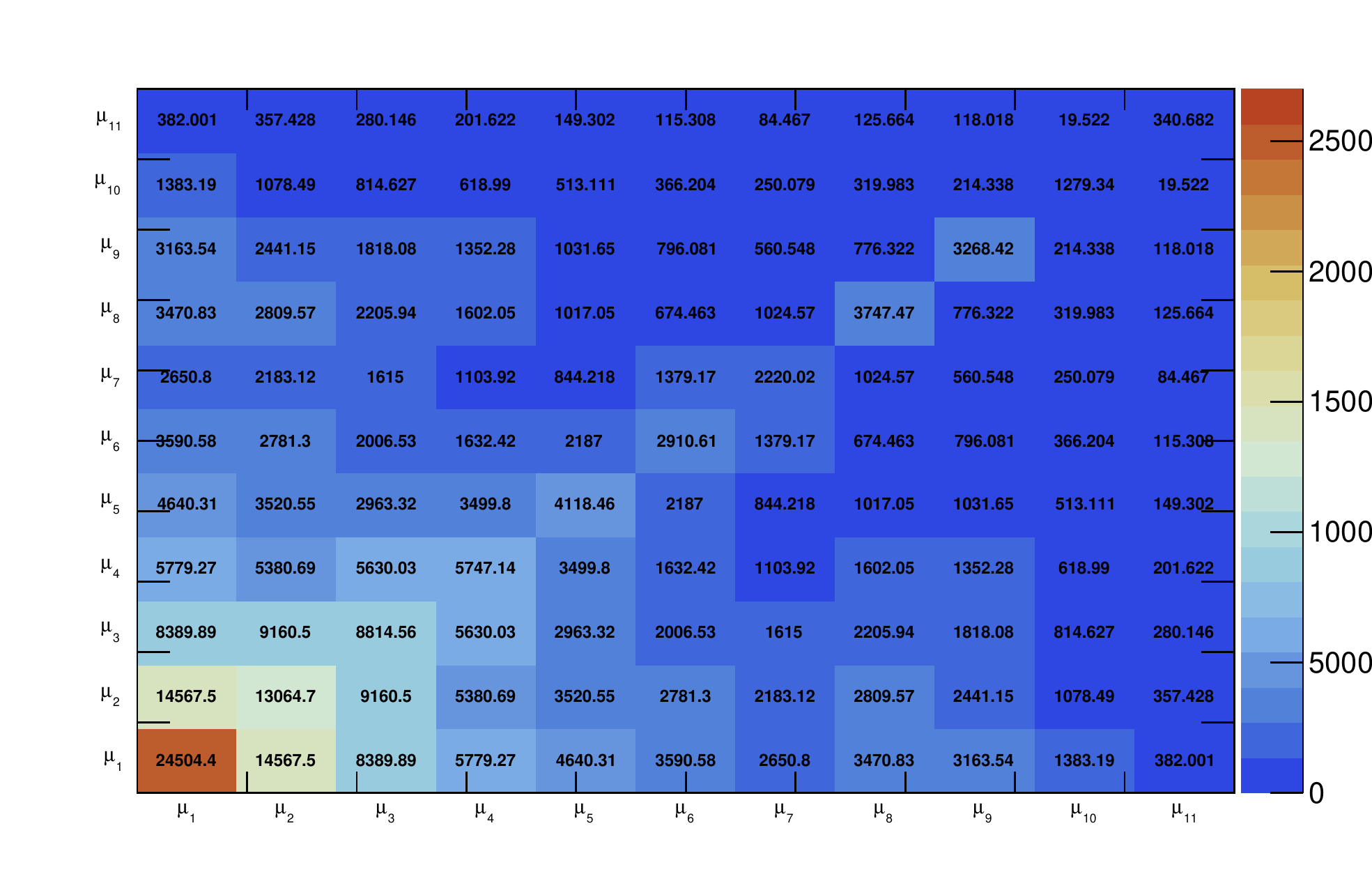}
  \caption{Hybrid $\tau=5\times 10^{-5}$}
\end{subfigure}
\begin{subfigure}{.33\textwidth}
  \includegraphics[width=1\linewidth]{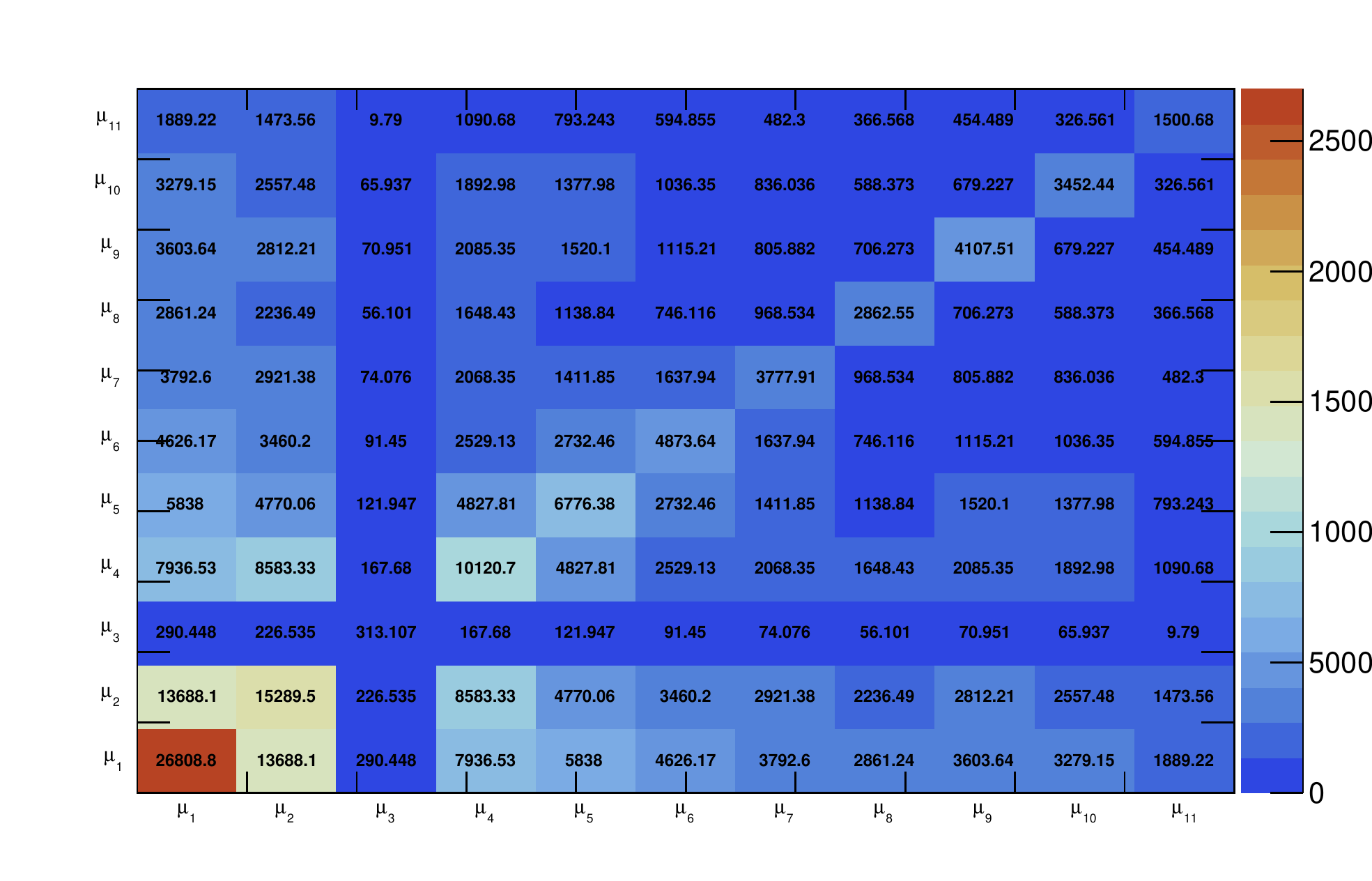}
  \caption{Hessian $\tau=5\times 10^{-5}$}
\end{subfigure}

\caption{Covariance matrix for the exponential distribution estimated with the frequentist pseudo-experiments, inverse hessian and frequentist-bayes hybrid pseudo-experiments method for various regularization strengths $\tau$}
\label{fig:expcovariances}
\end{figure}

\noindent
\subsection{Relative Differences}
In the next plots one can find the relative differences in percentages between the frequentist pseudo-experiment and the inverse hessian and between the frequentist pseudo-experiment and the frequentist-bayes hybrid pseudo-experiments.

\begin{figure}[h!]
\centering
\begin{subfigure}{.4\textwidth}
  \includegraphics[width=1\linewidth]{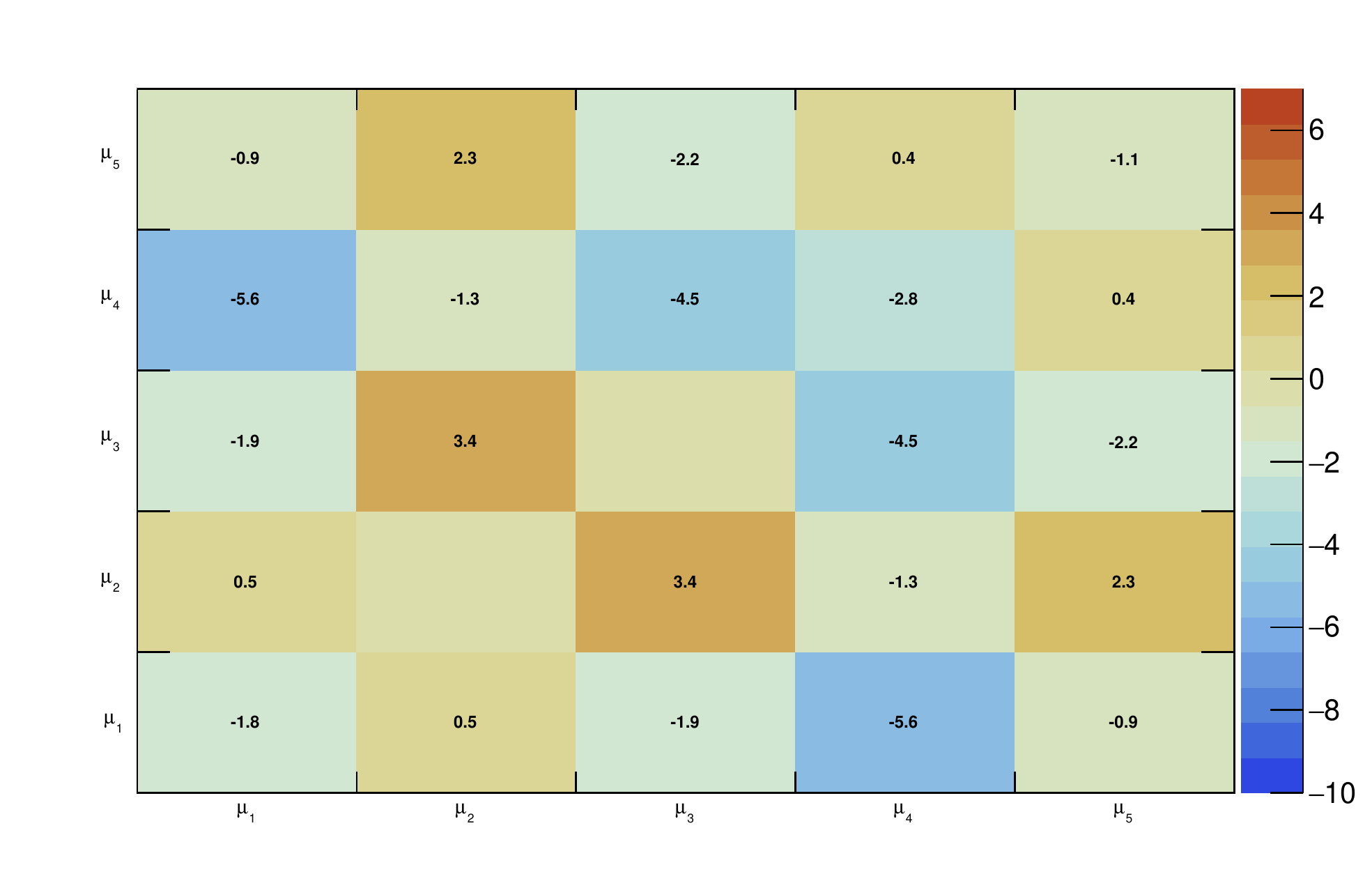}
  \caption{Hybrid $\tau=0.$}
\end{subfigure}
\begin{subfigure}{.4\textwidth}
  \includegraphics[width=1\linewidth]{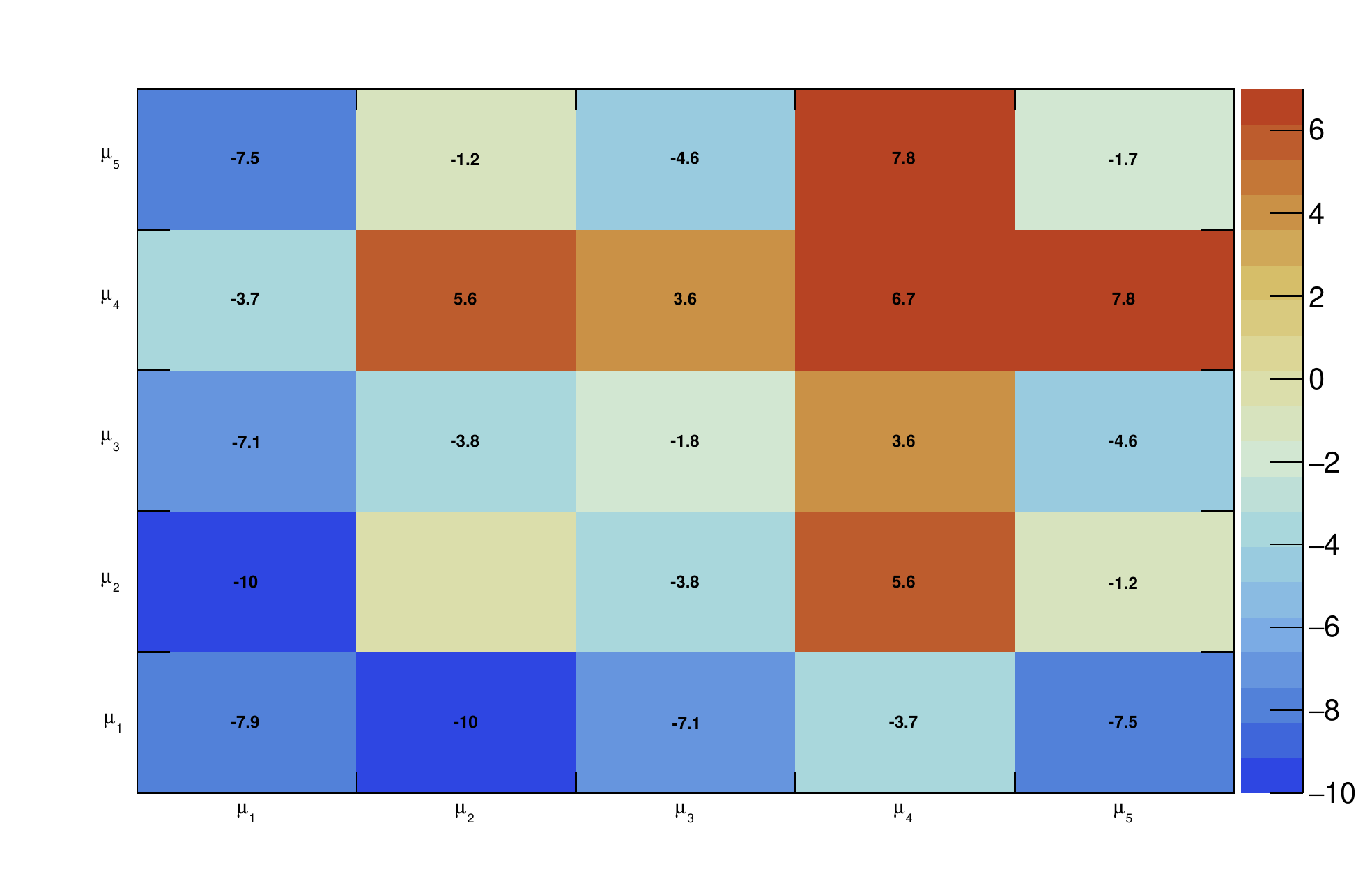}
  \caption{Hessian $\tau=0.$}
\end{subfigure}

\vskip\baselineskip

\begin{subfigure}{.4\textwidth}
  \includegraphics[width=1\linewidth]{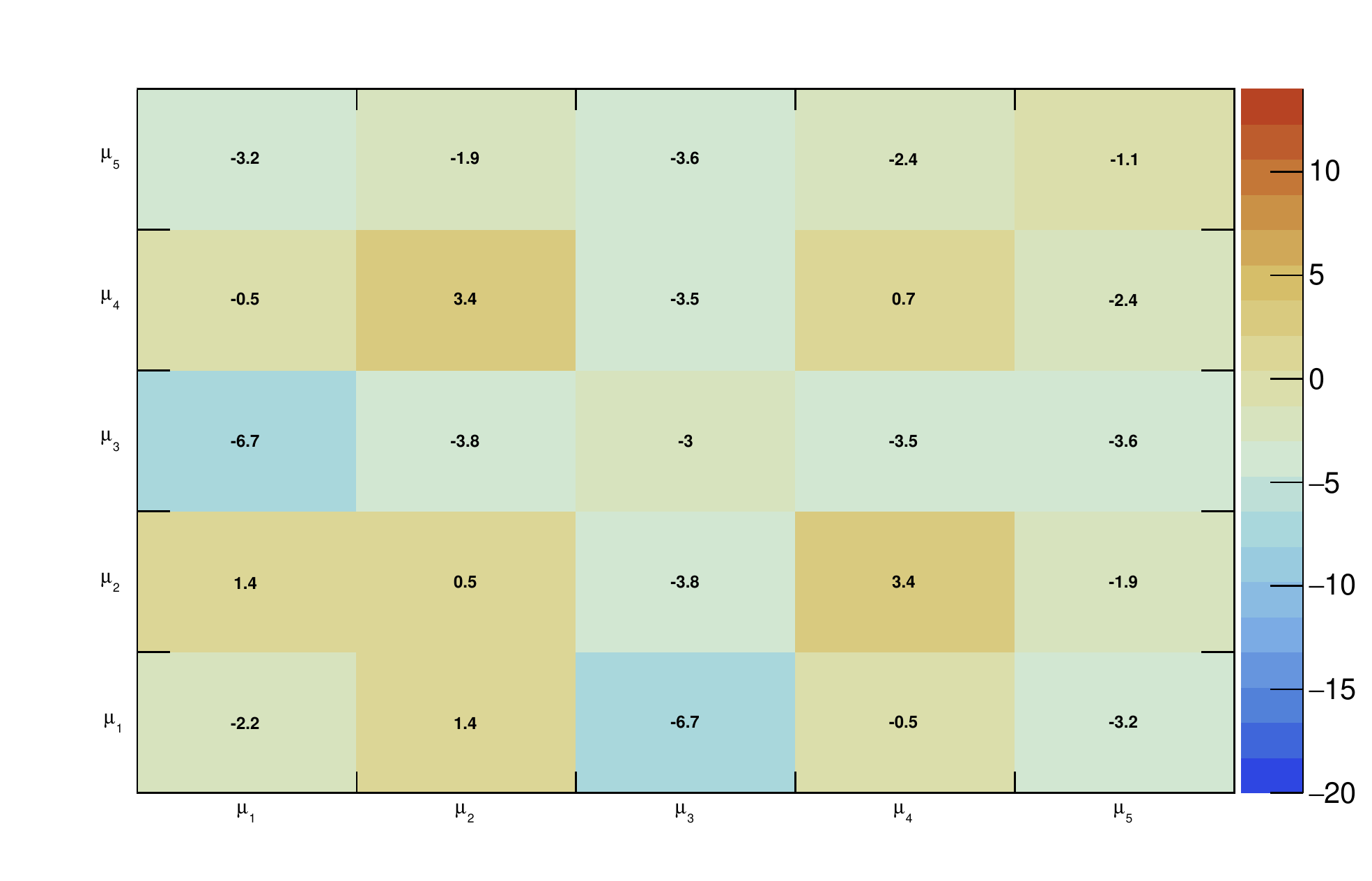}
  \caption{Hybrid $\tau=10^{-6}$}
\end{subfigure}
\begin{subfigure}{.4\textwidth}
  \includegraphics[width=1\linewidth]{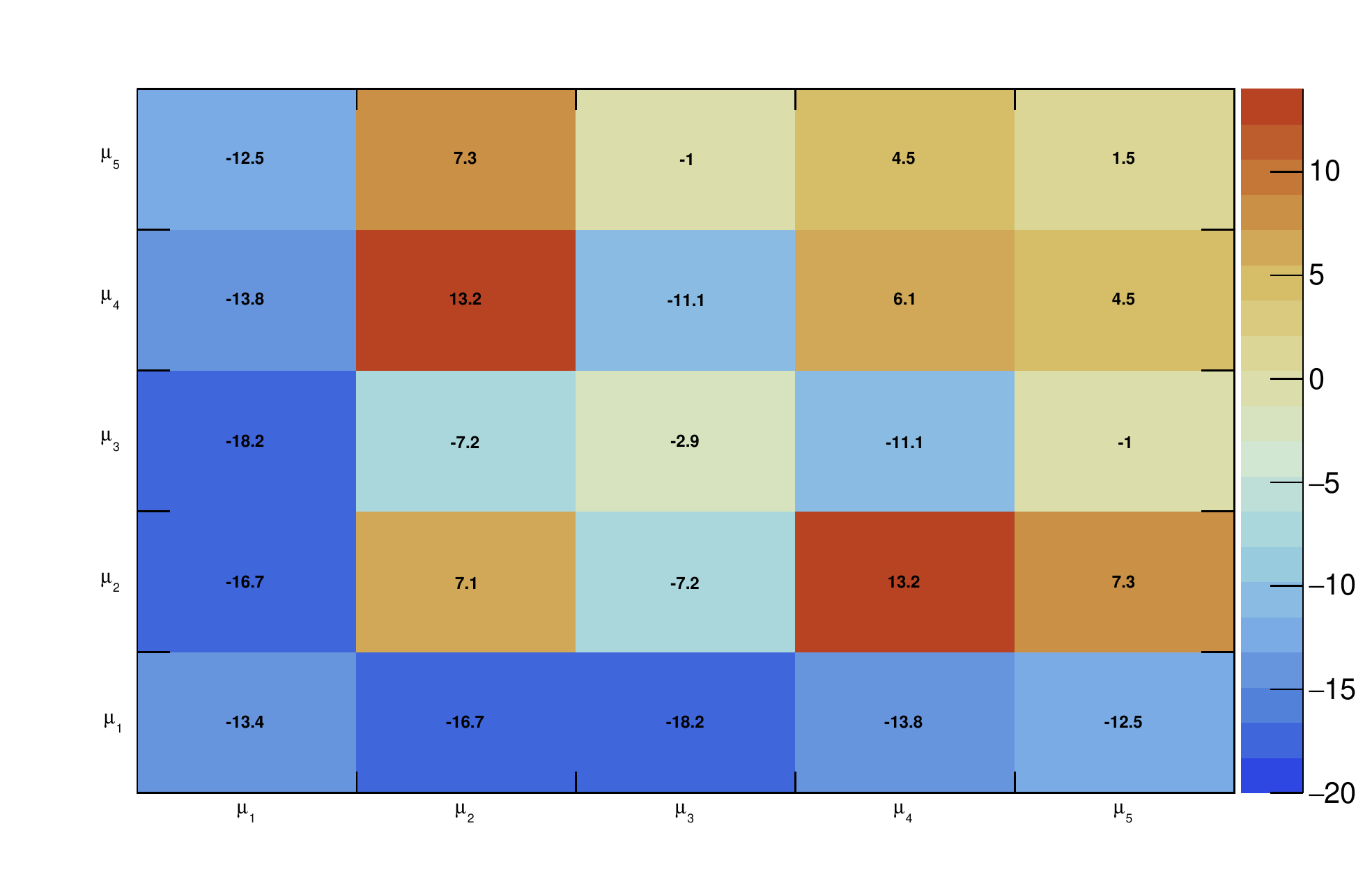}
  \caption{Hessian $\tau=10^{-6}$}
\end{subfigure}

\vskip\baselineskip

\begin{subfigure}{.4\textwidth}
  \includegraphics[width=1\linewidth]{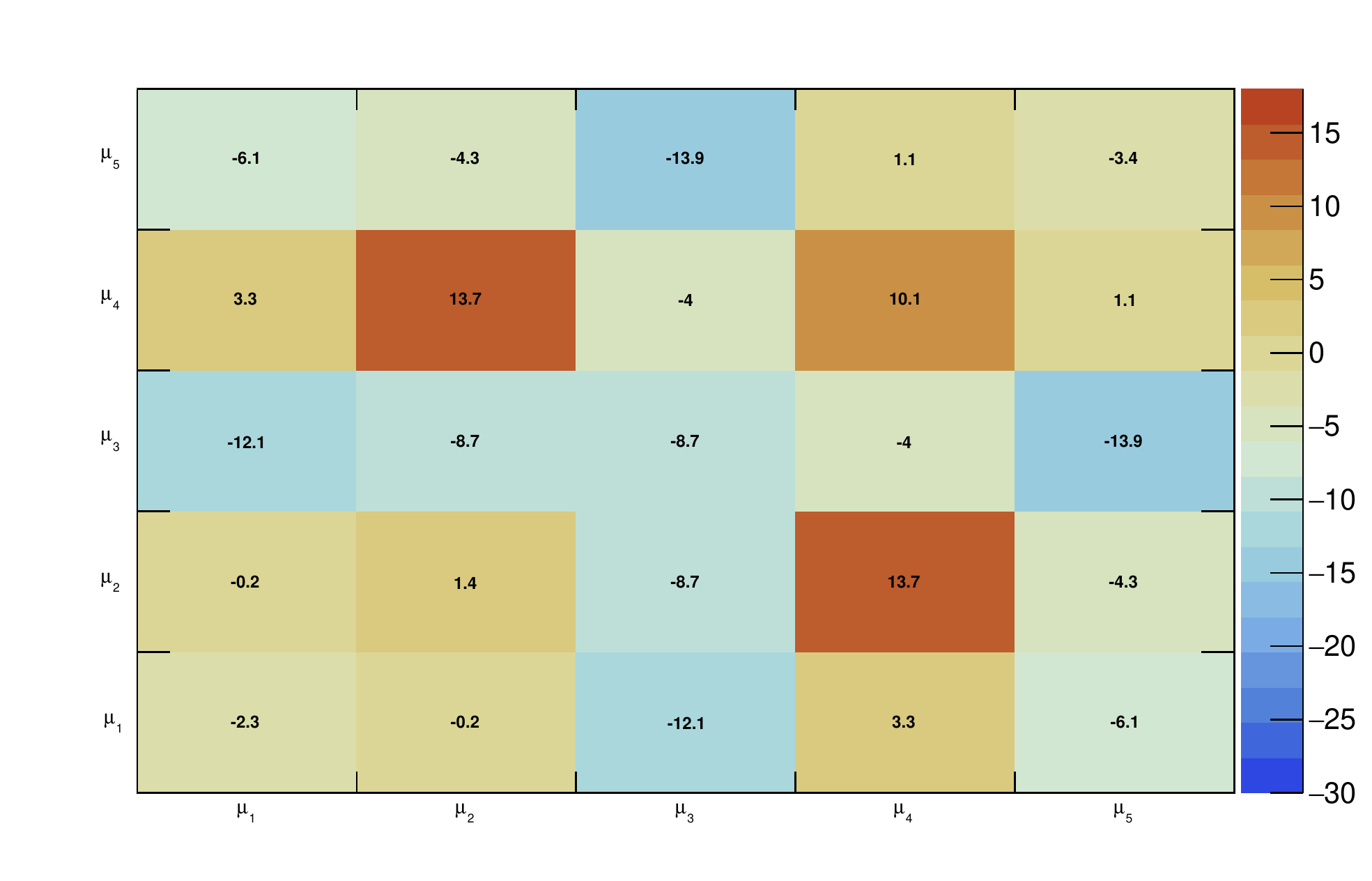}
  \caption{Hybrid $\tau=10^{-5}$}
\end{subfigure}
\begin{subfigure}{.4\textwidth}
  \includegraphics[width=1\linewidth]{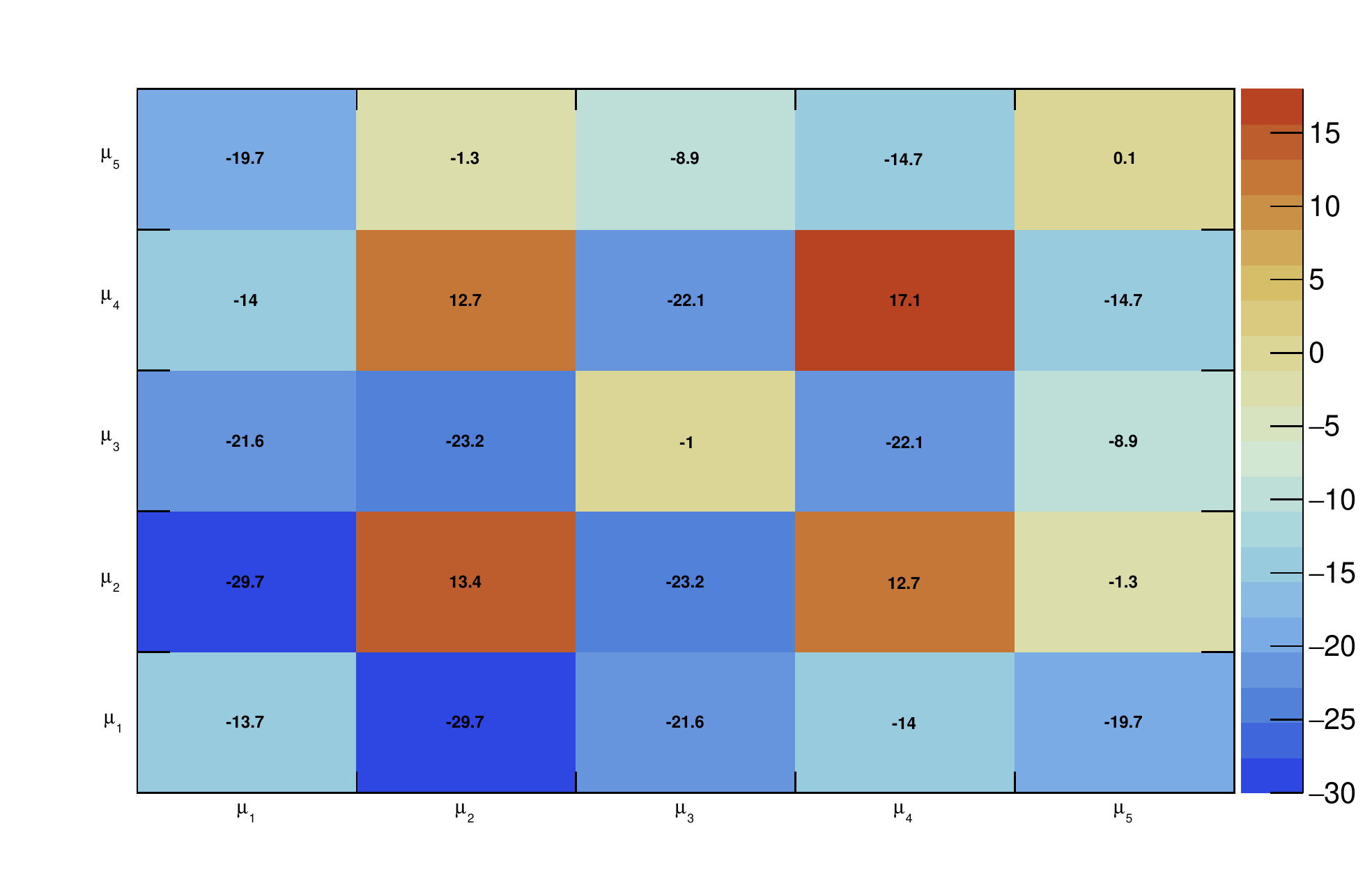}
  \caption{Hessian $\tau=10^{-5}$}
\end{subfigure}

\vskip\baselineskip

\begin{subfigure}{.4\textwidth}
  \includegraphics[width=1\linewidth]{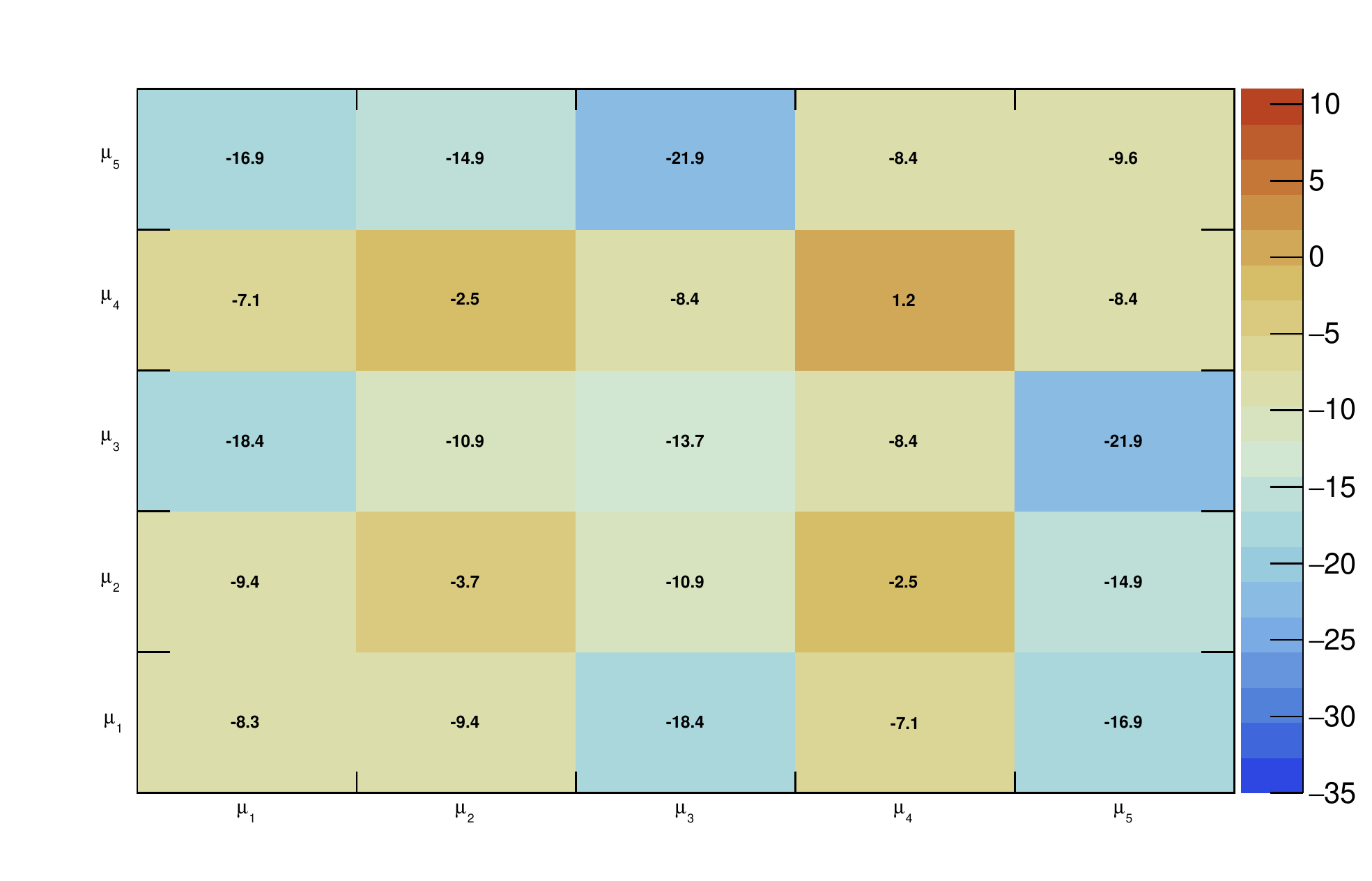}
  \caption{Hybrid $\tau=5\times 10^{-5}$}
\end{subfigure}
\begin{subfigure}{.4\textwidth}
  \includegraphics[width=1\linewidth]{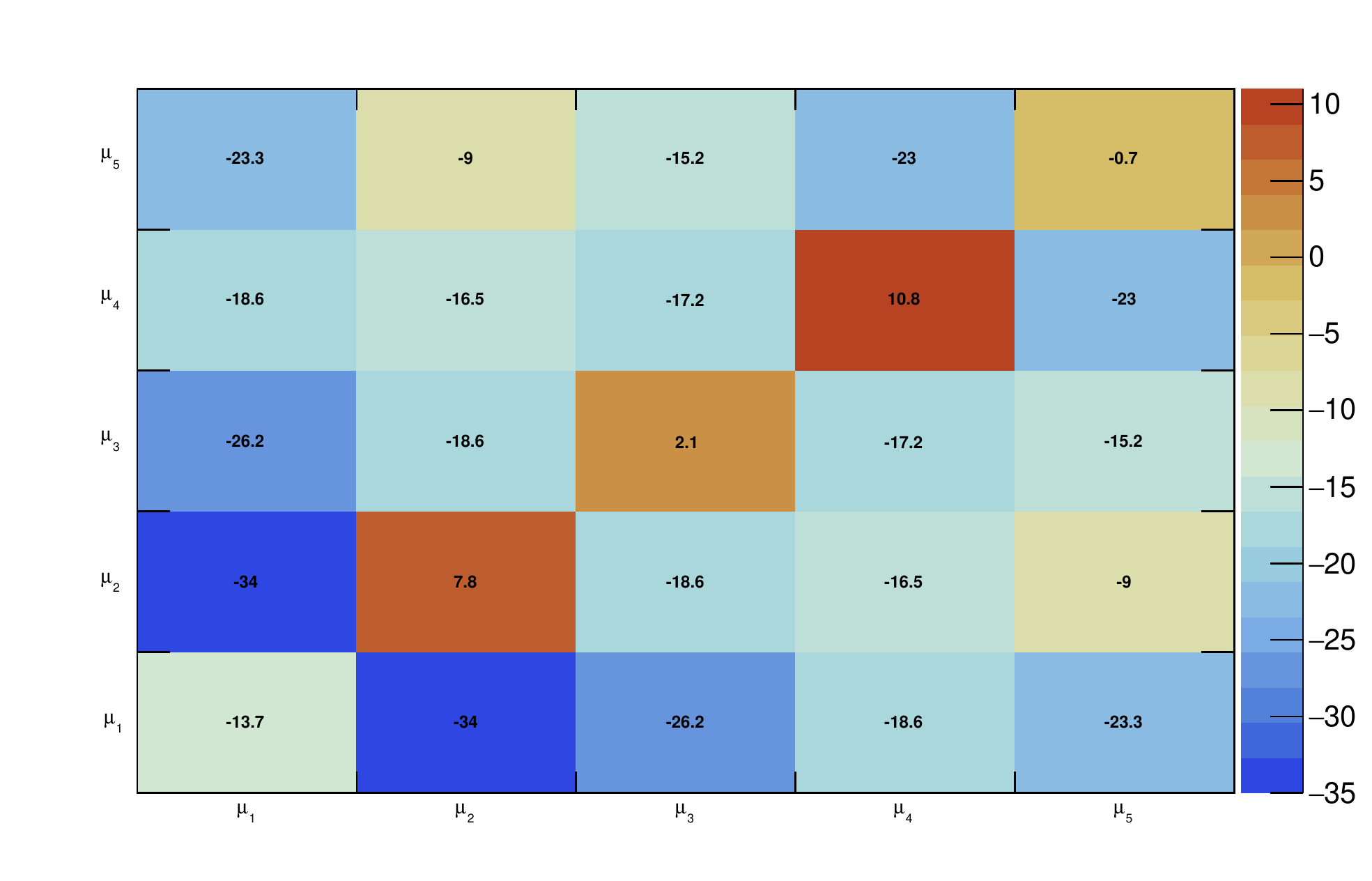}
  \caption{Hessian $\tau=5\times 10^{-5}$}
\end{subfigure}

\caption{Covariance matrix differences in percentage for the bimodal distribution between the fully frequentist toy method and the inverse hessian method(1nd column) and hybrid toy method(2nd column) for various regularization strengths $\tau$}
\label{fig:bimodaldiffcov}
\end{figure}

\newpage

\begin{figure}[h!]
\centering
\begin{subfigure}{.4\textwidth}
  \includegraphics[width=1\linewidth]{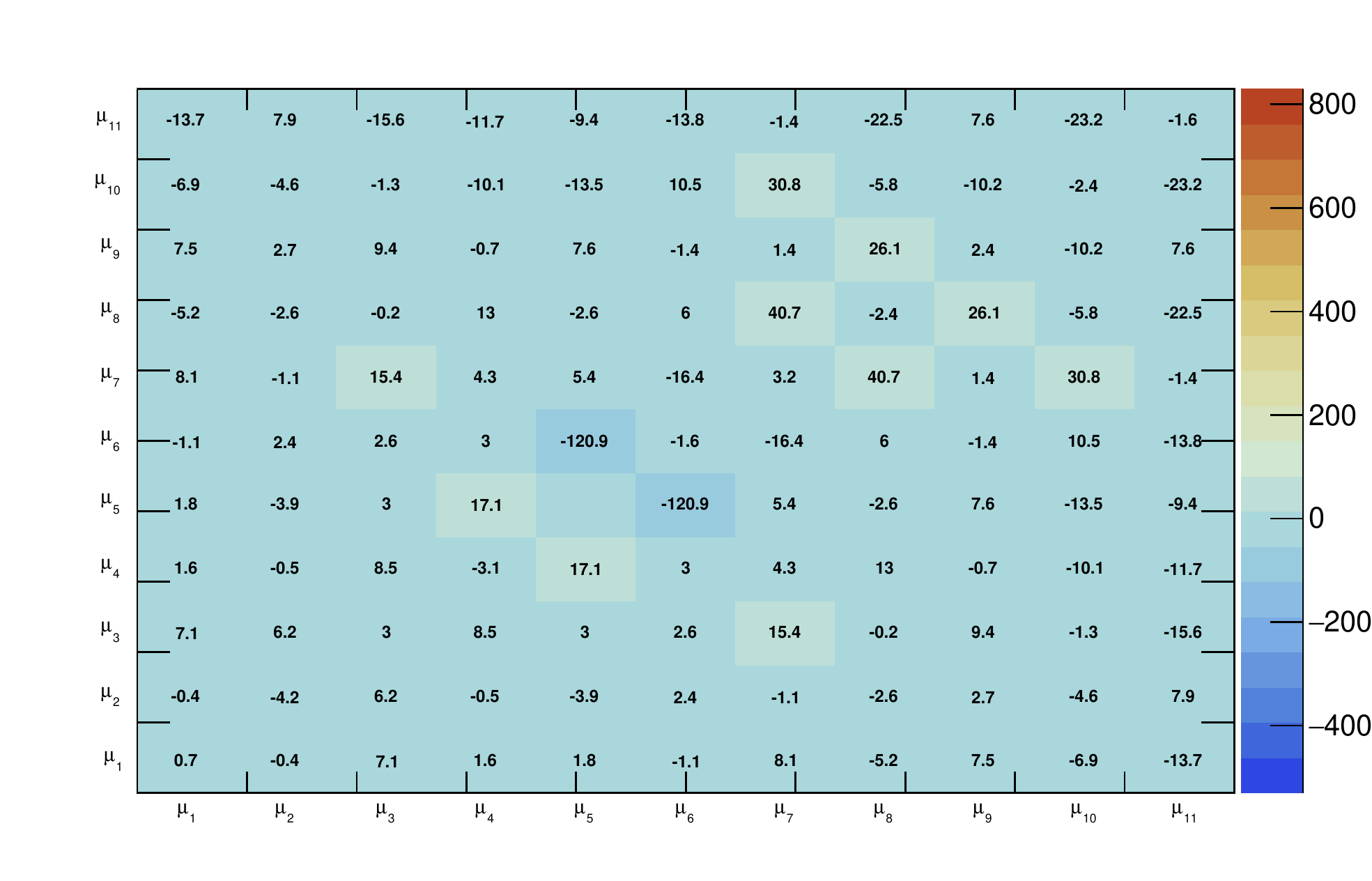}
  \caption{Hybrid $\tau=0.$}
\end{subfigure}
\begin{subfigure}{.4\textwidth}
  \includegraphics[width=1\linewidth]{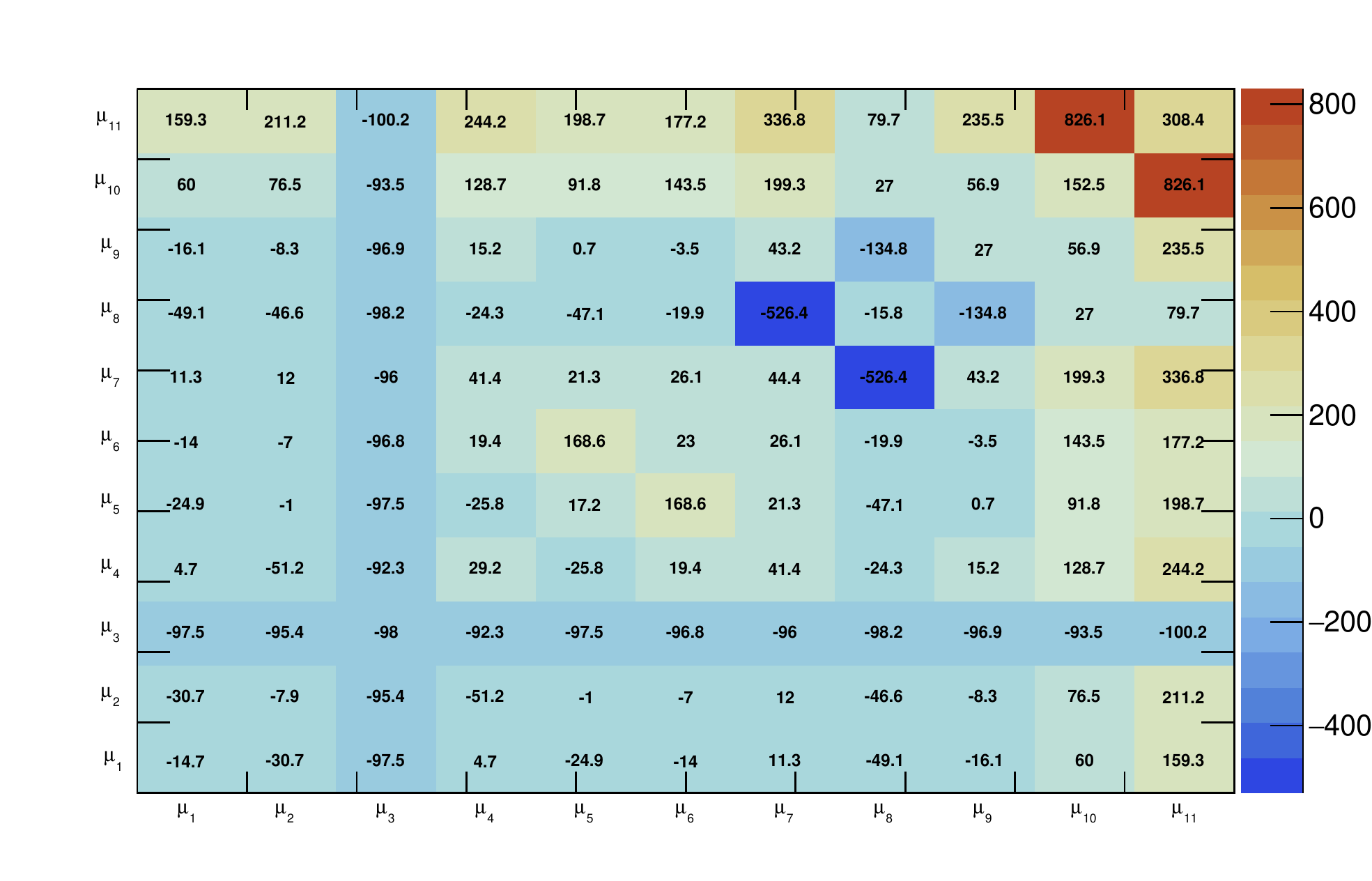}
  \caption{Hessian $\tau=0.$}
\end{subfigure}

\vskip\baselineskip

\begin{subfigure}{.4\textwidth}
  \includegraphics[width=1\linewidth]{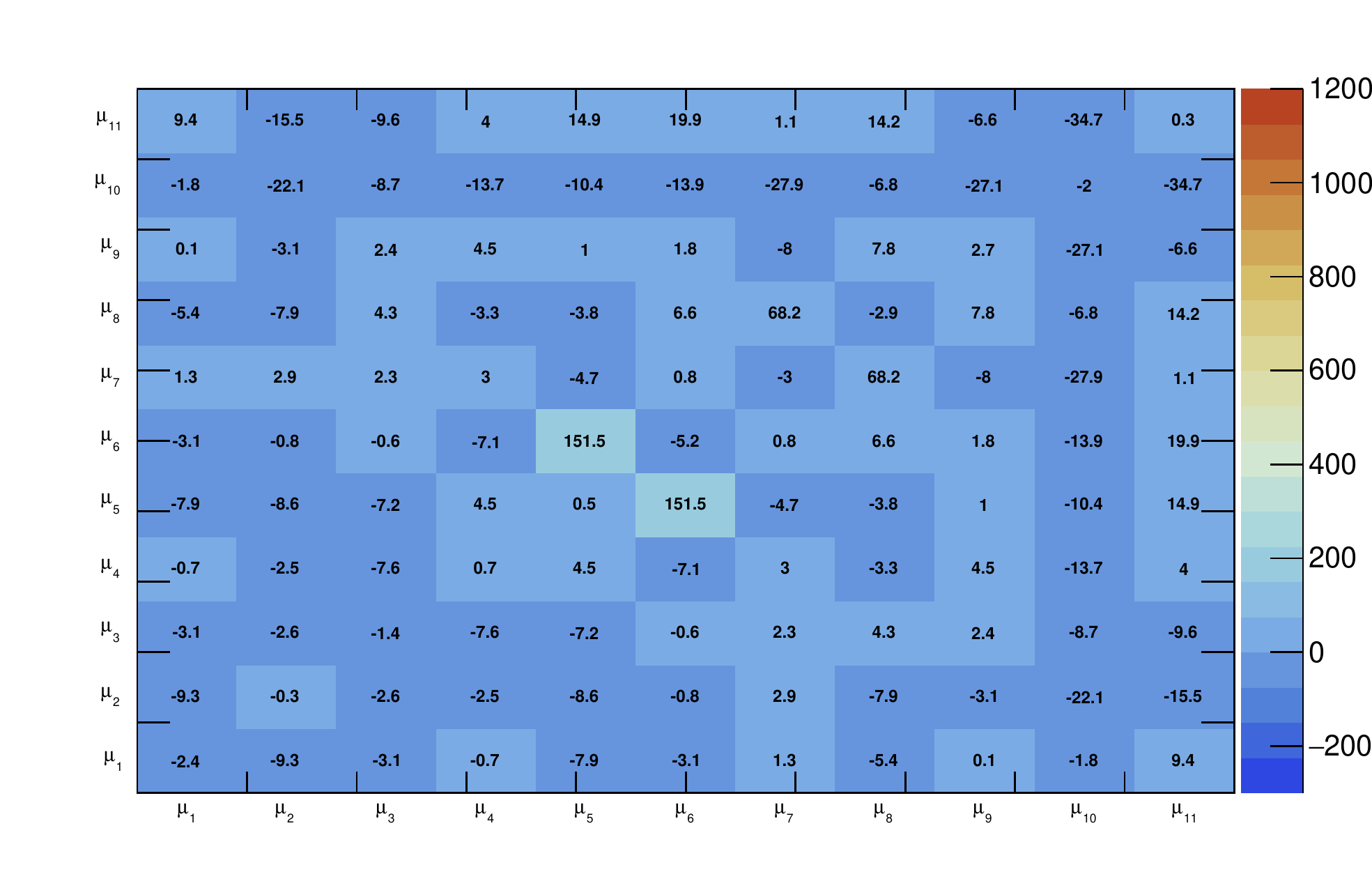}
  \caption{Hybrid $\tau=10^{-6}$}
\end{subfigure}
\begin{subfigure}{.4\textwidth}
  \includegraphics[width=1\linewidth]{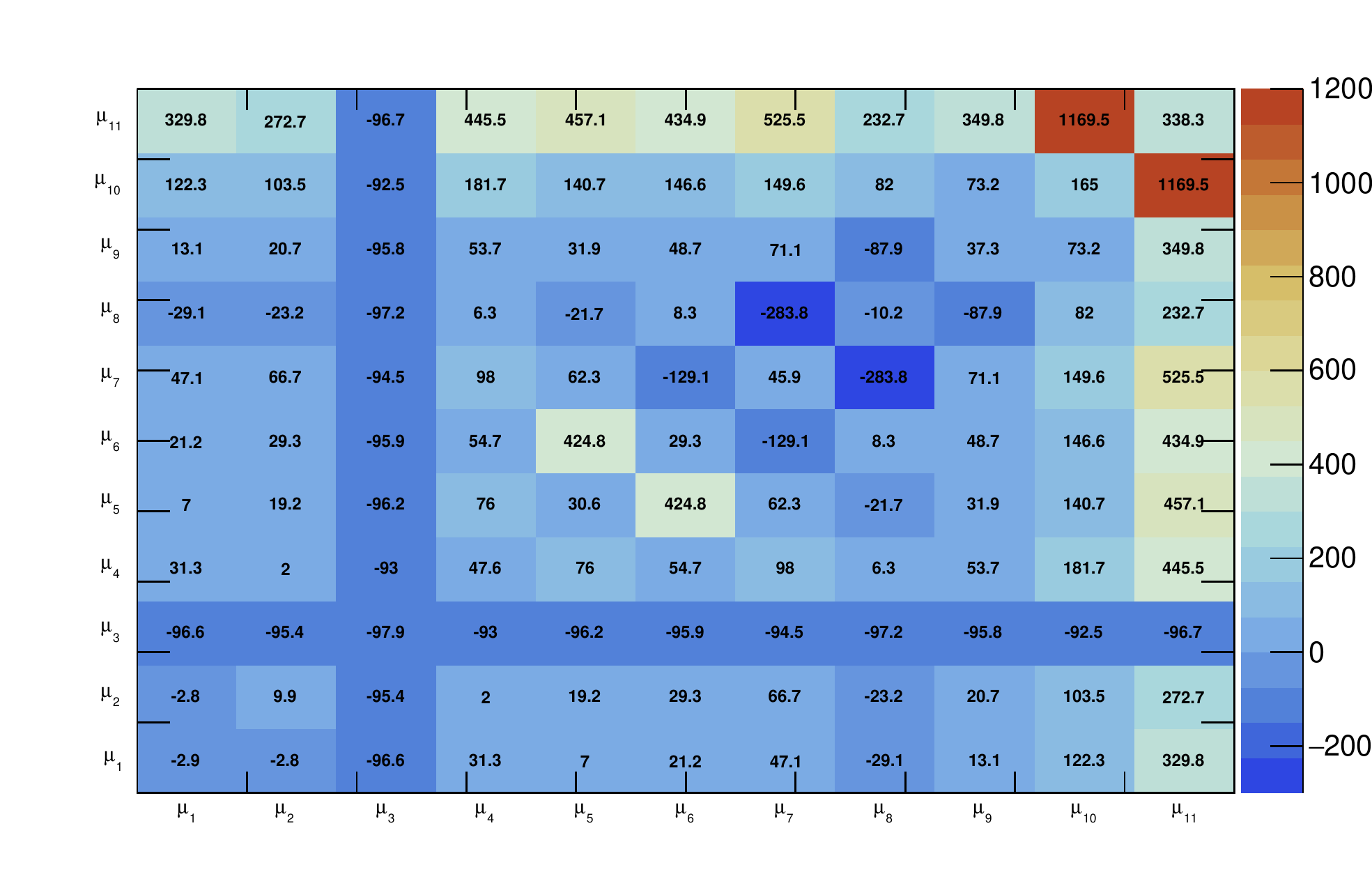}
  \caption{Hessian $\tau=10^{-6}$}
\end{subfigure}

\vskip\baselineskip

\begin{subfigure}{.4\textwidth}
  \includegraphics[width=1\linewidth]{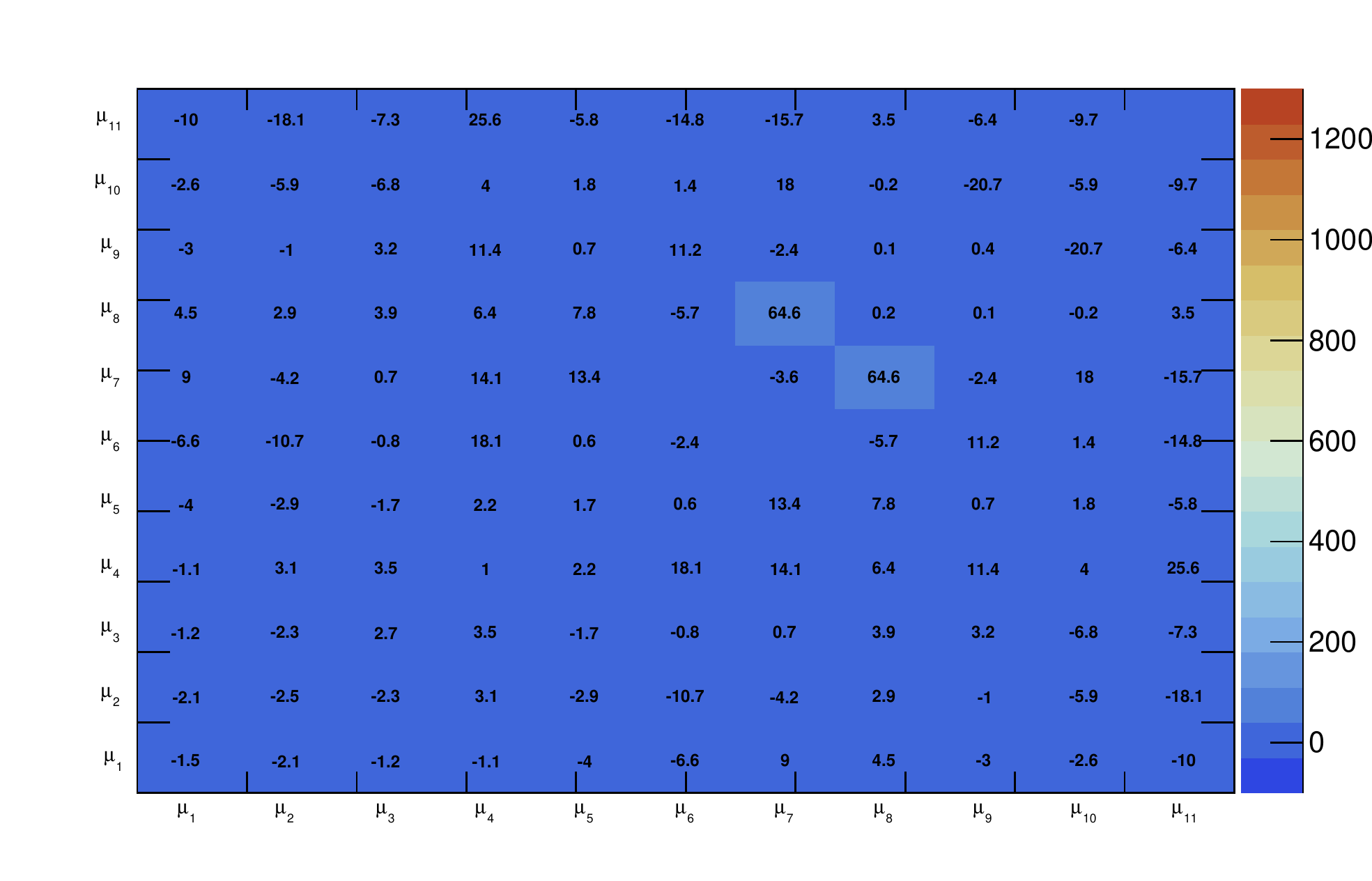}
  \caption{Hybrid $\tau=10^{-5}$}
\end{subfigure}
\begin{subfigure}{.4\textwidth}
  \includegraphics[width=1\linewidth]{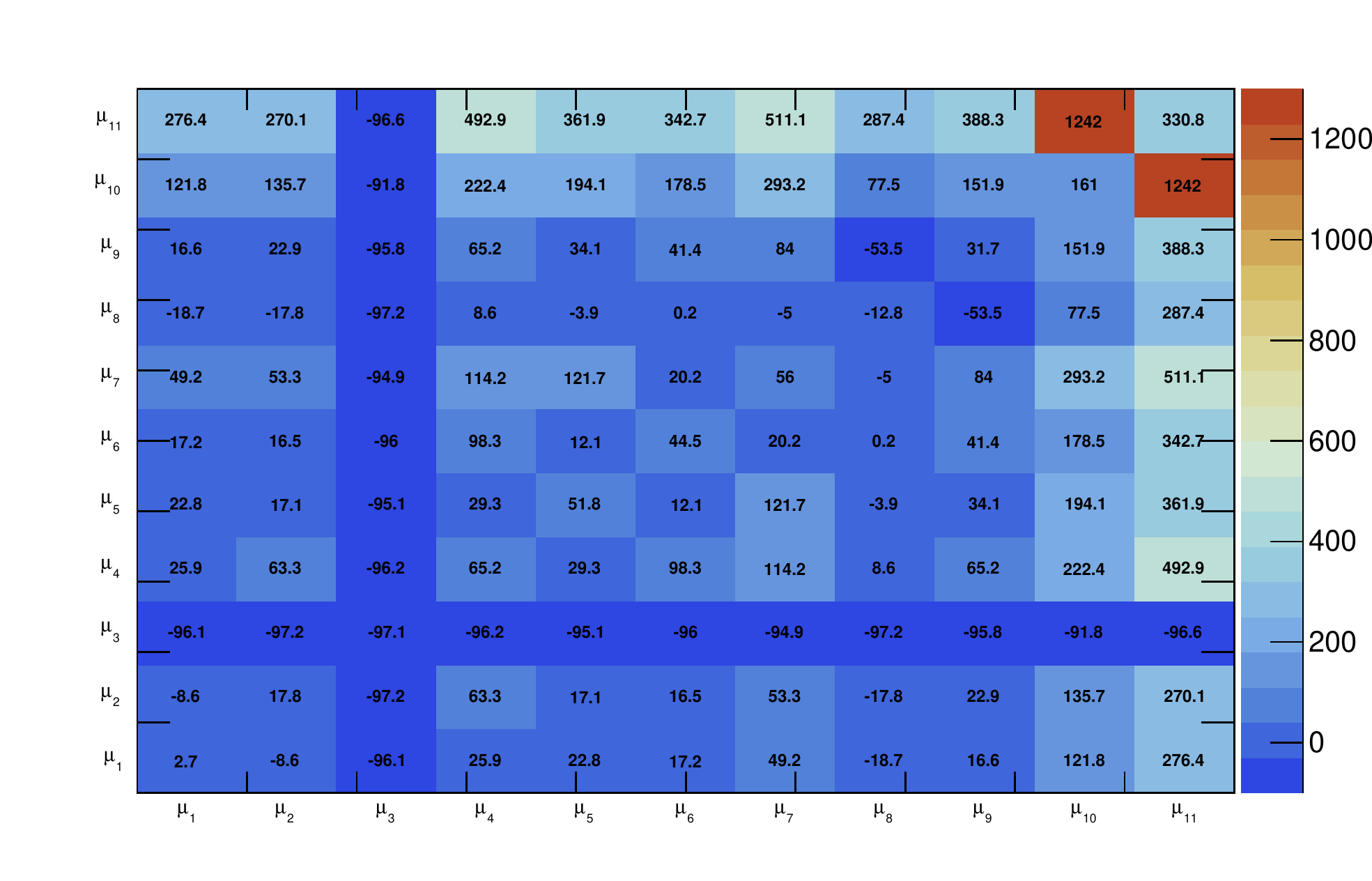}
  \caption{Hessian $\tau=10^{-5}$}
\end{subfigure}

\vskip\baselineskip

\begin{subfigure}{.4\textwidth}
  \includegraphics[width=1\linewidth]{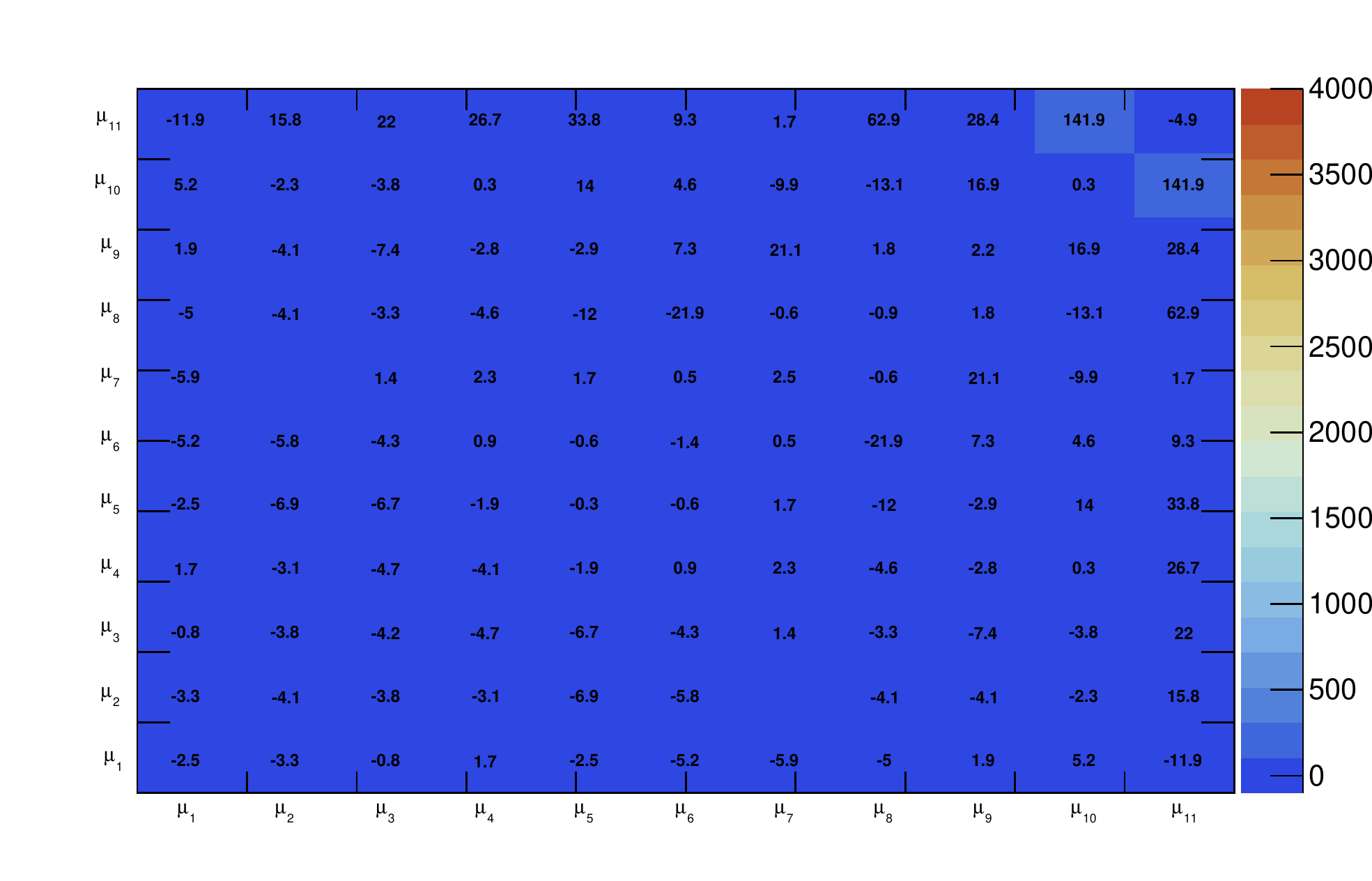}
  \caption{Hybrid $\tau=5\times 10^{-5}$}
\end{subfigure}
\begin{subfigure}{.4\textwidth}
  \includegraphics[width=1\linewidth]{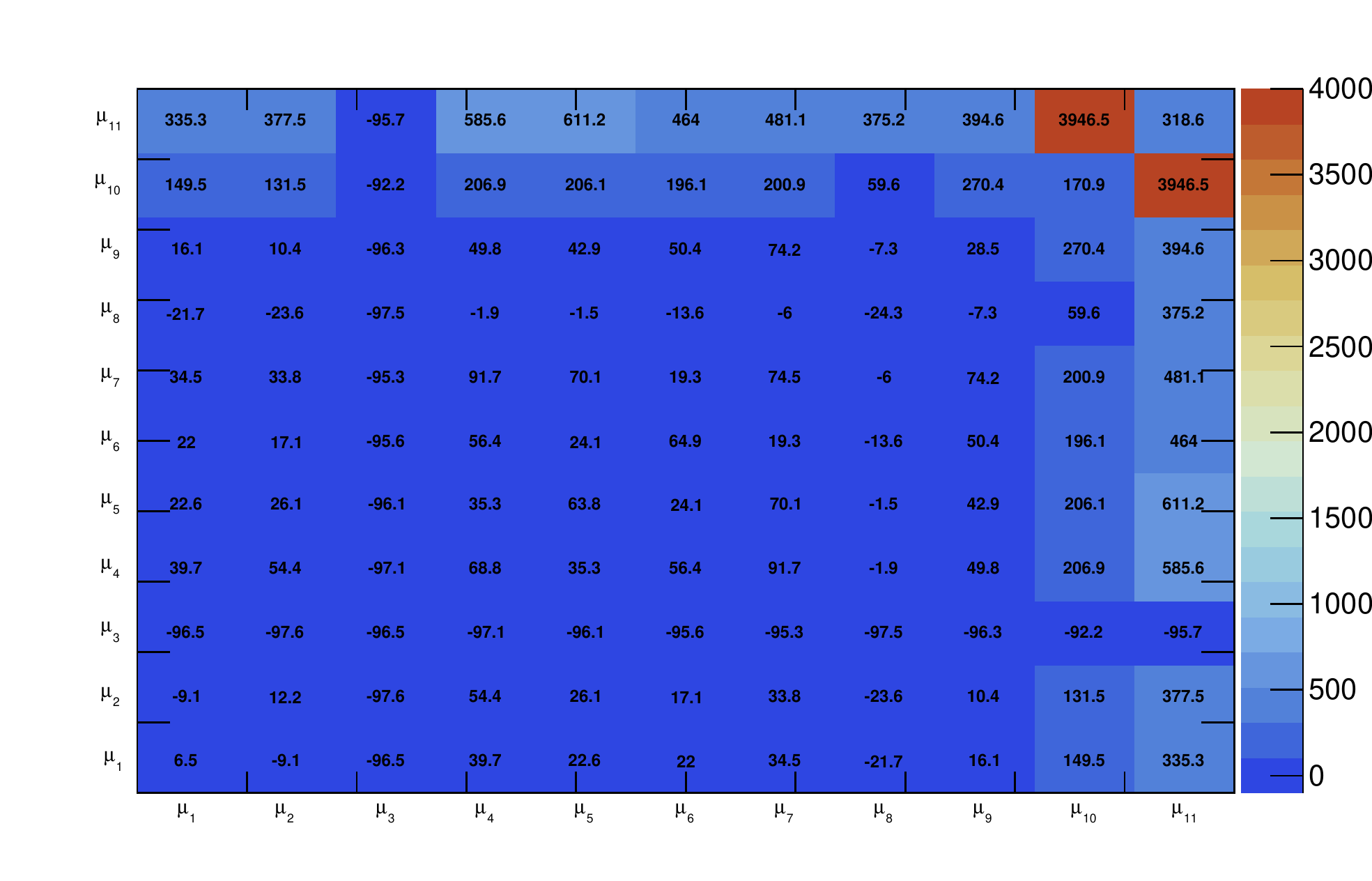}
  \caption{Hessian $\tau=5\times 10^{-5}$}
\end{subfigure}

\caption{Covariance matrix differences in percentage for the exponential distribution between the fully frequentist toy method and the inverse hessian method(1nd column) and hybrid toy method(2nd column) for various regularization strengths $\tau$}

\label{fig:expdiffcov}
\end{figure}

\newpage
\noindent
One can see that the differences between the frequentist and the hybrid method are substantially smaller than between the frequentist and the inverse Hessian method regardless of the amount of regularization, distribution shape or detector function. The inverse Hessian method shows big differences for the exponential distribution for all values of $\tau$. 

\subsection{Summary Statistics}
Covariance matrix estimates can be further compared with summary statistics that are common in HEP. These make the (dis)agreements more obvious and can have a practical uses for e.g. fine tuning the regularization parameter $\tau$\cite{schmittreview,Balasubramanian:2019itp}.

\subsubsection{Unfolding Errors}
The diagonal elements of the covariance matrix are commonly used as estimators for the errors on the truth bin estimators. An average is taken to make the comparison clearer. Additionally, the relative error is taken to ensure that low-count bins are weighted the same as high-count bins in the average. 

\begin{equation}\label{eq:averror}
    \mbox{Average } \sigma_{rel.} = \frac{1}{N} \sum_{i=1}^{N}\frac{\sqrt{V[\hat{\mu}_{i},\hat{\mu}_{i}]}}{\hat{\mu}_{i}}
\end{equation}

\subsubsection{Global Correlation Coefficients}
Global correlation coefficients\cite{schmittreview} estimate the correlation between an estimator $\hat{\mu}_{i}$ and a linear combination of estimators of the remaining truth bins. An average is again taken to make the comparison clearer.

\begin{equation}\label{eq:avglobcor}
    \mbox{Average } \rho = \frac{1}{N} \sum_{i=1}^{N}\sqrt{1-((V)_{ii}(V)^{-1}_{ii})^{-1}}
\end{equation}

\subsubsection{Chi-Square}
A chi-squared test is often used to quantify the agreement between the constructed estimators $\hat{\vec{\mu}}$ and a truth distribution $\vec{\mu}$.

\begin{equation}\label{eq:chindf}
    \chi^{2}/\mbox{n.d.f.} = \frac{1}{N}(\vec{\hat{\mu}} - \vec{\mu})V^{-1}(\vec{\hat{\mu}} - \vec{\mu})^{T}
\end{equation}

\newpage

\begin{figure}[h!]
\centering

\begin{subfigure}{.4\textwidth}
  \includegraphics[width=1\linewidth]{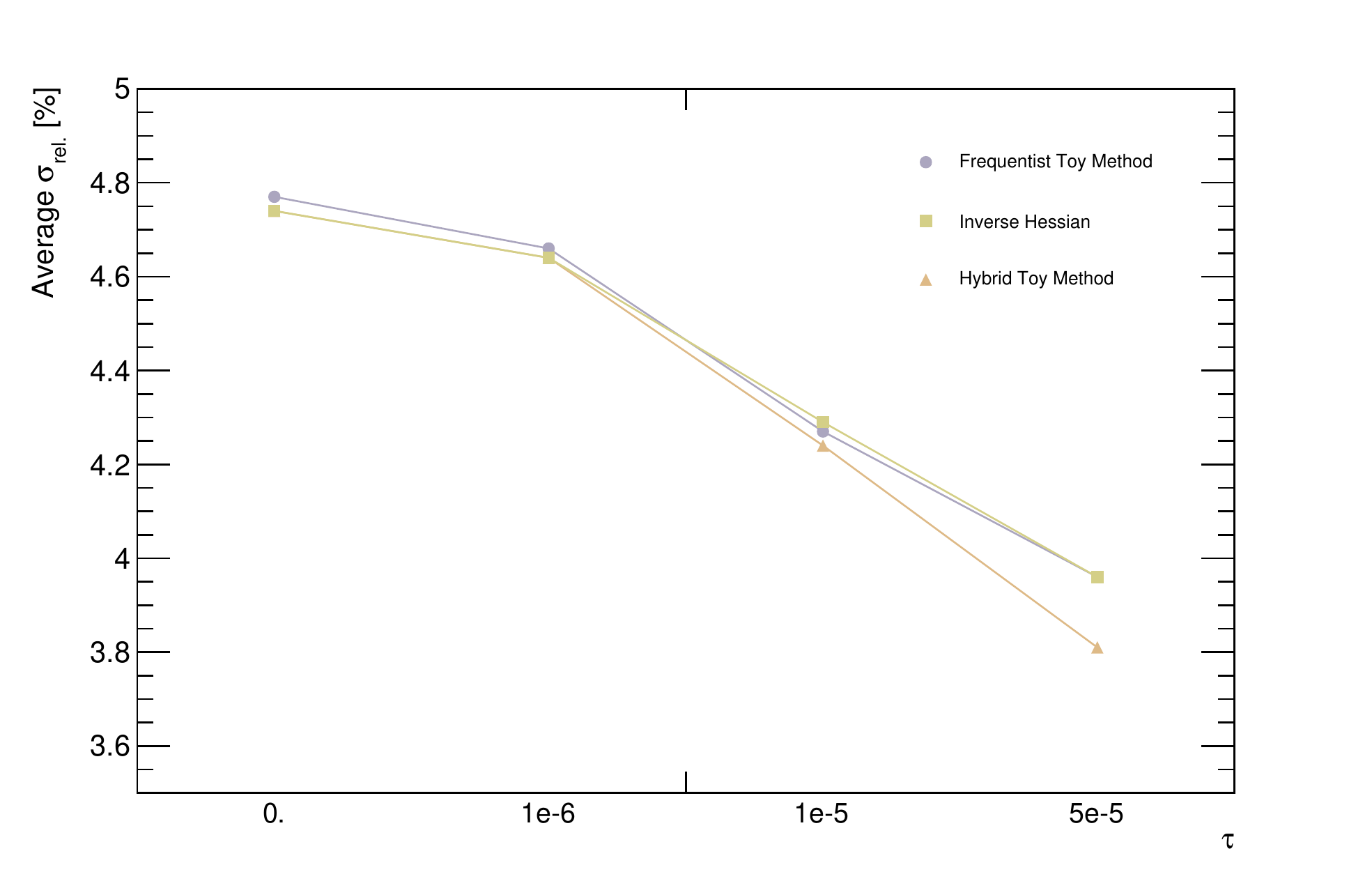}
  \caption{Double Gaussian distribution}
\end{subfigure}
\begin{subfigure}{.4\textwidth}
  \includegraphics[width=1\linewidth]{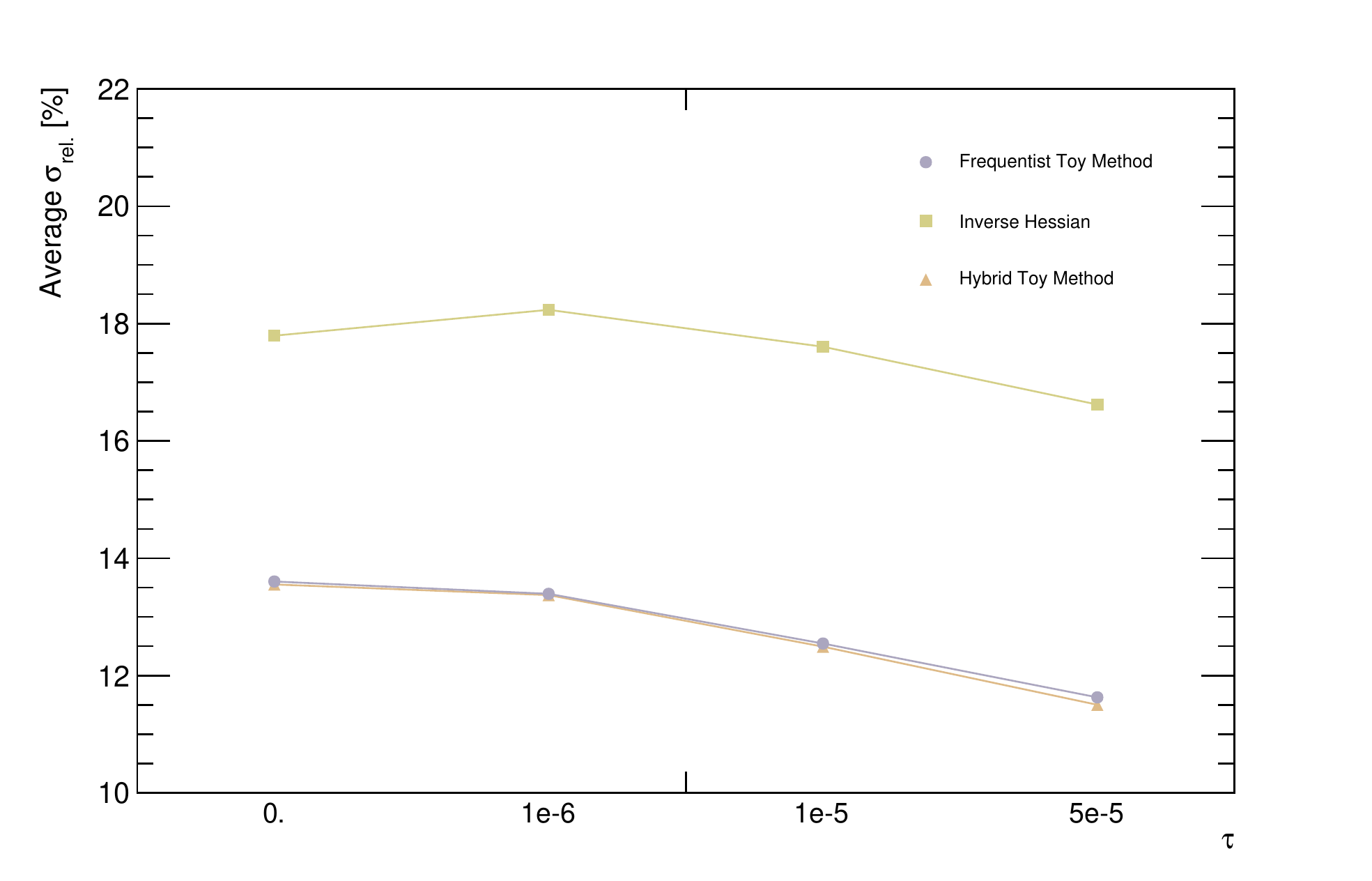}
  \caption{Exponential distribution}
\end{subfigure}

\caption{Average relative unfolding errors for a) the double Gaussian and b) the exponential distribution for all covariance estimation methods and various regularization strengths $\tau$}

\label{fig:relerror}
\end{figure}

\begin{figure}[h!]
\centering

\begin{subfigure}{.4\textwidth}
  \includegraphics[width=1\linewidth]{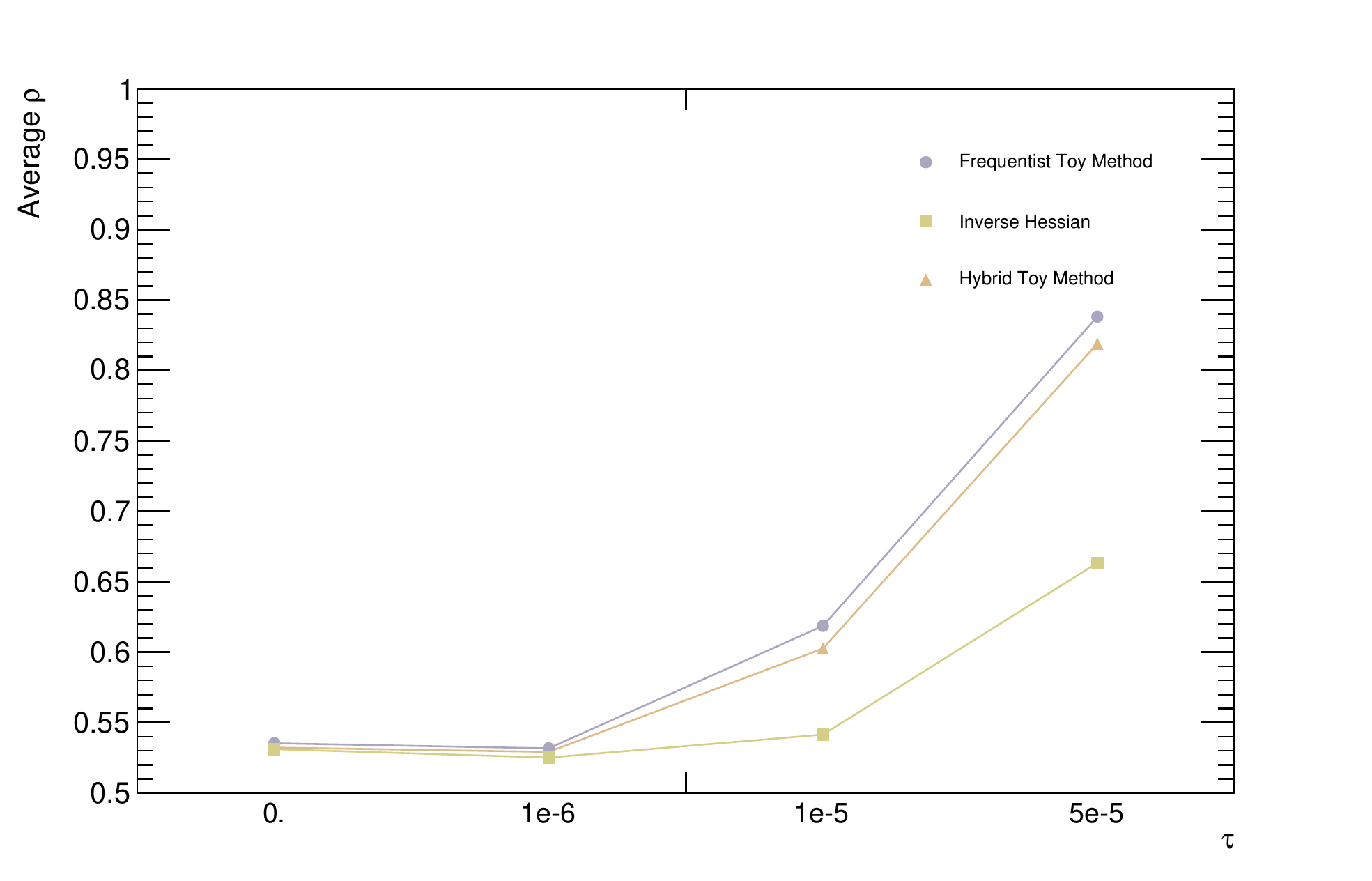}
  \caption{Double Gaussian distribution}
\end{subfigure}
\begin{subfigure}{.4\textwidth}
  \includegraphics[width=1\linewidth]{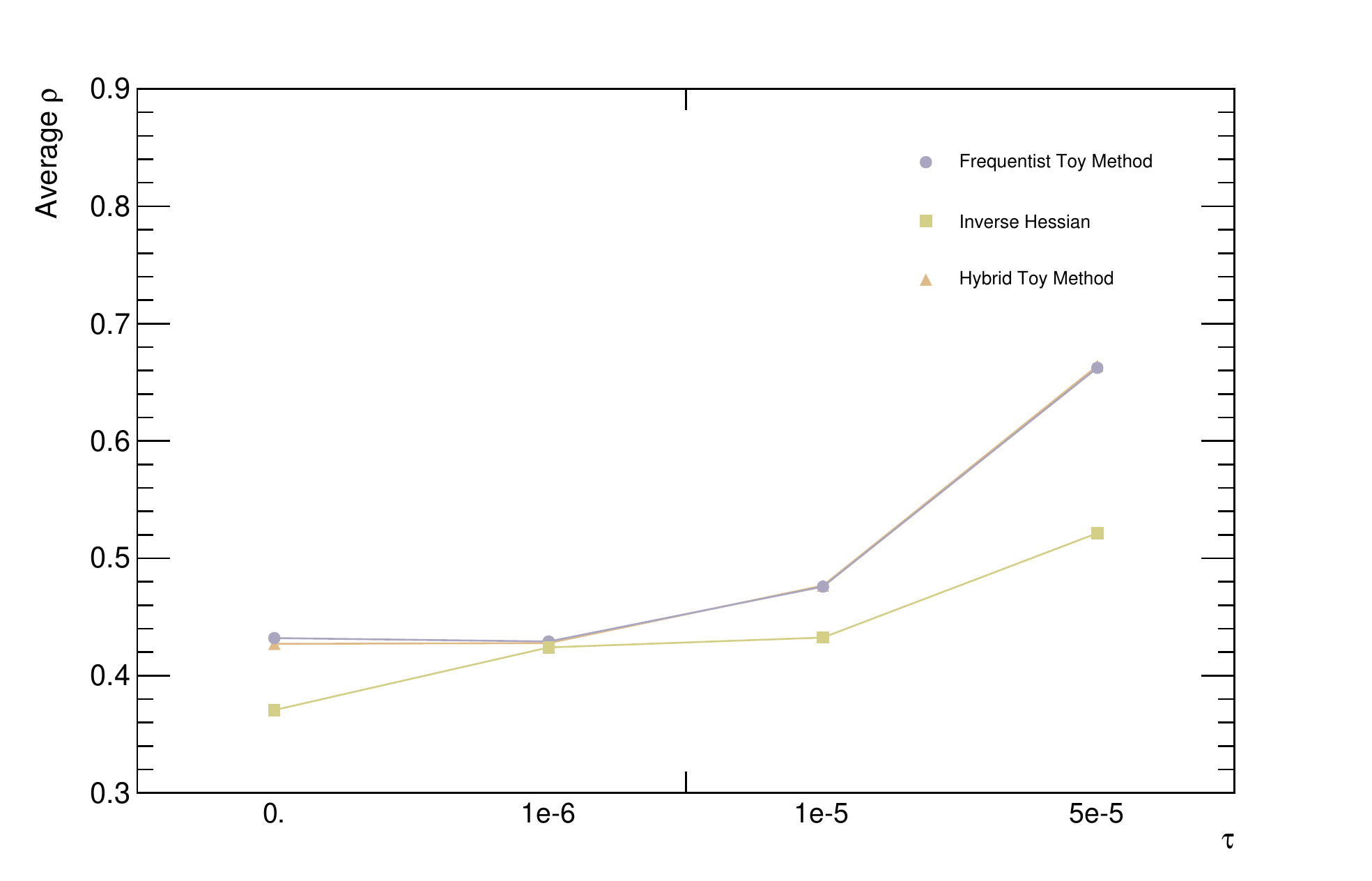}
  \caption{Exponential distribution}
\end{subfigure}

\caption{Average global correlation coefficient for a) the double Gaussian and b) the exponential distribution for all covariance estimation methods and various regularization strengths $\tau$}

\label{fig:globcorr}
\end{figure}

\begin{figure}[h!]
\centering

\begin{subfigure}{.4\textwidth}
  \includegraphics[width=1\linewidth]{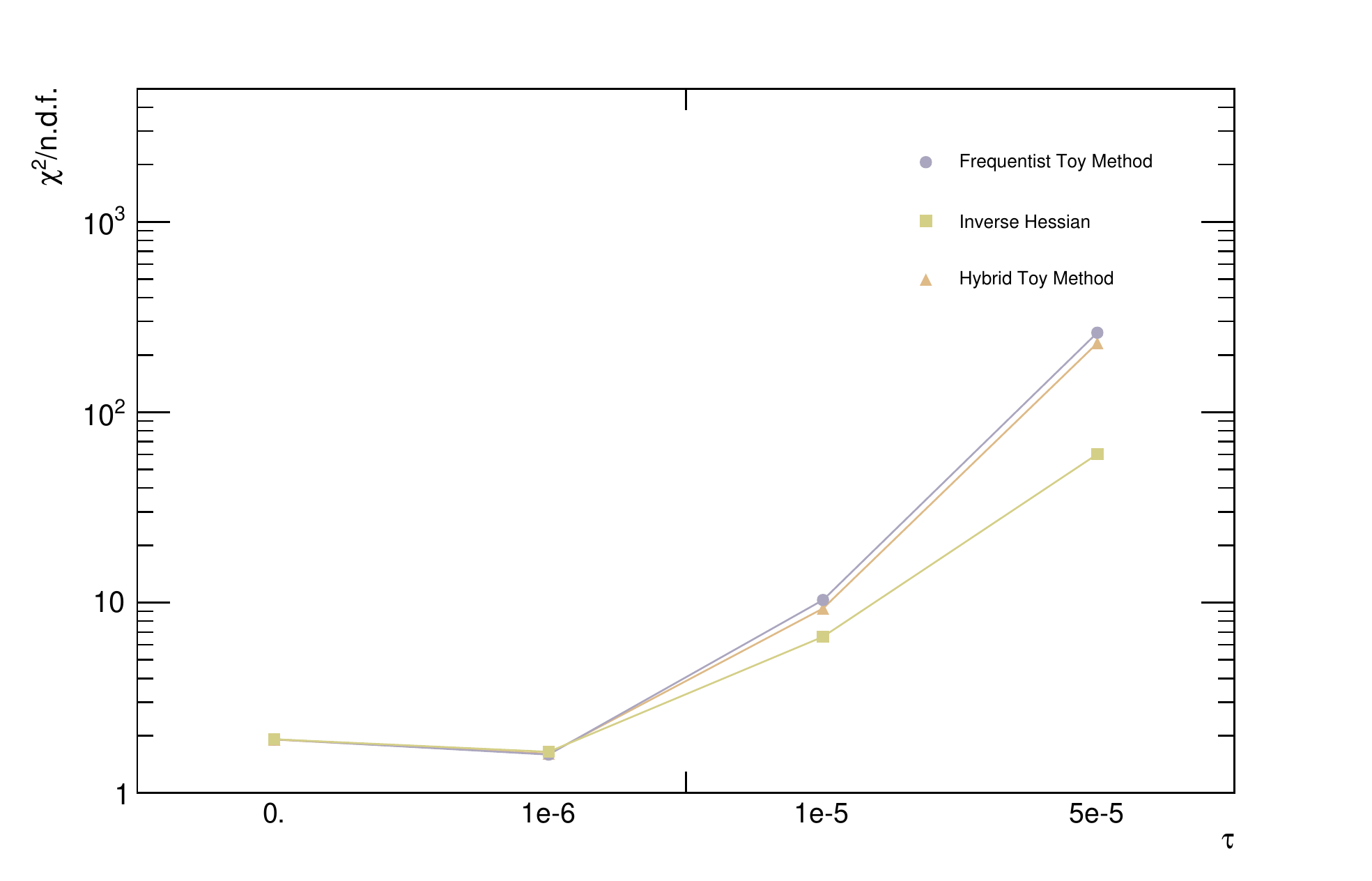}
  \caption{Double Gaussian distribution}
\end{subfigure}
\begin{subfigure}{.4\textwidth}
  \includegraphics[width=1\linewidth]{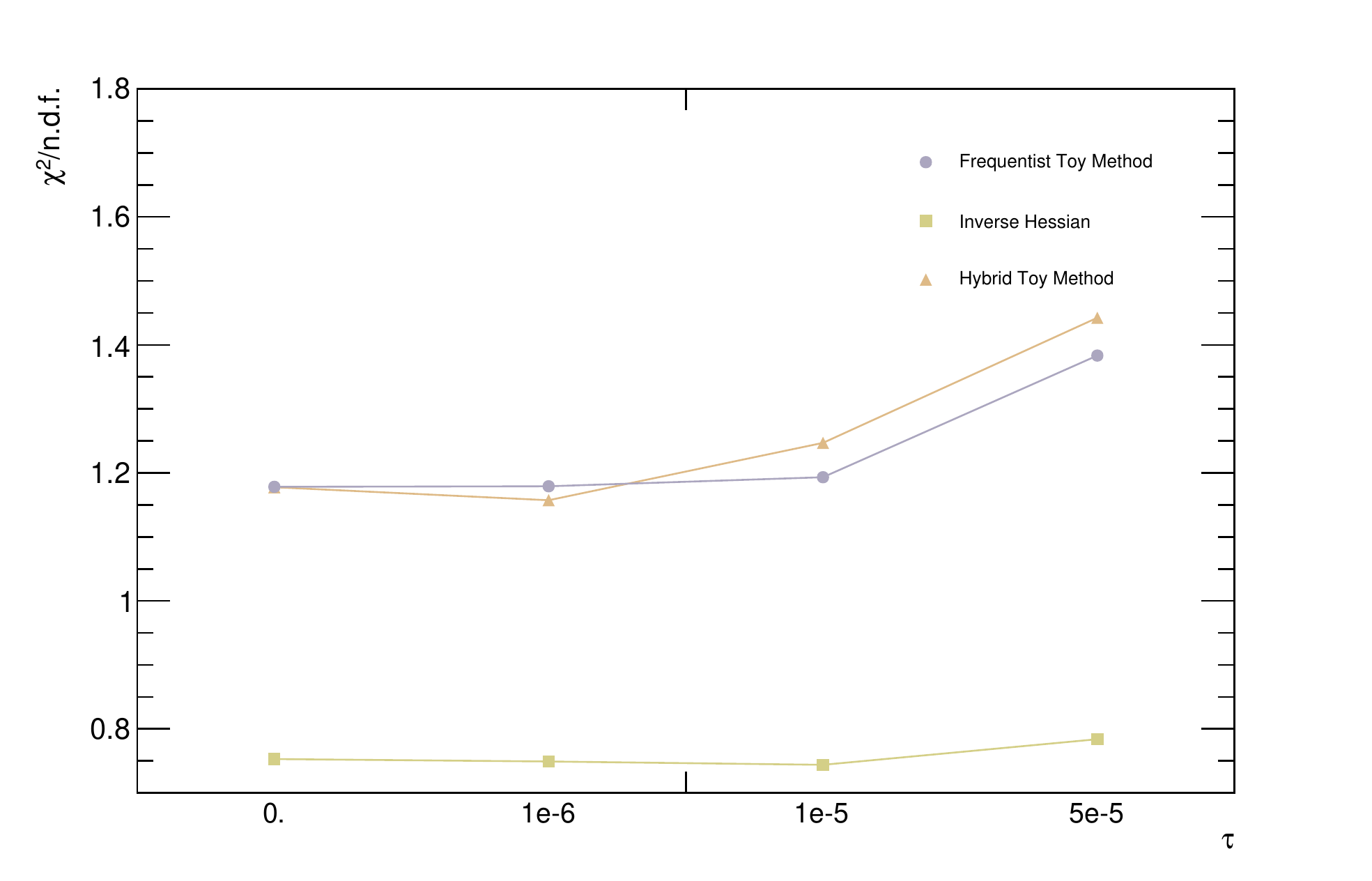}
  \caption{Exponential distribution}
\end{subfigure}

\caption{$\chi^{2}/\mbox{n.d.f.}$ for a) the double Gaussian and b) the exponential distribution for all covariance estimation methods and various regularization strengths $\tau$}

\label{fig:chindf}
\end{figure}

\newpage

\noindent
We observe that for the double Gaussian distribution with no regularization all of the methods seem to agree reasonably. However, as aspected, the inverse Hessian method diverges when regularization is introduced. The results on the exponential distribution show a complete disagreement of the inverse hessian method with the other two methods. The frequentist-bayes hybrid pseudo-experiments method agrees well on all quantities with the frequentist pseudo-experiments method.

\section{Summary and conclusions}\label{sec:conc}
A frequentist-Bayesian hybrid method has been presented for estimating covariances of unfolded distributions using pseudo-experiments. The method was compared with covariance estimation methods that use the unbiased Rao-Cramer Bound (RCB) and frequentist pseudo-experiments. The unfolding test scenarios showed that the RCB method, i.e. the inverse hessian method, diverges from the other two methods when regularization is introduced. This is because regularization inevitably introduces bias which needs to be taken into account in the RCB in the form of the bias matrix. However, the exponential example showed that even with no regularization the inverse hessian method can show disagreement with the other two methods. This could be caused by the non-parabolic shape of the likelihood around its maximum. The new hybrid method showed good agreement with the frequentist pseudo-experiments method for all distributions and different amounts of regularization. Lastly, unlike the frequentist pseudo-experiments method, the hybrid method does not need a specific likelihood definition which makes it suitable for a wider range of unfolding algorithms.

\bibliographystyle{plain}
\bibliography{references}
\end{document}